\documentclass[10pt,journal,compsoc,x11names]{IEEEtran}
\usepackage[switch]{lineno}
\usepackage{amsfonts}
\usepackage{amsmath, bm}
\usepackage{xspace}
\usepackage{enumitem}
\usepackage{amssymb}
\usepackage{booktabs}
\usepackage[hyphens]{url}

\usepackage{ragged2e}
\usepackage{hyperref}
\usepackage{multirow}
\usepackage{graphicx}
\usepackage{xcolor}
\usepackage{color}
\usepackage{colortbl}
\usepackage{tablefootnote}
\usepackage{pifont}
\usepackage{makecell}
\usepackage[most]{tcolorbox}
\usepackage{framed}
\usepackage{mdframed}
\usepackage{subfigure}
\usepackage{caption}
\usepackage{longtable}
\usepackage{float}
\usepackage{booktabs}
\usepackage{tikz}
\usepackage{fontawesome5}
\usepackage{url}
\usepackage{forest}
\usepackage[most]{tcolorbox}
\usepackage{makecell}

\usepackage{titlesec}
\titleformat{\paragraph}[runin]
  {\normalfont\normalsize\bfseries}{\theparagraph}{1em}{}
\titlespacing*{\paragraph}{0pt}{1ex}{1em}

\definecolor{perceptionColor}{RGB}{230,240,255}
\definecolor{explorationColor}{RGB}{255,240,230}
\definecolor{planningColor}{RGB}{230,255,230}
\definecolor{interactionColor}{RGB}{255,230,255}

\definecolor{perceptionBorder}{RGB}{100,150,255}
\definecolor{explorationBorder}{RGB}{255,150,100}
\definecolor{planningBorder}{RGB}{100,255,100}
\definecolor{interactionBorder}{RGB}{255,100,255}

\newcolumntype{H}{>{\setbox0=\hbox\bgroup}c<{\egroup}@{}}

\newcommand{\ignore}[1]{}

\definecolor{gold}{RGB}{205,133,63}
\definecolor{fGreen}{RGB}{34,139,34}
\definecolor{tOrange}{RGB}{255,215,0}
\definecolor{tBlue}{RGB}{135,206,250}
\definecolor{tPink}{RGB}{255,204,204}
\definecolor{tGreen}{RGB}{205,230,199}
\definecolor{tGold}{RGB}{255,215,0}

\usepackage[numbers,sort&compress]{natbib}

\ifCLASSINFOpdf
\else
\fi

\begin{document}
\title{A Survey on (M)LLM-Based GUI Agents}


\author{Fei Tang$^{1}$*, Haolei Xu$^{1}$*, Hang Zhang$^{1}$*, Siqi Chen$^{1}$*, Xingyu Wu$^{1}$*, Yongliang Shen$^{1, }$\\ Wenqi Zhang$^{1}$, Guiyang Hou$^{1}$, Zeqi Tan$^{1}$, Yuchen Yan$^{1}$, Kaitao Song$^{2}$, \\Jian Shao$^{1}$, Weiming Lu$^{1}$, Jun Xiao$^{1}$ and Yueting Zhuang$^{1}$

\IEEEcompsocitemizethanks{
\IEEEcompsocthanksitem Authors denoted by * contributed equally to this work. $^{1}$Zhejiang University, $^{2}$Microsoft Research Asia.
\IEEEcompsocthanksitem 
Email: syl@zju.edu.cn
\IEEEcompsocthanksitem 
Github Link: \url{https://github.com/zju-real/Awesome-GUI-Agents}
}
}

\markboth{}%
{Shell \MakeLowercase{\textit{et al.}}: Bare Advanced Demo of IEEEtran.cls for IEEE Computer Society Journals}

\IEEEtitleabstractindextext{%
\begin{abstract}
\justifying
Graphical User Interface (GUI) Agents have emerged as a transformative paradigm in human-computer interaction, evolving from rule-based automation scripts to sophisticated AI-driven systems capable of understanding and executing complex interface operations. This survey provides a comprehensive examination of the rapidly advancing field of (M)LLM-based GUI Agents, systematically analyzing their architectural foundations, technical components, and evaluation methodologies. We identify and analyze four fundamental components that constitute modern GUI Agents: (1) perception systems that integrate text-based parsing with multimodal understanding for comprehensive interface comprehension; (2) exploration mechanisms that construct and maintain knowledge bases through internal modeling, historical experience, and external information retrieval; (3) planning frameworks that leverage advanced reasoning methodologies for task decomposition and execution; and (4) interaction systems that manage action generation with robust safety controls. Through rigorous analysis of these components, we reveal how recent advances in large language models and multimodal learning have revolutionized GUI automation across desktop, mobile, and web platforms. We critically examine current evaluation frameworks, highlighting methodological limitations in existing benchmarks while proposing directions for standardization. This survey also identifies key technical challenges, including accurate element localization, effective knowledge retrieval, long-horizon planning, and safety-aware execution control, while outlining promising research directions for enhancing GUI Agents' capabilities. Our systematic review provides researchers and practitioners with a thorough understanding of the field's current state and offers insights into future developments in intelligent interface automation.
\end{abstract}

}

\maketitle

\IEEEdisplaynontitleabstractindextext

\IEEEpeerreviewmaketitle

\IEEEraisesectionheading{\section{Introduction}\label{sec:introduction}}

\IEEEPARstart{T}{he} ubiquity of graphical user interfaces (GUIs) in modern computing systems has led to an increasing demand for intelligent automation of user interface interactions \cite{wang2025guiagentsfoundationmodels,zhang2025largelanguagemodelbrainedgui,nguyen2024guiagentssurvey,yang2025magmafoundationmodelmultimodal}. Traditional automation approaches, primarily relying on rule-based scripts and screen recording/playback mechanisms, have proven inadequate for handling the complexity and dynamicity of modern interfaces \cite{nguyen2024guiagentssurvey}. The emergence of GUI Agents, which are autonomous systems capable of understanding, navigating, and interacting with digital interfaces \cite{zhang2025largelanguagemodelbrainedgui,nguyen2024guiagentssurvey,wang_mobile-agent_2024,zhang_appagent_2023,li_appagent_2024}, represents a significant advancement in human-computer interaction, offering the potential to bridge the gap between natural language instructions and complex interface operations \cite{wang_mobile-agent_2024,li_appagent_2024,cheng_seeclick_2024}.

The evolution of GUI automation technologies reflects the broader progress in artificial intelligence \cite{wang2025guiagentsfoundationmodels,zhang_ufo_2024,zhang2025largelanguagemodelbrainedgui,cheng_seeclick_2024,lin2024showuivisionlanguageactionmodelgui,jiang2025appagentxevolvingguiagents,he2024pcagentsleepai,zhang2025apiagentsvsgui,yang2025magmafoundationmodelmultimodal}. Early attempts at GUI automation were characterized by brittle, hand-crafted rules and simple pattern matching techniques, requiring extensive manual configuration and lacking adaptability to interface changes \cite{wang_mobile-agent_2024,zhang_appagent_2023,yang2025magmafoundationmodelmultimodal}. The introduction of computer vision and pattern recognition methods brought improved flexibility \cite{bai_uibert_2021} but still struggled with understanding context and handling dynamic elements \cite{wang_mobile-agent_2024,li_appagent_2024}. The recent integration of large language models (LLMs) \cite{touvron2023llamaopenefficientfoundation,bai2023qwentechnicalreport,openai2024gpt4technicalreport,glm2024chatglmfamilylargelanguage,deepseekai2025deepseekv3technicalreport} and multimodal language models (MLLMs) \cite{wang2024qwen2vlenhancingvisionlanguagemodels,liu2023visualinstructiontuning,bai2025qwen25vltechnicalreport,yang2025magmafoundationmodelmultimodal,lu2024deepseekvlrealworldvisionlanguageunderstanding,ye2024mplugowlmodularizationempowerslarge,chen2024internvlscalingvisionfoundation} marks a transformative moment in this field, enabling agents to comprehend natural language instructions, reason about complex tasks, and adapt to varying interface states with unprecedented sophistication \cite{cheng_seeclick_2024,lin2024showuivisionlanguageactionmodelgui,tang2025thinktwiceclickonce,shen2023hugginggptsolvingaitasks,zhang_ufo_2024}. Notably, with the release of OpenAI o1 and DeepSeek R1's \cite{deepseekai2025deepseekr1incentivizingreasoningcapability} successful replication of its capabilities, researchers have begun applying these reasoning models to GUI automation.

To address the limitations of earlier approaches and capitalize on recent (M)LLMs advancements \cite{yang2025magmafoundationmodelmultimodal,lu2024deepseekvlrealworldvisionlanguageunderstanding,ye2024mplugowlmodularizationempowerslarge}, contemporary GUI Agents have evolved into sophisticated systems built upon four fundamental technical components. These components work in concert to enable comprehensive interface understanding, knowledge management, intelligent planning, and reliable execution.


\begin{figure}
    \centering
    \includegraphics[width=1.0\linewidth]{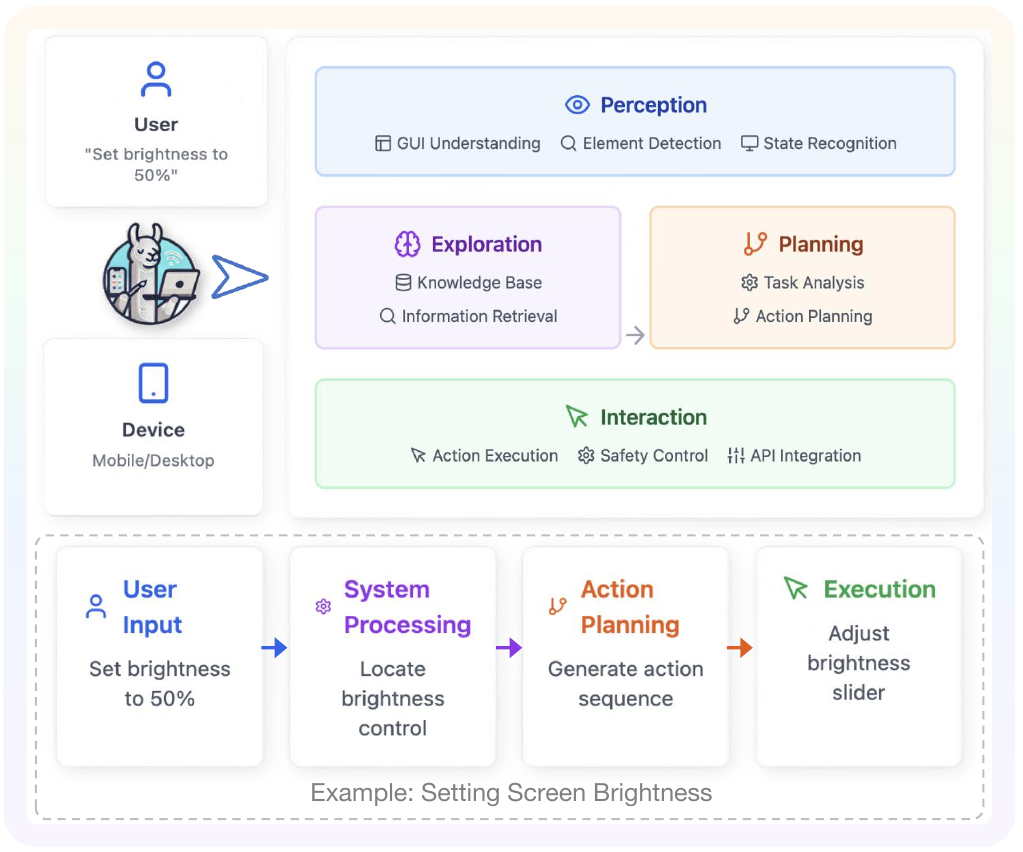}
    \caption{Overview of GUI Agents}
    \label{fig:enter-label}
\end{figure}

\begin{itemize}
    \item \textbf{Perception:} The agent's ability to understand interfaces has evolved from simple text parsing to sophisticated multimodal comprehension. Recent advances in perception can be categorized into text-based parsing (leveraging DOM/HTML structures) \cite{zhang_appagent_2023,li_appagent_2024,wang_mobile-agent_2024,xu_androidlab_2024} and multimodal understanding (utilizing MLLMs and specialized UI models) \cite{cheng_seeclick_2024,lin2024showuivisionlanguageactionmodelgui,nong_mobileflow_2024,tang2025thinktwiceclickonce,liu2025infiguiagentmultimodalgeneralistgui}. However, challenges remain in accurate element localization, dynamic content tracking, and resolution adaptation.
    \item \textbf{Exploration:} Knowledge acquisition and management have become crucial for effective GUI automation \cite{wen2024autodroidllmpoweredtaskautomation,zhang_appagent_2023}. Agents now build comprehensive knowledge bases incorporating internal understanding (UI functions, element properties), historical experience (task trajectories, skill libraries), and external information (API documentation, web resources) \cite{zhang_appagent_2023,zhu2024mobatwolevelagentsystem,agashe_agent_2024,lu2024guiodysseycomprehensivedataset}. The challenge lies in effectively organizing and retrieving this knowledge to guide decision-making.
    The integration of advanced reasoning frameworks has significantly improved agents' ability to handle complex tasks \cite{wei2023chainthoughtpromptingelicitsreasoning,zeng2022socraticmodelscomposingzeroshot}. Modern approaches leverage various thought chain methodologies and reactive frameworks for systematic planning \cite{yao2023reactsynergizingreasoningacting,yang2023mmreactpromptingchatgptmultimodal}. Key challenges include long-horizon planning, error recovery, and maintaining consistency across multiple interaction paths.
    \item \textbf{Interaction:} The action space has expanded from basic GUI operations to sophisticated API integrations while maintaining safety and reliability \cite{wu_os-copilot_2024}. Contemporary agents employ diverse strategies for action generation and execution, with an increasing focus on safety controls and error handling mechanisms.
\end{itemize}

These four components - perception, exploration, planning, and interaction - form an integrated pipeline that enables GUI Agents to process natural language instructions, understand interface contexts, plan appropriate actions, and execute them safely. The perception component provides the foundational understanding of the interface, while exploration builds and maintains the knowledge necessary for intelligent decision-making. The planning component leverages this knowledge to formulate effective strategies, which are then carried out through the interaction component. This architectural framework has proven effective across various applications, though each component continues to present unique challenges that drive ongoing research.

Modern GUI Agents operate across a diverse ecosystem of platforms, including desktop applications, mobile devices, and web browsers, each presenting unique challenges and opportunities. Desktop environments demand precise control over complex application interfaces \cite{he2024pcagentsleepai}, while mobile platforms require adaptation to touch-based interactions and varying screen sizes \cite{cheng_seeclick_2024,wang2024mobileagentbench}. Web browsers introduce additional complexity through dynamic content and diverse interface implementations \cite{deng2024mind2web,zhou2023webarena}. This variety of deployment contexts has driven the development of increasingly sophisticated agent architectures that can generalize across different platforms while maintaining reliability and safety.

To systematically evaluate the capabilities of GUI Agents and track progress in the field, researchers have developed various datasets and benchmarks. These evaluation frameworks span different platforms and scenarios, from mobile app navigation to web browsing and desktop software operation \cite{xu_androidlab_2024,rawles_androidworld_2024,cheng_seeclick_2024,wang2024mobileagentbench,deng2024mind2web}. Static benchmarks focus on assessing specific component capabilities, such as element localization accuracy or task planning efficiency \cite{rawles2024androidworld,cheng_seeclick_2024,li2024screenspot-pro}, while dynamic benchmarks evaluate end-to-end performance in real-world scenarios \cite{zhou2023webarena,deng2024mind2web,koh2024visualwebarena}. However, the development of comprehensive evaluation methodologies remains challenging due to the diversity of interfaces, the complexity of user interactions, and the need to assess both functional correctness and user experience aspects. The establishment of standardized evaluation frameworks is crucial for comparing different approaches and guiding future research directions.

This survey provides a comprehensive analysis of GUI Agents research, with the following contributions:

\begin{itemize}
    \item A systematic taxonomy of GUI Agents technologies, organized around the four core components: perception, exploration, planning, and interaction.
    \item A detailed examination of various technical approaches, from fundamental building blocks to state-of-the-art methodologies.
    \item A critical analysis of current evaluation methodologies and benchmarks, highlighting their strengths and limitations.
    \item An in-depth discussion of challenges and opportunities, providing insights for future research directions.
\end{itemize}

The remainder of this survey systematically explores these aspects, beginning with an overview of GUI Agents architectures in Section \ref{sec:overview}. Sections \ref{sec:perception} through \ref{sec:interaction} delve into the four core components: perception mechanisms, exploration strategies, planning frameworks, and interaction methods. Section \ref{sec:application} examines applications across different platforms, while Section \ref{sec:datasets} discusses evaluation methodologies and benchmarks. Section \ref{sec:future} analyzes challenges and future directions, followed by conclusions in Section \ref{sec:conclusion}. Throughout this comprehensive examination, we aim to provide researchers and practitioners with a thorough understanding of the current state of GUI Agents while highlighting promising directions for future research and development.

Through this survey, we seek to illuminate the rapid progress and remaining challenges in GUI Agents development, providing a foundation for future advances in this increasingly important field. As digital interfaces continue to evolve and proliferate, the role of GUI Agents in bridging the gap between human intent and computer execution becomes increasingly crucial, making this systematic examination of the field both timely and essential.
\section{Architecture}
\subsection{Overview}
\label{sec:overview}

\begin{figure*}
    \scriptsize
    \centering
\tikzset{
    basic/.style  = {draw, text width=.6cm, align=center, font=\sffamily, rectangle},
    root/.style   = {basic, rounded corners=5pt, thin, align=center, fill=gray!10, minimum height=10em, line width=1pt},
    pnode/.style = {basic, rounded corners, thin, align=center, draw=blue!60, line width=1pt, 
                   fill=blue!15, text width=7em},
    p1node/.style = {basic, rounded corners, thin, align=center, 
                    fill=blue!8, text width=31.5em, draw=blue!8, line width=1pt},
    enode/.style = {basic, rounded corners, thin, align=center, draw=blue!40!purple!60, line width=1pt, 
                   fill=blue!20!purple!15, text width=7em},
    e1node/.style = {basic, rounded corners, thin, align=center, 
                    fill=blue!10!purple!8, text width=31.5em, draw=blue!10!purple!8, line width=1pt},
    rnode/.style = {basic, rounded corners, thin, align=center, draw=purple!60!red!70, line width=1pt, 
                   fill=purple!30!red!15, text width=7em},
    r1node/.style = {basic, rounded corners, thin, align=center, 
                    fill=purple!15!red!8, text width=31.5em, draw=purple!15!red!8, line width=1pt},
    inode/.style = {basic, rounded corners, thin, align=center, draw=orange!70!brown!60, line width=1pt, 
                   fill=orange!25!brown!15, text width=7em},
    i1node/.style = {basic, rounded corners, thin, align=center, 
                    fill=orange!15!brown!8, text width=31.5em, draw=orange!15!brown!8, line width=1pt},
    edge from parent/.style={draw=gray!60, edge from parent fork right}
}

\begin{forest} for tree={
    grow=east,
    growth parent anchor=west,
    parent anchor=east,
    child anchor=west,
    edge={gray, thick},
    l sep=25pt,
    s sep=7pt,
    reversed=true,
    edge path={
        \noexpand\path [\forestoption{edge}]
        (!u.parent anchor) -- +(10pt,0) |- (.child anchor)\forestoption{edge label};
    },
}
[\rotatebox{90}{GUI Agents}, root, l sep=5mm,
    [Perception, pnode, l sep=10mm,
        [Text-based Parsing, pnode
            [DOM/HTML Analysis:\tiny~Tree Structure Parsing (WebArena'23 \cite{zhou2023webarena}); HTML Simplification (AutoDroid'22 \cite{wen_autodroid_2024}); Semantic Extraction (AiTW'23 \cite{rawles2024androidinthewild}), p1node]
            [XML Processing:\tiny~Layout Understanding (AppAgent'23 \cite{zhang_appagent_2023}); Element Hierarchy (MobileAgent'23 \cite{ding_mobileagent_2024}); Attribute Analysis (AssistGUI'23 \cite{gao2023assistgui}), p1node]
            [Tool Integration:\tiny~PaddleOCR Recognition (Mobile-Agent'24 \cite{wang_mobile-agent_2024}); IconNet Detection (AppAgent-v2'24 \cite{li_appagent_2024}); Visual Grounding (SoM'24 \cite{yang_set--mark_2023}), p1node]
        ]
        [Multimodal Perception, pnode
            [General MLLMs:\tiny~GPT-4V Direct Understanding (AppAgent'23); Claude-3 Screen Analysis (Agent S'24 \cite{agashe_agent_2024}); GPT-4o Game Scene (Cradle'24 \cite{tan_cradle_2024}), p1node]
            [Specialized Models:\tiny~UI-BERT Feature Learning (UIBert'21 \cite{bai_uibert_2021}); Position Prediction (CogAgent'23 \cite{hong_cogagent_2023}); Element Grounding (SeeClick'24 \cite{cheng_seeclick_2024}; UI-TARS'25 \cite{qin2025uitarspioneeringautomatedgui}); Dual System (Focus'25 \cite{tang2025thinktwiceclickonce}), p1node]
            [Tool Collaboration:\tiny~DINO+SAM Integration (SoM'24 \cite{yang_set--mark_2023}); Multi-tool Pipeline (OmniParser'24 \cite{lu_omniparser_2024}; Mobile-Agent-E'25 \cite{wang2025mobileagenteselfevolvingmobileassistant}); Hybrid Perception (Mobile-Agent-v2'24 \cite{wang_mobile-agent-v2_2024}); OmniParser-based Perception (AppAgentX'25 \cite{jiang2025appagentxevolvingguiagents}), p1node]
        ]]
    [Exploration, enode, l sep=10mm,
        [Knowledge Types, enode
            [Internal Knowledge:\tiny~UI State Mapping (AutoDroid'22 \cite{wen_autodroid_2024}); Element Function Repository (AppAgent'23 \cite{zhang_appagent_2023}); Interaction Graph (MobA'24 \cite{zhu2024mobatwolevelagentsystem}), e1node]
            [Historical Records:\tiny~Success Trajectories (Agents'24 \cite{zhou2023agentsopensourceframeworkautonomous}); CoT Reflection (Reflexion'23 \cite{shinn2023reflexionlanguageagentsverbal}); Skill Database (VOYAGER'23 \cite{wang2023voyageropenendedembodiedagent}), e1node]
            [External Sources:\tiny~API Documentation (OS-Copilot'24 \cite{wu_os-copilot_2024}); Web Search Integration (Mobile-Agent'24 \cite{wang_mobile-agent_2024}); Manual Analysis (Agent S'24 \cite{agashe_agent_2024}), e1node]
        ]
        [Knowledge Engineering, enode
            [Static Building:\tiny~Page Transition Graph (AutoDroid'22 \cite{wen_autodroid_2024}); Task-oriented Mining (AppAgent'23 \cite{zhang_appagent_2023}); Rule Extraction (KnowAgent'24 \cite{zhu2024knowagentknowledgeaugmentedplanningllmbased}), e1node]
            [Dynamic Learning:\tiny~Real-time Memory Update (MobA'24 \cite{zhu2024mobatwolevelagentsystem}); Adaptive Knowledge Base (Agent S'24 \cite{agashe_agent_2024}); Tool Evolution ((OS-Copilot'24 \cite{wu_os-copilot_2024}), e1node]
            [Knowledge Retrieval:\tiny~Embedding Similarity Search (AutoDroid'22 \cite{wen_autodroid_2024}); Context-based Query (VOYAGER'23 \cite{wang2023voyageropenendedembodiedagent}); State Matching (WKM'24 \cite{qiao2024agentplanningworldknowledge}), e1node]
        ]]
    [Planning, rnode, l sep=10mm,
        [Reasoning, rnode
            [Sequential Reasoning:\tiny~Zero-shot Prompting (ZS-CoT'22 \cite{zeng2022socraticmodelscomposingzeroshot}); Few-shot Learning (FS-CoT'22 \cite{wei2023chainthoughtpromptingelicitsreasoning}); Self-Consistency Verification (SC'23 \cite{wang2023selfconsistencyimproveschainthought}), r1node]
            [Structured Reasoning:\tiny~Tree Search (ToT'23 \cite{yao2023treethoughtsdeliberateproblem}); Monte Carlo Planning (ReST-MCTS'24 \cite{zhang2024restmctsllmselftrainingprocess}); Graph Reasoning (GoT'24 \cite{besta2024graph}), r1node]
            [Interactive Reasoning:\tiny~Action-Observation Loop (ReAct'23 \cite{yao2023reactsynergizingreasoningacting}); Action Chain (CoA'24 \cite{zhang_you_2024}); Thought Integration (CoAT'24 \cite{zhang_android_2024}); InfiGUIAgent: Native Reasoning and Reflection (InfiGUIAgent'25 \cite{liu2025infiguiagentmultimodalgeneralistgui}); System 2 Reasoning (UI-TARS'25 \cite{qin2025uitarspioneeringautomatedgui}), r1node]
        ]
        [Task Planning, rnode
            [Iterative Planning:\tiny~Step-by-step Execution (AppAgent'23 \cite{zhang_appagent_2023}); Self-Planning (Mobile-Agent'24 \cite{wang_mobile-agent_2024}); DAG-based Planning (OS-Copilot'24 \cite{wu_os-copilot_2024}), r1node]
            [Task Decomposition:\tiny~Hierarchical Planning (Agent S'24 \cite{agashe_agent_2024}); Subtask Management (MobA'24 \cite{zhu2024mobatwolevelagentsystem}); Graph-based Coordination (VillagerAgent'24 \cite{dong2024villageragentgraphbasedmultiagentframework}), r1node]
        ]
        [Verification, rnode
            [Self-Assessment:\tiny~Task Progress Tracking (VOYAGER'23 \cite{wang2023voyageropenendedembodiedagent}); Strategy Reflection (Reflexion'24 \cite{shinn2023reflexionlanguageagentsverbal}); Plan Adjustment (UFO'24 \cite{zhang_ufo_2024}), r1node]
            [External Feedback:\tiny~Environment State Check (ReAct'23 \cite{yao2023reactsynergizingreasoningacting}); Tool Response Analysis (OS-Copilot'24 \cite{wu_os-copilot_2024}); User Confirmation (UFO'24 \cite{zhang_ufo_2024}), r1node]
        ]]
    [Interaction, inode, l sep=10mm,
        [Action Generation, inode
            [Interface Analysis:\tiny~Screen Understanding (ScreenAgent'24 \cite{2024arXiv240207945N}); Element Detection (CoCo-Agent'24 \cite{ma_coco-agent_2024}); Navigation Logic (WebVoyager'24 \cite{2024arXiv240113919H}), i1node]
            [Generation Methods:\tiny~Memory-guided Generation (Mobile-Agent'24); Plan-based Execution (OS-Copilot'24 \cite{wu_os-copilot_2024}); State Space Search (LASER'24 \cite{2023arXiv230908172M}), i1node]
        ]
        [Action Space, inode
            [User Simulation:\tiny~Basic UI Operations (AppAgent'23 \cite{zhang_appagent_2023}); Game Controls (Cradle'24 \cite{tan_cradle_2024}); Complex Workflows (OS-ATLAS'24 \cite{wu_os-atlas_2024}); Evolving Actions (AppAgentX'25 \cite{jiang2025appagentxevolvingguiagents}), i1node]
            [API Integration:\tiny~System Function Calls (OS-Copilot'24 \cite{wu_os-copilot_2024}); Tool API Wrapper (SheetCopilot'24 \cite{li2023sheetcopilotbringingsoftwareproductivity}); Service Integration (UFO'24 \cite{zhang_ufo_2024}), i1node]
        ]
        [Safety Control, inode
            [Operation Safety:\tiny~Sensitive Action Detection (UFO'24 \cite{zhang_ufo_2024}); Permission Management (Mobile-Agent-v2'24 \cite{wang_mobile-agent-v2_2024}); Risk Assessment (OS-Copilot'24 \cite{wu_os-copilot_2024}), i1node]
            [Error Resolution:\tiny~Exception Handling (Agent S'24 \cite{agashe_agent_2024}); Recovery Planning (OS-Copilot'24 \cite{wu_os-copilot_2024}); Fallback Strategies (UFO'24 \cite{zhang_ufo_2024}), i1node]
        ]]
    ]
]
\end{forest}
    \caption{A Comprehensive Taxonomy of GUI Agents Components}
    \label{fig:taxonomy}
\end{figure*}
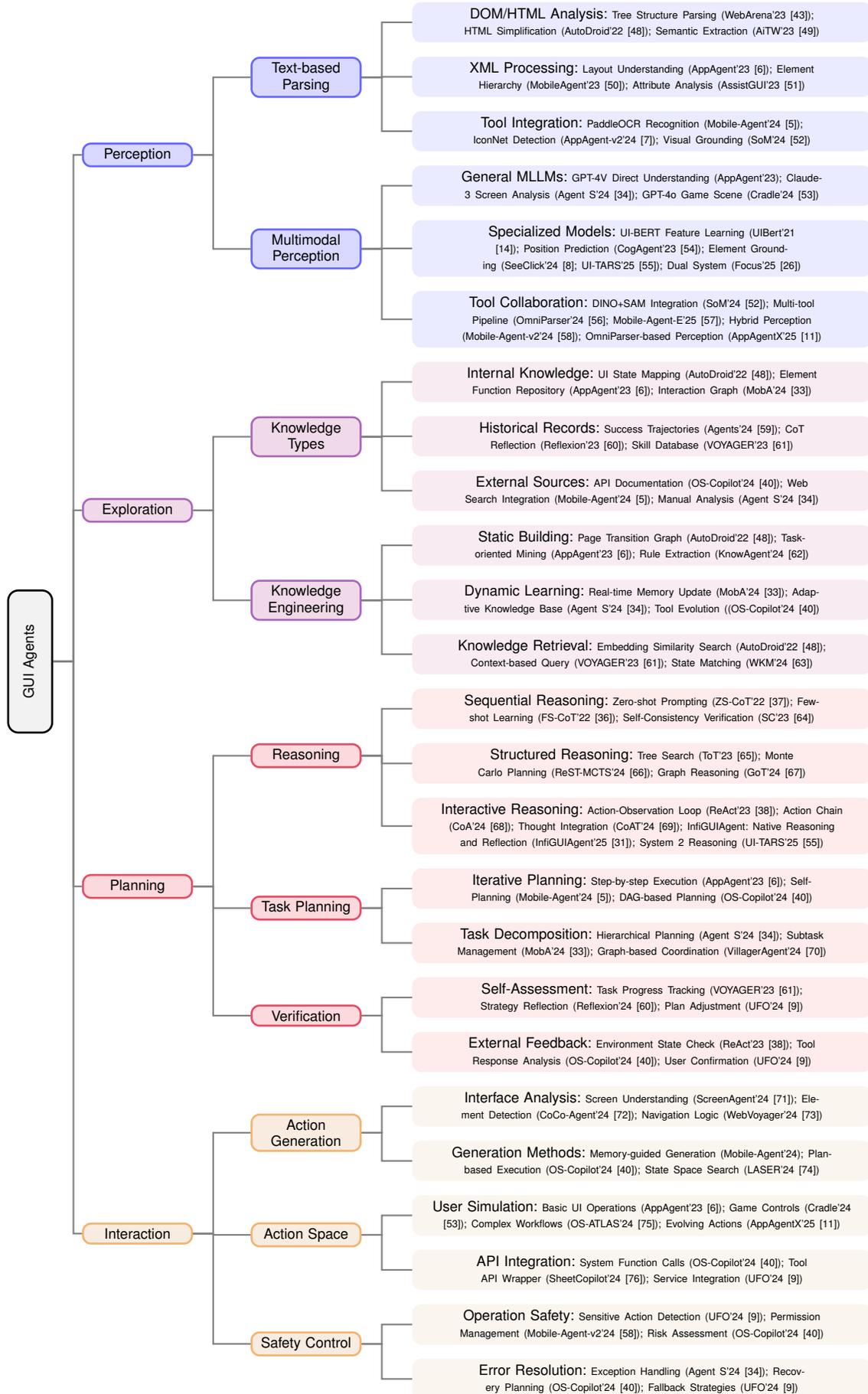

In this section, we present a systematic framework for understanding the architecture of contemporary GUI Agents. As illustrated in Figure \ref{fig:taxonomy}, we decompose these intelligent systems into four fundamental modules: perception, exploration, planning, and interaction. This functional decomposition not only provides analytical clarity but also reflects the natural progression of information processing within GUI Agents from sensory input to knowledge acquisition, decision making, and ultimately physical action.

Through comprehensive analysis of recent literature (2021-2025), we classify GUI Agent architectures into four primary categories: General MLLM-based approaches that leverage off-the-shelf multimodal models; Training MLLM approaches that specifically fine-tune multimodal models for GUI tasks; Multi-Agent frameworks that distribute specialized functions across collaborative agents; and LLM-Based methodologies that rely primarily on text-based representation of interfaces. Figure \ref{fig:222} tracks the chronological development and publication volume across these architectural paradigms, revealing the field's rapid evolution and growing research interest.

The information flow within GUI Agents begins with the perception module, which serves as the system's primary sensory interface (\S\ref{sec:perception}). This module encompasses text-based perception (parsing DOM trees or XML structures) (\S\ref{sec:text-based}), multimodal perception (directly processing screenshots) (\S\ref{sec:mllmbased}), and specialized tools for screen understanding (OCR, object detection) (\S\ref{sec:tool}). Similar to human visual processing, perception provides the fundamental interface understanding upon which all subsequent processes depend.
Building on this perceptual foundation, the exploration module enables GUI Agents to acquire, organize, and utilize diverse knowledge sources (\S\ref{sec:exploration}). We identify three distinct knowledge acquisition strategies (\S\ref{sec:Categorization}): internal exploration (examining interface structures and functionalities), external exploration (retrieving supplementary information from documentation or web resources), and historical exploration (leveraging past interactions and successful trajectories). We also discussed how GUI Agents utilize knowledge in (\S\ref{sec:Operation}). This comprehensive knowledge management system enables agents to navigate complex interfaces with contextual awareness and adaptability.

The planning module constitutes the cognitive core of GUI Agents, enabling systematic reasoning about tasks and decision-making (\S\ref{sec:planning}). Within our taxonomy, we analyze planning capabilities across three dimensions: the underlying reasoning frameworks (whether based on LLMs (\S\ref{sec:llmreason}), MLLMs (\S\ref{sec:mllmreason}), or advanced reasoning models like o1-series systems (\S\ref{sec:o1likes}), task planning methodologies (including iterative and decomposition-based approaches), and verification mechanisms for ensuring plan reliability (\S\ref{sec:taskplanning}). This sophisticated planning architecture allows agents to handle multi-step tasks while maintaining consistency and adapting to environmental feedback.

Finally, the interaction module enables GUI Agents to execute planned actions within the target environment (\S\ref{sec:interaction}). We present a systematic classification framework based on action types, generation strategies, and safety control mechanisms. These interaction capabilities represent the culmination of the agent's processing pipeline, translating internal representations and plans into concrete actions that manipulate the interface.

Together, these four modules—perception, exploration, planning, and interaction—form an integrated information processing pipeline that enables GUI Agents to receive user instructions, comprehend interface contexts, develop appropriate action plans, and execute them reliably. The perception module provides the foundational understanding of the environment, while exploration builds and maintains the knowledge necessary for intelligent decision-making. The planning module leverages this knowledge to formulate effective strategies, which are then implemented through the interaction module. This architectural framework has demonstrated effectiveness across diverse applications, though each component presents unique research challenges that continue to drive innovation in the field.

In the following sections, we examine each of these core components in detail, analyzing their theoretical foundations, implementation approaches, and technical challenges. Through this systematic decomposition, we aim to provide a comprehensive understanding of current GUI Agent architectures while highlighting opportunities for future architectural innovations.

\subsection{Perception}
\label{sec:perception}

\begin{figure}[htbp]
    \centering
    \includegraphics[width=0.48\textwidth]{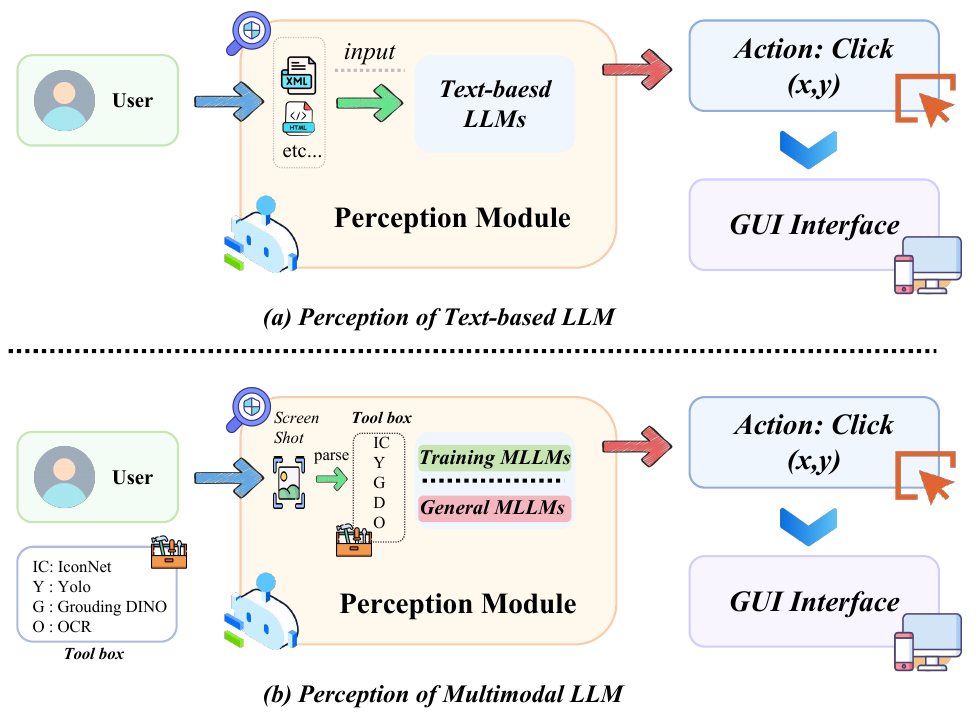}
    \caption{Comparison between Text-based and MLLM-based GUI Agents. Text-based agents process textual information such as XML or DOM trees, while MLLM-based agents utilize screen screenshots. Some MLLM-based GUI Agents additionally employ specialized tools for element localization, including IconNet, YOLO, Grounding DINO, and OCR.}
    \label{fig:image_label}
\end{figure}

The eye is an important part of the human body, capable of collecting visual information that is then reflected to the brain for decision-making. Similarly, the perception of a GUI Agents is also a crucial module that acquires multimodal information from the screen or the internal structure of the current screen, thereby aiding in task planning and execution. After extensive literature research, we can categorize the perception mechanisms of GUI Agents into text-based and multimodal-based approaches, as shown in Figure \ref{fig:image_label}. For instance, text-based GUI Agents \cite{zheng_gpt-4vision_2024, zhou_webarena_2024, xu_androidlab_2024, zhang_appagent_2023} perceive the current screen by parsing web HTML or system-provided XML files into structured textual representations. In contrast, multimodal GUI Agents \cite{li_ferret-ui_2024, hong_cogagent_2023} process screenshots directly, utilizing specialized models pre-trained and fine-tuned on GUI grounding data to accurately localize and interpret interface elements from visual information. 

In the following sections, we delve deeper into these two perception mechanisms: Section (\S\ref{sec:text-based}) provides a detailed examination of text-based perception methods, elucidating how these techniques leverage structured data to understand interface semantics; Section (\S\ref{sec:mllmbased}) explores multimodal perception technologies that enable direct visual understanding of interfaces. Additionally, we investigate the emerging area of tool-enhanced perception (\S\ref{sec:tool}), exploring how specialized utilities and frameworks augment a GUI Agent's ability to interpret complex screen elements beyond the limitations of either text-based or vision-based approaches alone.
\subsubsection{Text-based LLM GUI Agent}
\label{sec:text-based}
Text-based GUI Agents rely solely on textual information as input and predominantly utilize closed-source large language models such as GPT-4o, Claude-3.5-Sonnet. For instance, AutoDroid \cite{wen2024autodroidllmpoweredtaskautomation} proposed its own perception method, parsing the UI page into HTML format for input to the large model. AiTW \cite{rawles_android_2023} employed the PaLM \cite{chowdhery2023palm} model, with inputs also resembling HTML structures. Notably, it also utilized OCR tools to recognize icons and text for information enhancement. MobileAgent \cite{ding_mobileagent_2024} is a text-based GUI Agent that employs standard program operations for context learning, enabling the model to understand what to do next. Its input is a prompt that includes task objectives, operation history, and the current page's DOM information. Like AiTW, it also uses OCR and icon detection tools. AssistGUI \cite{gao_assistgui_2024} conducted experiments related to productivity tools using the closed-source large model GPT-3.5 \cite{openai2021chatgpt}. Specifically, it learns from instructional videos for productivity tools such as Photoshop and Premiere Pro. For AssistGUI \cite{gao_assistgui_2024}, it employs various tools like Google OCR and YOLOv8 to parse the current screen and integrate the information into the prompt. WebArena \cite{zhou_webarena_2024} is a Web Agent that operates on browsers, creating a virtual environment focused on real-world scenarios by utilizing task instructions and DOM trees derived from the structure of current webpages.

However, Text-based GUI Agents still face several limitations as below: 
\begin{itemize}
\item \textbf{Redundancy:} Text-based GUI Agents must process extensive HTML/DOM trees or XML files, much of which contains redundant information not relevant to the current task \cite{zhang2025apiagentsvsgui,qin2025uitarspioneeringautomatedgui}.
\item \textbf{Noise:} Textual representations often include implementation details, styling information, and other noise that can distract from the core interactive elements \cite{wang_mobile-agent_2024,qin2025uitarspioneeringautomatedgui}.
\item \textbf{Availability:} On various platforms, comprehensive textual information (complete XML or HTML structures) may be inaccessible or incomplete, limiting the agent's understanding of the interface \cite{zhang2025apiagentsvsgui,qin2025uitarspioneeringautomatedgui}.
\end{itemize}

To address these challenges, researchers have begun using multimodal large language models for GUI screen understanding \cite{xu_androidlab_2024, wang_mobile-agent_2024,yang2025magmafoundationmodelmultimodal,lin2024showuivisionlanguageactionmodelgui,cheng_seeclick_2024}, which enables agents to directly interpret visual elements of interfaces, circumventing the limitations of purely textual representations and enhancing their ability to navigate and interact with graphical environments.

\subsubsection{Multimodal LLM-based GUI Agent}
\label{sec:mllmbased}
Due to the limitations of text-based LLM GUI Agents, researchers increasingly leverage MLLMs to enhance agent perception. Recent studies on multimodal GUI Agents can be categorized into three approaches:
(1) \textit{General MLLM Approaches}: These directly utilize general-purpose models like GPT-4V for screen understanding without specialized training for GUI tasks.
(2) \textit{Training MLLM Approaches}: These involve training dedicated multimodal models for deeper UI comprehension, exemplified by CogAgent \cite{hong_cogagent_2023}.
(3) \textit{Tool-augmented MLLM Approaches}: These employ multimodal models enhanced with supplementary tools such as OCR and Grounding DINO \cite{liu2024groundingdinomarryingdino} for comprehensive UI interpretation.

\paragraph{\textbf{Understanding UI with General Multimodal Large Models}}
With the advancement of multimodal technology, powerful models such as GPT-4V, GPT-4o, and Claude-3.5 Sonnet have emerged. Most research leveraging these general multimodal large models for screen understanding relies on closed-source systems.

Several notable implementations demonstrate this approach. AppAgent \cite{zhang_appagent_2023} employs GPT-4 to perceive virtual environments through screenshots and XML files, storing learned information in relevant documents. OS-Copilot \cite{wu_os-copilot_2024} similarly utilizes GPT-4 in its three-module system (planner, configurator, executor) but distinguishes itself by generating code segments to operate software, such as using python-ppt for PowerPoint manipulation. The work by \cite{zheng_gpt-4vision_2024} represents a web-focused implementation that leverages GPT-4 for perceiving screens through element attributes, text options, and image annotations. Mobile-Agent \cite{wang_mobile-agent_2024} addresses the limitations of text-based GUI Agents' dependence on XML and HTML by integrating GPT-4 with supplementary tools like OCR and Grounding DINO \cite{liu2024groundingdinomarryingdino}. Similarly, UFO [26] combines MLLMs for Windows platform operations, while Mobile-Agent-V2 \cite{wang_mobile-agent-v2_2024} expands on its predecessor by implementing a multi-agent framework that utilizes both text-based GPT-4 and GPT-4V.

Recent implementations have further specialized these approaches. Cradle \cite{tan_cradle_2024} creates a game-playing framework using GPT-4o, enhanced with SAM \cite{kirillov2023segment} and Grounding DINO for improved perception. AppAgent v2 \cite{li_appagent_2024} builds upon AppAgent by enhancing UI understanding capabilities through XML processing and integrating OCR and detection models. Agent S \cite{agashe_agent_2024} specifically targets positioning capabilities using GPT-4o and Claude-3.5-Sonnet, incorporating PaddleOCR \cite{du2020ppocrpracticalultralightweight} for screenshot understanding. Mobile-Agent-E \cite{wang2025mobileagenteselfevolvingmobileassistant} employs GPT-4o for planning while utilizing OCR, icon grounding, and icon captioning for GUI comprehension. The Operator framework \cite{openai_2025_operator} leverages GPT-4o's visual capabilities alongside reinforcement learning for screen interaction, while Deep Research integrates OpenAI's o3 model to enable efficient multi-step reasoning for task completion.

GUI Agents based on general large models often integrate various tools to enhance their capabilities. However, these approaches face several significant limitations:

\begin{itemize}
    \item \textit{Localization Precision}: Despite claims of element localization capabilities \cite{zheng_gpt-4vision_2024, li2024ferretui2masteringuniversal, cheng_seeclick_2024}, general MLLMs struggle with the precise identification and targeting of specific GUI components, particularly in complex or densely populated interfaces.
    \item \textit{Domain-Specific Understanding}: GUI interfaces employ unique visual languages, interaction patterns, and structural hierarchies that differ significantly from natural images. General models lack sufficient pre-training on GUI-specific datasets, resulting in limited comprehension of interface semantics, component relationships, and functional affordances.
    \item \textit{API Dependencies}: Most implementations rely on closed-source models requiring API access, introducing substantial computational costs and potential rate limitations.
\end{itemize}
To address these limitations, researchers have developed specialized approaches focused on training multimodal large models specifically for GUI understanding.
\paragraph{\textbf{Understanding UI with Trained Multimodal Large Models}}
To address these fundamental limitations, researchers have developed training models specifically for GUI understanding.

 Early work based on multimodal training to understand screens, such as UIBert \cite{bai_uibert_2021}, is a Transformer-based \cite{vaswani2023attentionneed} model that learns general feature representations of UI interfaces through UI images, text, and view distributions, and is designed with five pre-training tasks to achieve a general model. Another early work is META-GUI \cite{sun_meta-gui_2022}, which uses Faster R-CNN \cite{ren2016fasterrcnnrealtimeobject} to extract image features from UI pages and BERT \cite{devlin2019bertpretrainingdeepbidirectional} to extract text features, merging different modal features for training.

 However, existing multimodal models still face issues with inaccurate UI localization and are affected by resolution, leading to potential failure in recognizing smaller icons \cite{hong_cogagent_2023, cheng_seeclick_2024}. CogAgent \cite{hong_cogagent_2023} addressed localization issues by pre-training to output element coordinates and fusing high-resolution and low-resolution images. SeeClick \cite{cheng_seeclick_2024} constructed the multi-platform ScreenSpot dataset, training with Qwen2-VL \cite{wang2024qwen2} to improve screen perception. CoCo-Agent \cite{ma_coco-agent_2024} leveraged LLaVA \cite{liu2024visual} to output precise element coordinates while incorporating OCR and icon recognition tools. MobileFlow \cite{nong_mobileflow_2024} tackled Chinese understanding and privacy concerns using ViT \cite{dosovitskiy2021imageworth16x16words} and LayoutLMv3 \cite{huang2022layoutlmv3pretrainingdocumentai} for image encoding with an MoE-based \cite{bi2024deepseek} LLM. AutoGLM \cite{liu_autoglm_2024} separated localization and planning tasks for Android and Web interfaces, proposing self-evolution methods using ChatGLM \cite{glm2024chatglm}. OS-ATLAS \cite{wu_os-atlas_2024} utilized large-scale datasets with screen screenshots, element instructions, and coordinates for GUI perception. ShowUI \cite{lin2024showuivisionlanguageactionmodelgui} employed a lightweight vision-language-action framework for element localization. Aguvis \cite{xu2024aguvisunifiedpurevision} enhanced localization accuracy through pure vision architecture and large-scale GUI dataset pre-training.  Aria-UI \cite{yang2024ariauivisualgroundinggui} leverages a Mixture-of-Experts (MoE) approach as a purely visual multimodal model, strengthening contextual awareness and localization through text-image hybrid operation histories. InfiGUIAgent \cite{liu2025infiguiagentmultimodalgeneralistgui} proposes a two-stage supervised fine-tuning paradigm: the initial stage focuses on foundational visual-semantic understanding, while the subsequent stage enhances multi-step reasoning capabilities to optimize GUI interaction logic. Notably, InfiGUIAgent \cite{liu2025infiguiagentmultimodalgeneralistgui} introduces a hierarchical expectation-reflection reasoning framework coupled with dual-phase training mechanisms to deepen GUI semantic parsing.  PC-Agent \cite{he2024pcagentsleepai} improved localization robustness by combining Molmo \cite{deitke2024molmopixmoopenweights} multimodal capabilities with external feedback from system APIs for self-verification. UI-TARS \cite{qin2025uitarspioneeringautomatedgui} enhances screen perception by utilizing extensive GUI corpora for training.
 AppAgentX \cite{jiang2025appagentxevolvingguiagents} perceives screens by using OmniParser \cite{lu_omniparser_2024} to detect UI elements in screenshots before passing the annotated images to an LLM for action planning.

Despite these specialized approaches, training models specifically for GUI understanding still faces several notable limitations:
\begin{itemize}
    \item \textit{Cross-Platform Inconsistency}: GUI elements vary significantly in style, structure, and interaction patterns across web, mobile, and desktop platforms, making it difficult for a single model to generalize effectively.
    \item \textit{Resolution Constraints}: GUI images typically require high resolution for accurate parsing, yet many models struggle to identify and interact with small icons and elements crucial for completing tasks.
    \item \textit{Training Resource Intensity}: Developing dedicated GUI Agents demands extensive computational resources and specialized datasets that are costly to create and maintain.
\end{itemize}
\subsubsection{Multi-Tool Collaborative UI Parsing}
\label{sec:tool}
\begin{figure*}[htbp]
    \centering
    \includegraphics[width=1\textwidth]{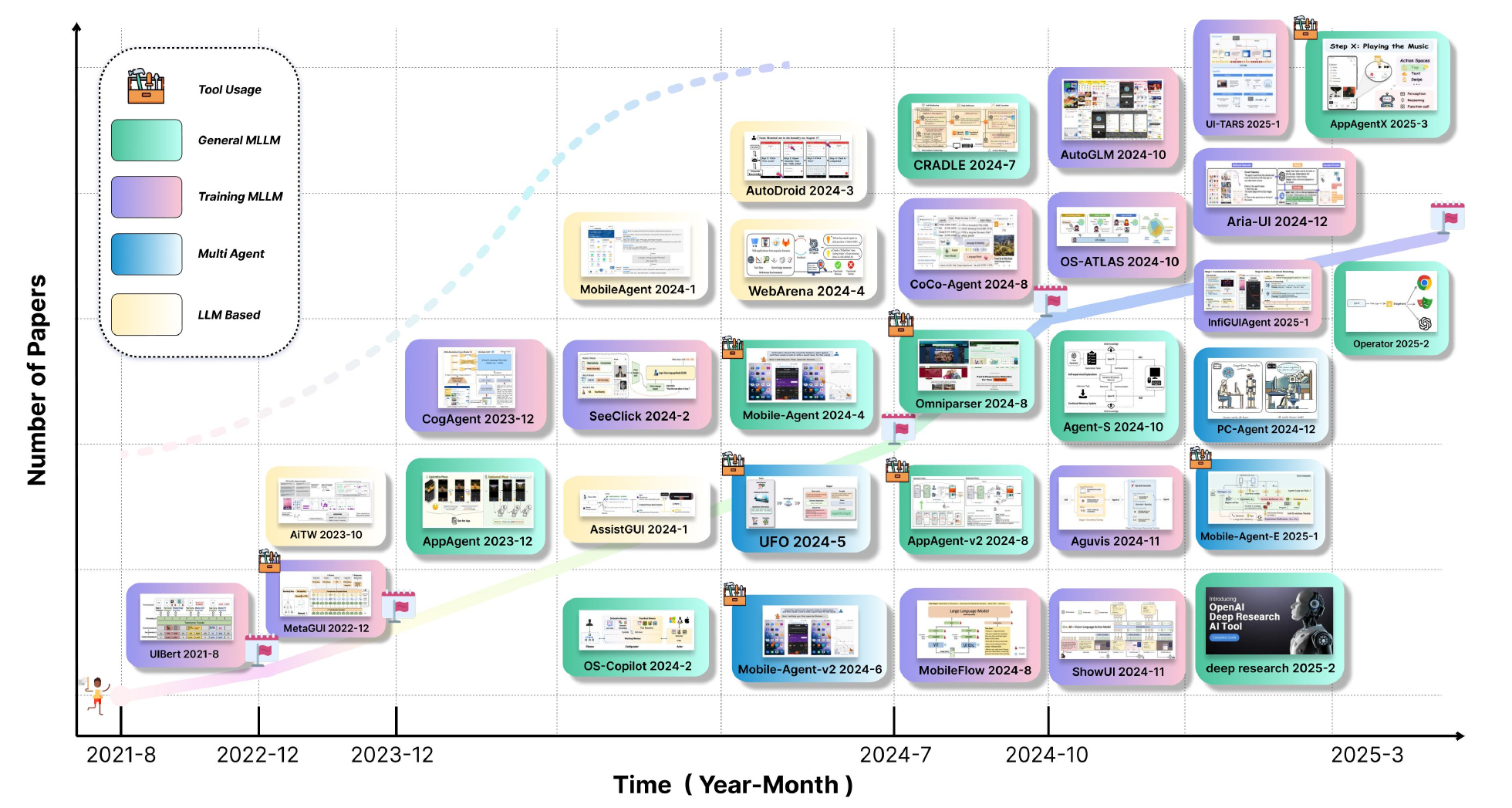}
    \caption{Recent work summary and classification of GUI Agents (2021-2025). The visualization presents the evolution of approaches across four primary categories: General MLLM, Training MLLM, Multi-Agent, and LLM-Based methodologies. The horizontal axis tracks chronological development (Year-Month), while the vertical axis indicates publication volume.}
    \label{fig:222}
\end{figure*}

Due to generalization limitations in both text-based and multimodal GUI Agents, researchers have increasingly adopted tool-based approaches to enhance screen information recognition. As shown in Figure \ref{fig:222}, several specialized tools have emerged in this domain:
SoM \cite{yang_set--mark_2023} combines Grounding DINO \cite{liu2024groundingdinomarryingdino} and SAM \cite{kirillov2023segment} to identify and semantically tag screen objects, then integrates these annotations with GPT-4V for effective localization. Similarly, OmniParser \cite{lu_omniparser_2024} employs multiple specialized models—including object detection, OCR, and icon recognition—to parse screen elements independently before consolidating this information for GPT-4V comprehension.
These approaches typically leverage a diverse toolkit including OCR systems, YOLO detectors, Grounding DINO, and IconNet for comprehensive GUI parsing. By using specialized tools to extract relevant information as prompt enhancements, these methods significantly reduce the perceptual burden on GUI Agents.
However, this tool-assisted paradigm sacrifices end-to-end processing efficiency. The necessity to employ multiple recognition algorithms before perception introduces substantial computational overhead and resource requirements, creating a trade-off between enhanced accuracy and system performance.

\subsection{Exploration}
\label{sec:exploration}

In recent years, the emergence of LLMs and MLLMs has endowed intelligent agents with robust natural language understanding and multimodal perception capabilities. These advancements have not only improved the efficiency of human-agent interactions but also significantly expanded the applicability of such agents in complex task scenarios. However, intelligent agents fundamentally operate within the constraints of their internal knowledge when making decisions. When confronted with heterogeneous user interface architectures and continuously evolving content, these systems frequently demonstrate performance limitations that fail to meet practical requirements. To address these limitations, researchers have increasingly explored graphical interface integration approaches, focusing on GUI Agents that can better perceive and navigate visual environments, thereby enhancing knowledge acquisition for more efficient task planning and decision-making, as shown in Figure \ref{fig:enter-label}.

In the following sections, we examine how GUI Agents process knowledge through various approaches. Section (\S\ref{sec:Categorization}) presents three exploration types (Internal, Historical, and External), while Section (\S\ref{sec:Operation}) covers the operational aspects of Knowledge Construction, Storage, and Application.
\begin{figure}[h]
    \centering
    \includegraphics[width=0.9\linewidth]{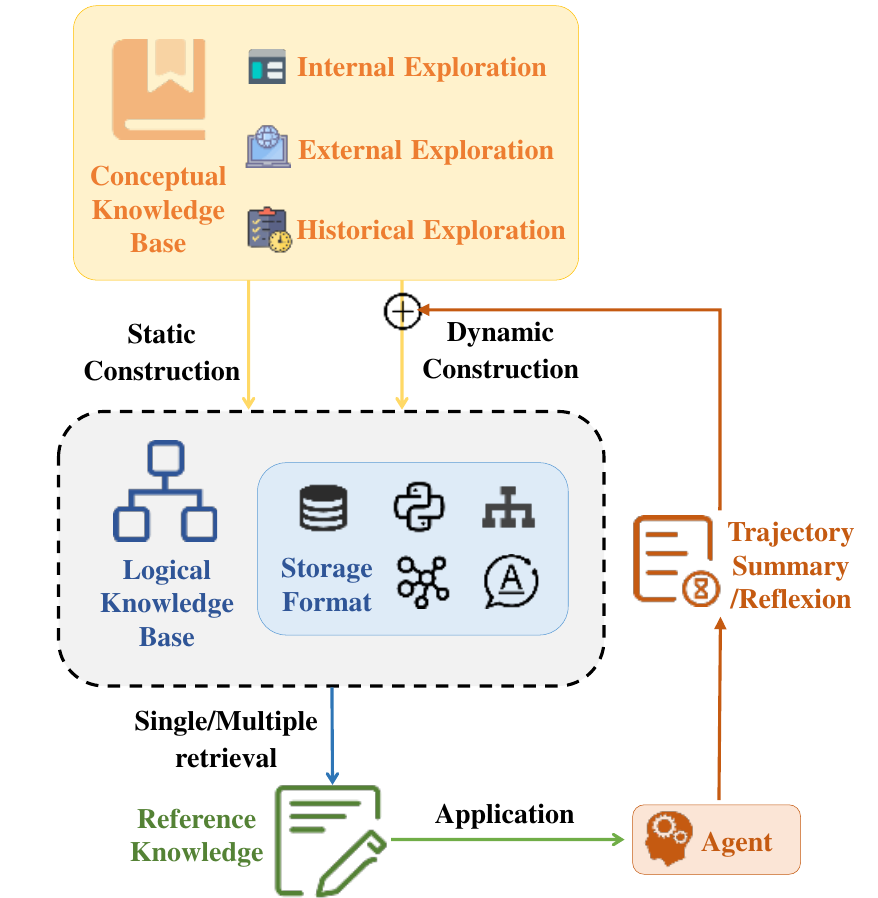}
    \caption{Exploration of GUI Agents. The diagram illustrates the knowledge acquisition process through three exploration types (internal, external, and historical), showing how conceptual knowledge is constructed and stored in logical knowledge bases before being retrieved and applied by agents.}
    \label{fig:enter-label}
\end{figure}
\subsubsection{Knowledge Acquisition}
\label{sec:Categorization}
When navigating new environments, humans leverage a variety of information sources to acquire knowledge and solve problems. Similarly, GUI Agents designed to interpret and interact with graphical interfaces must incorporate diverse information sources for effective operation. Through analysis of relevant research from the past three years, we identify three distinct approaches to knowledge acquisition:
\begin{itemize}
    \item \textit{Internal Exploration}: GUI Agents leverage built-in knowledge representations and reasoning capabilities.
    \item \textit{Historical Exploration}: GUI Agents analyze successful examples from past task trajectories.
    \item \textit{External Exploration}: GUI Agents retrieve supplementary knowledge and rules from external sources such as the internet.
\end{itemize}

\paragraph{\textbf{Internal Exploration}} focuses on gathering a range of information about user interface functionalities, page element purposes, and application overviews. For instance, AutoDroid \cite{wen2024autodroidllmpoweredtaskautomation} conducts offline random exploration of UI pages, summarizing element functionalities and page states to construct a knowledge base. Similarly, AppAgent \cite{zhang_appagent_2023} and AppAgent v2 \cite{li_appagent_2024} employ multimodal agent frameworks to analyze the roles of elements and actions used during interactions, documenting the resulting knowledge as reference materials. To further enhance the agent's understanding of user needs, MobA \cite{zhu2024mobatwolevelagentsystem} utilizes LLMs to summarize user preferences based on daily activity trajectories, establishing a user memory. Additionally, it records application functionality descriptions and visited pages as procedural memory, aiding the agent in acquiring application-specific knowledge and improving generalization capabilities.
Internal exploration enables GUI Agents to systematically build comprehensive knowledge representations by examining interface structures, analyzing element functionalities, and mapping application behaviors within their internal frameworks.
\sloppy
\paragraph{\textbf{Historical Exploration}} focuses on gathering a range of information about Historical Task Trajectories, Environmental Feedback, and Skill Repositories, which leverages past interactions and outcomes to inform current decision-making processes.

\textbf{Environmental Feedback.} Simliar to human short-term memory, environmental feedback refers to the temporary storage of information from the previous steps or the immediately preceding decision. This feedback is then applied to guide the agent's subsequent decision-making process. In Reflexion \cite{shinn2023reflexionlanguageagentsverbal}, the agent collects language-based feedback to improve a language agent by transforming binary or scalar environmental feedback into textual summaries. These summaries allow the agent to reflect on failed planning attempts, storing its reflections in an episodic memory buffer for inclusion in the context of subsequent decisions, thus enabling better planning. In OS-Copilot \cite{wu_os-copilot_2024}, the working memory gathers real-time execution feedback, which it uses to adjust the stored declarative and procedural memories, preventing the agent from repeating the same mistakes.

\textbf{Task Trajectories.} Agents \cite{zhou2023agentsopensourceframeworkautonomous}, Synapse \cite{zheng2024synapsetrajectoryasexemplarpromptingmemory}, and RaDA \cite{kim-etal-2024-rada} store successful trajectories collected during the training of agents on foundational tasks to supplement agent knowledge. Similarly, OS-Copilot \cite{wu_os-copilot_2024} maintains three types of memory: declarative memory (semantic knowledge including past trajectories), procedural memory, and working memory. To address complex control tasks, the Agents \cite{agashe_agent_2024} framework stores both narrative and episodic knowledge. Narrative knowledge summarizes complete task experiences for assisting in complex task decomposition, while episodic knowledge provides hierarchical task strategies for subtask planning. Due to the sensitivity of input prompts and the context length limitations of large language model-based agents, Expel \cite{zhao2023expelllmagentsexperiential} and AutoGuide \cite{fu2024autoguideautomatedgenerationselection} extract knowledge from success and failure trajectories to generate state-dependent guidelines. These guidelines are used in real-time state to predict optimal actions. OdysseyAgent, trained through GUI Odyssey \cite{lu2024guiodysseycomprehensivedataset}, is a multimodal agent for cross-application navigation. Given that task-related contexts often involve numerous screenshots and extended operational steps, the framework employs a historical replay module that compresses historical image annotations as inputs for decision-making. UFO \cite{zhang_ufo_2024} specializes in UI-focused tasks using a dual-agent framework. It records the decision history and execution outcomes of both agents to regulate interactions, achieving seamless integration with the Windows operating system.

\textbf{Skill Repositories.} Voyager \cite{wang2023voyageropenendedembodiedagent}, an embodied agent developed for Minecraft, maintains a growing skill repository that stores effective skill codes for tasks posed by an automated curriculum module, enabling continuous exploration of the game world. Similarly, GITM \cite{zhu2023ghostminecraftgenerallycapable} maintains a text memory library of common plans for each encountered task, providing referential experiential knowledge. Task completion often follows a sequence governed by certain rules. KnowAgent \cite{zhu2024knowagentknowledgeaugmentedplanningllmbased} explores different task types, summarizing corresponding action and rule-based knowledge. This knowledge is used to fine-tune large models, enhancing their ability to understand and utilize task-relevant information.

\paragraph{\textbf{External Exploration}} involves acquiring information from the external environment using tools such as search engines. In OS-Copilot \cite{wu_os-copilot_2024}, which emulates the cognitive architecture of the human brain, procedural memory constitutes a critical component of the system's long-term memory structure. This memory framework stores a comprehensive collection of facts and events gathered from the internet, user interactions, and operating system outputs to support the agent's decision-making processes. 
In GITM \cite{zhu2023ghostminecraftgenerallycapable}, researchers construct a comprehensive textual knowledge base drawing from sources such as the Minecraft Wiki and detailed crafting/smelting recipes. This external repository serves as an essential reference for the agent when performing complex task decomposition within the Minecraft environment. To improve decision-making efficiency and avoid rigid operational directives, Mobile-Agent \cite{wang_mobile-agent_2024} searches the internet to summarize standard operating procedures for various tasks, which are then used as blueprints for agent input commands. In Agents \cite{agashe_agent_2024}, when decomposing complex tasks, the agent searches the web for similar categorized examples and classification criteria. This knowledge is combined with narrative memory, or historical knowledge, to provide a more comprehensive reference for decision-making.

\subsubsection{Knowledge Operation}
\label{sec:Operation}
GUI Agents form conceptual knowledge bases through information exploration from various sources. This knowledge operation framework comprises three interconnected processes for functioning across environments like Android or Windows. The process begins with transforming concepts into logical repositories (\S\ref{sec:knowledge_construction}), followed by establishing optimal structural organization for effective representation (\S\ref{sec:knowledge_storage}), and concludes with implementing efficient retrieval mechanisms that balance access speed and planning effectiveness (\S\ref{sec:knowledge_application}). These integrated components enable agents to operationalize knowledge in practical applications.

\paragraph{\textbf{Knowledge Construction}} 
\label{sec:knowledge_construction}
is a crucial component in the knowledge processing pipeline. A well-designed knowledge construction process maximizes the extraction of valuable information, thereby significantly enhancing intelligent agent performance. Based on temporal characteristics, we categorize knowledge construction processes into two types: static construction and dynamic construction.

\textbf{Static Construction.} This approach involves offline exploration of GUI pages to gather information, as demonstrated by AutoDroid \cite{wen2024autodroidllmpoweredtaskautomation}, which randomly explores GUI pages to generate transition graphs and, through a memory generator, traverses the graphs to create element function descriptions and page function summaries. Unlike AutoDroid's random exploration, AppAgent \cite{zhang_appagent_2023} and AppAgent v2 \cite{li_appagent_2024} focus on task-driven exploration, guiding the agent to more efficiently operate specific elements. In this process, the agent learns the functional relationships between different UI elements and actions during interactions, while also supplementing the agent's knowledge with complex functions observed from manual operations. MobileAgent \cite{ding_mobileagent_2024} further performs offline knowledge extraction on task-driven path trajectories to derive more abstract knowledge of standard operating procedures. Expel \cite{zhao2023expelllmagentsexperiential} and AutoGuide \cite{fu2024autoguideautomatedgenerationselection} extract knowledge from historical task trajectories to generate state-dependent guidelines. First, they construct a dataset of successful and failed trajectory pairs from the collected task decision data. They then extract trajectories before deviations occur in the paired dataset and use state summarization by LLM to generate corresponding guidelines. KnowAgent \cite{zhu2024knowagentknowledgeaugmentedplanningllmbased} for a specific domain task, performs large-scale offline knowledge construction, where GPT-4 generates an initial knowledge base of task exploration actions and action rules, which is then manually optimized. Due to the lack of real-time interaction and feedback mechanisms, static knowledge bases often struggle to capture contextual information in complex scenarios, leading them less adaptable to dynamically changing GUI environments. Recognizing this limitation, researchers have shifted toward dynamic construction methods.

\textbf{Dynamic Construction.} MobA \cite{zhu2024mobatwolevelagentsystem} implements three types of memory that are initially configured by human experts and continuously updated in real time during interactions. WKM \cite{qiao2024agentplanningworldknowledge} serves as a world knowledge model to assist GUI Agents, which extracts expert trajectories from manually labeled datasets, comparing preferences with trajectories generating by agent. These preferences are input into the agent to summarize task knowledge, and each step is associated with a status summary knowledge based on expert trajectories. OS-Copilot \cite{wu_os-copilot_2024} utilizes procedural memory that is initially custom-defined by the user. For subsequent tasks, the agent retrieves applicable tools from memory. When memory lacks suitable tools, the tool generator activates to create and execute new tools, collecting feedback until reaching a termination condition. Upon task completion, the reflection module evaluates each tool's utility and determines whether to update the memory base accordingly. Agents \cite{agashe_agent_2024} includes both initial memory construction and continuous memory updates. In the initial phase, tasks are generated based on a dataset into two categories: environment-independent and environment-aware tasks, which are then explored by an agent supplemented with network knowledge. The agent collects experiences from complete tasks and subtasks, continually updating the knowledge base as it interacts with new tasks. Dynamic construction enhances adaptability and contextual understanding by enabling real-time knowledge refinement based on interactive feedback, addressing the limitations of static approaches.

\paragraph{\textbf{Knowledge Storage}}
\label{sec:knowledge_storage}
The format of knowledge storage media constitutes a critical dimension in exploration module research, with formats ranging from natural language to structured databases. Each storage paradigm offers unique advantages tailored to specific application contexts, reflecting the diverse approaches to knowledge representation in autonomous systems.

\textbf{Natural Language.} serves as a foundational knowledge storage medium that enables intuitive representation and human-interpretable storage of diverse information types. AppAgent \cite{zhang_appagent_2023} and AppAgent v2 \cite{li_appagent_2024} represent knowledge through natural language, which is then stored in documents. Reflexion \cite{shinn2023reflexionlanguageagentsverbal} and Voyager \cite{wang2023voyageropenendedembodiedagent} both leverage natural language for knowledge representation, with the former maintaining experiential feedback within a sliding window mechanism while the latter encodes Minecraft skills directly in memory, enabling efficient skill retrieval and utilization during gameplay.

\textbf{Database.} provides systematic organization and efficient retrieval mechanisms for complex agent knowledge, enabling structured storage with sophisticated indexing capabilities.
AutoDroid \cite{wen2024autodroidllmpoweredtaskautomation} implements a dual-table knowledge architecture comprising simulation task tables with UI interaction element function descriptions and user page function tables that document current page functionalities. The Agents framework \cite{zhou2023agentsopensourceframeworkautonomous} incorporates memory components from Zhou et al. \cite{zhou2023recurrentgptinteractivegenerationarbitrarily}, utilizing VectorDB for efficient knowledge storage, while Agashe et al. \cite{agashe_agent_2024} distinguishes between narrative memory keyed by query tasks and situational memory indexed by both query tasks and contextual subtask information within their database structure.

\textbf{Data Structures.} offer versatile frameworks for organizing complex relationships between knowledge elements, enabling efficient representation of hierarchical, associative, and sequential information for GUI Agents. The task memory of MobA \cite{zhu2024mobatwolevelagentsystem} is based on the ICE \cite{qian2024investigateconsolidateexploitgeneralstrategyintertask} strategy, where tasks and subtasks are organized into a hierarchical tree structure. This structure is suitable for sequences of ordered subtasks and captures detailed information about each task, including task status, environmental context, and relevant subtasks. Expel \cite{zhao2023expelllmagentsexperiential} and AutoGuide \cite{fu2024autoguideautomatedgenerationselection} employ dictionary data structures with key-value pairs to efficiently represent the one-to-one correspondence between states and guidelines. KnowAgent \cite{zhu2024knowagentknowledgeaugmentedplanningllmbased} adopts a graph structure for its knowledge base to represent complex action relationships and sequences that mirror the intricacy of social networks. Within this framework, actions function as nodes (containing action names and definitions) while rules serve as connecting edges (encoding rule descriptions), effectively capturing the interdependencies between different actions in the system. The state knowledge in WKM \cite{qiao2024agentplanningworldknowledge} is also stored in a graph structure, with nodes representing actions in expert trajectories and edges representing state knowledge between sequential actions. GITM \cite{zhu2023ghostminecraftgenerallycapable} stores the action lists of subgoals in a hierarchical tree structure, clearly capturing the relationship between goals and corresponding plans.

\textbf{Code.} Executable representations offer a direct encoding of procedural knowledge, enabling agents to perform complex operations through functional libraries. The procedural memory involved in OS-Copilot \cite{wu_os-copilot_2024} consists of a series of manually created tool libraries stored in dual formats as both API services and Python files. Voyager \cite{wang2023voyageropenendedembodiedagent} stores skills as values within its architecture, forming a comprehensive skill knowledge base that facilitates direct execution of learned capabilities.

\paragraph{\textbf{Knowledge Application}}
\label{sec:knowledge_application}
The purpose of memory retrieval is to extract meaningful information from memory to enhance GUI Agents's actions, such as utilizing previously successful actions to achieve similar goals. The critical challenge in memory retrieval centers on identifying and extracting valuable information from the agent's historical action repository. This process can be categorized based on retrieval scope and methodology, with approaches ranging from targeted single retrieval to comprehensive multiple extraction strategies.

\textbf{Single Retrieval.} 
AppAgent \cite{zhang_appagent_2023}, AppAgent v2 \cite{li_appagent_2024}, and Agents \cite{zhou2023agentsopensourceframeworkautonomous} retrieve knowledge based on task relevance, serving as additional input to assist the agent. AutoDroid \cite{wen2024autodroidllmpoweredtaskautomation} employs the embedding model Instructor-XL \cite{Instructor-XL} to map knowledge and task descriptions, utilizing a similarity function (sim) to compare both and retrieve knowledge for agent planning. The retrieved functional descriptions are applied by adding new onclick attributes to the corresponding HTML framework UI elements on the current page. Agashe et al. \cite{agashe_agent_2024} retrieve knowledge from the knowledge database based on task descriptions and hierarchical subtask information. Voyager \cite{wang2023voyageropenendedembodiedagent} utilizes GPT-3.5 to generate environmental feedback based on task descriptions and retrieves the top-5 skill codes from the skill database through feedback embedding. KnowAgent \cite{zhu2024knowagentknowledgeaugmentedplanningllmbased}, which operates on text-processing LLMs, initially retrieves the relevant knowledge stored in a graph and converts it into text format. This serves dual purposes: during the self-learning process, each generated path is filtered and merged to fine-tune the agent, and subsequently provided directly as a prompt in downstream tasks.
Qiao et al. \cite{qiao2024agentplanningworldknowledge} utilize task knowledge and state knowledge to train the WKM model, which simulates the human process of knowledge building by progressing from general overviews to specific details. The WKM initially assists the agent in establishing a general planning path and subsequently generates the current state in real-time during execution. This state functions as a key to retrieve knowledge from the database, identifying n possible next actions and selecting the one with the highest probability based on the agent's predicted action probabilities. Despite these advances, single retrieval methods may present limitations when knowledge bases store multi-source information. Such approaches can be one-sided with poor adaptability, potentially missing critical information and leading to significant deviations in decision-making paths. To address these limitations, multiple retrieval approaches have been proposed in recent research.
\textbf{Multiple Retrieval.} OS-Copilot \cite{wu_os-copilot_2024} retrieves different stored procedural memory through two distinct mechanisms: API calls via POST requests and Python file information that matches specific calls. AutoGuide \cite{fu2024autoguideautomatedgenerationselection} integrates key-value pair knowledge during the agent's decision-making process, where the state summarization module first summarizes the current state in real-time for state matching, followed by the guideline selection module for further refinement when more than k matches are identified. MobA \cite{zhu2024mobatwolevelagentsystem} implements two complementary retrieval methods tailored to diverse task requirements. Their relation-based retrieval targets action modules by extracting task descriptions from both the previous action node and its parent node within the tree-structured memory, thereby enhancing the agent's decision-making capacity for subsequent actions. Concurrently, their content-based retrieval leverages cosine similarity to search through historical task experiences, which facilitates the hierarchical decomposition of complex tasks.



\subsection{Reasoning and Planning}
\label{sec:planning}
Planning in GUI Agents necessitates sophisticated reasoning capabilities to effectively navigate complex interfaces and accomplish user goals. This section examines the theoretical underpinnings and practical implementations of planning mechanisms. We begin by analyzing the reasoning frameworks employed in contemporary LLMs (\S\ref{sec:llmreason}) and MLLMs (\S\ref{sec:mllmreason}), establishing the foundation for understanding advanced planning techniques. We then explore emerging reasoning architectures in state-of-the-art o1-like models that integrate multi-step reasoning and complex decision trees (\S\ref{sec:o1likes}). Subsequently, we provide a comprehensive analysis of GUI Agents-specific planning architectures (\S\ref{sec:taskplanning}), focusing on planner design, verification methodologies, and feedback integration systems that collectively enable robust task execution in graphical user environments.

\subsubsection{LLMs Reasoning}
\label{sec:llmreason}
In recent years, LLMs have demonstrated outstanding performance in context modeling, which has significantly contributed to their exceptional reasoning abilities in natural language processing tasks \cite{min2022metaicllearninglearncontext}. Inspired by this, the Chain-of-Thought (CoT) \cite{wei2023chainthoughtpromptingelicitsreasoning} prompting method, which introduces intermediate reasoning steps to guide LLMs through a step-by-step reasoning process, has successfully addressed multi-step reasoning problems \cite{zeng2022socraticmodelscomposingzeroshot}.
The intermediate reasoning steps in CoT provide a more transparent reasoning pathway, which aids in the interpretability and evaluation of LLMs \cite{xia-etal-2024-aligning}. This method not only enhances the model's reasoning capabilities but also provides significant insights for Agent planning in multi-step reasoning tasks. In the following, we will introduce several common variants of CoT prompting.

\textbf{CoT.} is a chain structure that breaks down complex tasks into a series of simpler steps, executed sequentially.
The CoT structure can be represented as $\langle \textit{input}, \textit{thought}_1, \textit{thought}_2, \dots, \textit{thought}_n, \textit{output} \rangle$, where \textit{input} refers to the task received by the model, \textit{thought} represents the intermediate reasoning steps, $n$ denotes the number of such steps, and \textit{output} is the final answer or solution. From the perspective of sample prompts, CoT can be categorized into two types:
 (1) \textit{zero-shot CoT}, which prompts reasoning using a phrase like "Let's think step by step" \cite{zeng2022socraticmodelscomposingzeroshot}, and (2) \textit{few-shot CoT}, where a small number of sample prompts guide the model through incremental reasoning \cite{wei2023chainthoughtpromptingelicitsreasoning}. For example, Yuan et al. \cite{yuan2024reversalthoughtenhancinglarge} utilized a preference-guided reverse reasoning strategy to generate sample templates that enhance the logical reasoning ability of LLMs.

\textbf{ToT.} Tree-of-Thoughts (ToT) \cite{yao2023treethoughtsdeliberateproblem} extends chain-based reasoning by using a tree structure, generating multiple reasoning thoughts at each step of the reasoning process and conducting self-evaluation to prune and plan for the optimal path. Recent studies have further extended the tree structure by combining Monte Carlo Tree Search (MCTS) \cite{feng2024alphazeroliketreesearchcanguide} with LLMs, further enhancing the model’s reasoning ability \cite{ding2024everythingthoughtsdefyinglaw,zhang2024restmctsllmselftrainingprocess,gao2024interpretablecontrastivemontecarlo,zhou2024languageagenttreesearch}.
For example, ReST-MCTS* integrates MCTS with reinforcement self-training \cite{zhang2024restmctsllmselftrainingprocess}, and SC-MCTS \cite{gao2024interpretablecontrastivemontecarlo} improves the accuracy and speed of LLMs in complex reasoning tasks through enhanced reward models, node selection strategies, and backpropagation methods. Additionally, RAP \cite{hao2023reasoninglanguagemodelplanning} constructs a world model and uses MCTS to effectively balance exploration and exploitation. LATS \cite{zhou2024languageagenttreesearch} follows the ReAct framework and applies MCTS to LM Agents, achieving a more detailed and adaptive problem-solving mechanism through a collaborative process of search, interaction, and reflection.

\textbf{GoT.} Graph of Thoughts \cite{besta2024graph} extends the traditional tree-based reasoning structure by adopting a graph structure by utilizing a graph structure that introduces cycles and many-to-one connections. This allows for more effective modeling of subproblem aggregation and self-assessment.

\textbf{ReAct.} Reasoning and Acting (ReAct) \cite{yao2023reactsynergizingreasoningacting} builds on chain-based reasoning by adding an action space $\hat{A} = \mathcal{A} \cup \mathcal{L} $ to facilitate the synergy between reasoning and actions, where \textit{L} represents the reasoning trajectory and \textit{A} is the set of allowable actions. Specifically, the agent first generates a thought based on the assigned task, then executes an action based on this thought, observes the result, and iterates the process until the task is completed.

As the reasoning capabilities of LLMs become increasingly complex, it is becoming more important and necessary to effectively evaluate their complex reasoning paths. Some research, such as Self-Consistency \cite{wang2023selfconsistencyimproveschainthought,zhou2024languageagenttreesearch}, Best-of-N \cite{sun2024fastbestndecodingspeculative,wang2024mathshepherdverifyreinforcellms,lightman2023letsverifystepstep} , Process Supervision Reward Model (PRM) \cite{lightman2023letsverifystepstep,gao2024interpretablecontrastivemontecarlo,shang2024synergythoughtselicitingefficientreasoning,setlur2024rewardingprogressscalingautomated}, and Outcome Supervision Reward Model (ORM) \cite{zhang2024generativeverifiersrewardmodeling,zheng2023judgingllmjudgemtbenchchatbot,qi2024mutualreasoningmakessmaller}, aims to improve the reasoning capabilities of large language models by evaluating intermediate reasoning steps or final outcomes. Meanwhile, some research solves complex problems by decomposing them into simpler subproblems \cite{zhou2023leastmostpromptingenablescomplex,gao2024interpretablecontrastivemontecarlo,touvron2023llama2openfoundation,huang2024understandingplanningllmagents,yuan2024selfrewardinglanguagemodels}. For example, Huang et al. \cite{huangqdmrbasedplanningsolvingpromptingcomplex} decompose questions into a directed acyclic graph, and use the Question Decomposition Meaning Representation (QDMR) prompting method to perform step-by-step reasoning based on the dependency relationships between nodes in the graph. In addition, some research suggests that LLMs are capable of self-improvement during the reasoning process through self-refinement \cite{zhang2024restmctsllmselftrainingprocess,zelikman2022starbootstrappingreasoningreasoning,hosseini2024vstartrainingverifiersselftaught,madaan2023selfrefineiterativerefinementselffeedback,yuan2023scalingrelationshiplearningmathematical,zelikman2024quietstarlanguagemodelscan,zhang2024chainpreferenceoptimizationimproving,yuan2024selfrewardinglanguagemodels,andukuri2024stargateteachinglanguagemodels}. For example, Quiet-STaR \cite{zelikman2024quietstarlanguagemodelscan} uses a method in which the language model learns to generate a reasoning chain for each token.

\subsubsection{Multimodal LLMs Reasoning}
\label{sec:mllmreason}
In the context of MLLMs, chain-of-thought reasoning is not confined to natural language but can also integrate and analyze other perceptual information such as images, videos, and audio, thereby extending textual chain-of-thought reasoning to multimodal chain-of-thought reasoning \cite{zhang2024multimodalchainthoughtreasoninglanguage,zeng2022socraticmodelscomposingzeroshot,wu2023rolechainthoughtcomplexvisionlanguage}. 
Lu et al. \cite{lulearnexplainmultimodalreasoning} proposed the ScienceQA dataset, laying the foundation for the subsequent introduction of the concept of Multimodal Chain-of-Thought reasoning (MCoT) \cite{zhang2024multimodalchainthoughtreasoninglanguage}. 
To further enhance the reasoning capabilities of MCoT, researchers have explored various approaches.
KAM-CoT \cite{mondal2024kamcotknowledgeaugmentedmultimodal} enhances the reasoning ability of MCoT by integrating chain-of-thought, knowledge graphs, and multimodal information. 
CCoT \cite{mitracompositionalchainthoughtpromptinglarge} utilizes scene graph representations as prompts to encourage LLMs to generate structured descriptions as reasoning steps. Moreover, some works such as MM-REACT \cite{yang2023mmreactpromptingchatgptmultimodal}, VisCoT \cite{shao2024visualcotadvancingmultimodal}, and VoCoT \cite{li2024vocotunleashingvisuallygrounded} further integrate visual expert capabilities, enhancing multimodal reasoning and action capabilities. 
For example, VoCoT \cite{li2024vocotunleashingvisuallygrounded} represents object concepts visually through multimodal cross-alignment, thereby enabling an object-oriented chain-of-thought reasoning process that is guided by multi-step visual support.
Similarly, CoCoT \cite{zhang2024cocotcontrastivechainthoughtprompting} leverages information across multiple images to perform cross-image analysis.
To mitigate potential errors and hallucinations during the reasoning process, DDCoT \cite{zheng2023ddcotdutydistinctchainthoughtprompting} employs negative space cues to reveal uncertainties in subproblem decomposition. 
Meanwhile, TextCoT \cite{luan2024textcotzoomenhancedmultimodal} utilizes the image description capabilities of MLLMs to grasp the global context of images and leverages localization abilities to examine local text regions.
MC-CoT \cite{tanboostingpowersmallmultimodal} employs a voting mechanism similar to CoT-SC, but unlike CoT-SC, it introduces voting in both the generation reasoning and answer reasoning stages, thereby enhancing the robustness of the LMM’s reasoning.
CogCoM \cite{qi2024cogcomtrainlargevisionlanguage} simulates human visual reasoning processes, inferring correct answers through incremental manipulation steps.
WoT \cite{menon2024whiteboardthoughtthinkingstepstepmodalities} introduces an innovative mechanism that converts intermediate reasoning step results into an image and then further reasons about the image using the visual input capabilities of MLLMs.
In specific application domains, Wei et al. \cite{wei2024mccotmodularcollaborativecot} applied MCoT to the medical field, improving reasoning and information extraction effectiveness by integrating medical knowledge and task-specific guidance. \textsc{Focus} \cite{tang2025thinktwiceclickonce} proposed a dual-system framework based on MLLMs to locate GUI elements, which can generate explicit multi-stage CoT to enhance grounding capability.

Furthermore, in the Text-to-Speech (TTS) domain, there are also works related to chain-of-thought \cite{xin2024rallerobustcodeclanguage,gong2024seamlessexpressivelmspeechlanguagemodel,hu2024chainthoughtpromptingspeechtranslation}.
RALL-E \cite{xin2024rallerobustcodeclanguage} enhances the robustness of LLM-based TTS by decomposing tasks into simpler steps using CoT.
Meanwhile, Hu et al. \cite{hu2024chainthoughtpromptingspeechtranslation} enhanced the performance of automatic speech translation (AST) by using the automatic speech recognition(ASR)-transcribed text generated by a speech encoder together with the encoded speech as CoT prompts.
In addition, Gong et al. \cite{gong2024seamlessexpressivelmspeechlanguagemodel} improved the translation quality and parameter efficiency of expressive speech-to-speech translation (S2ST) by decomposing complex source-to-target speech mappings into intermediate generation steps through chain-of-thought prompts.

\begin{figure}[h]
    \centering
    \includegraphics[width=0.48\textwidth]{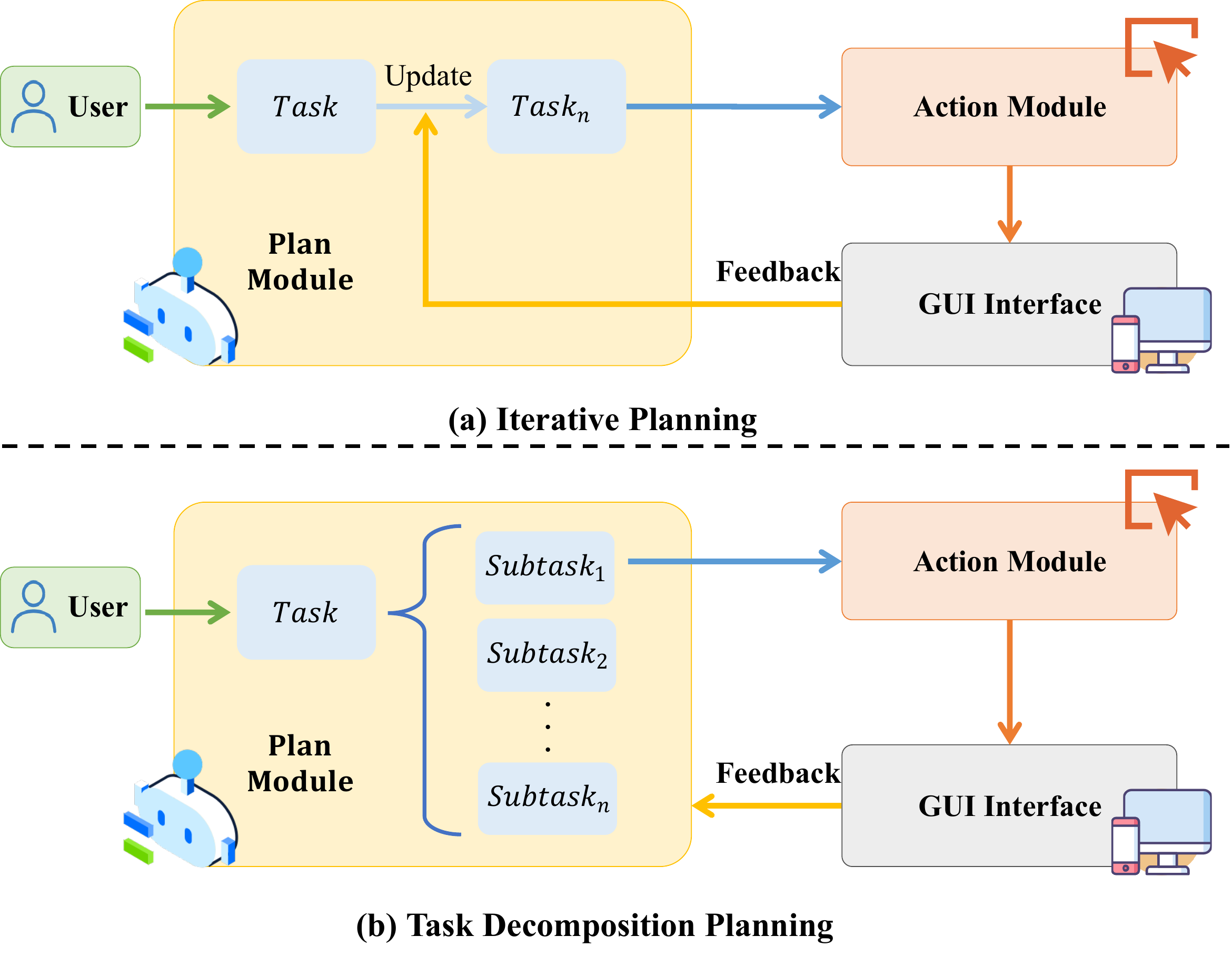}
    \caption{Different task planning strategies for GUI agents: (a) iterative planning, and (b) task decomposition planning.}
    \label{fig:planning}
\end{figure}

\subsubsection{o1-likes Reasoning Models}
\label{sec:o1likes}
Since the release of OpenAI's o1, researchers have been dedicated to replicating its exceptional reasoning capabilities. The o1 model demonstrates superior reasoning performance by engaging in deep thinking before providing answers. As a significant breakthrough in this field, DeepSeek-R1 \cite{deepseekai2025deepseekr1incentivizingreasoningcapability} stands as a representative model for replicating o1, successfully reproducing similar capabilities through reinforcement learning and verifiable rule-based rewards. Specifically, it employs the Group Relative Policy Optimization (GRPO) \cite{shao2024deepseekmathpushinglimitsmathematical} algorithm instead of the traditional PPO algorithm \cite{schulman2017proximalpolicyoptimizationalgorithms} for training, thereby achieving an "aha moment" (a phenomenon where the model reevaluates the answer and reflects previous reasoning path. Inspired by DeepSeek-R1's success, a wave of R1 replication efforts has emerged within the open-source community.
These replication efforts primarily focus on two directions: text-based models and multimodal reasoning models. In terms of text-based models, Logic-RL \cite{xie2025logicrlunleashingllmreasoning} successfully reproduced the "aha moment" using a model of only 7B parameters, demonstrating that smaller-scale models can also acquire powerful reasoning capabilities through reinforcement learning. Meanwhile, R1-Searcher \cite{song2025r1searcherincentivizingsearchcapability} innovatively proposed a two-stage reinforcement learning approach enabling LLMs to autonomously invoke external retrieval systems during reasoning, a method that effectively overcomes the inherent knowledge limitations of models.
On the other hand, in the field of multimodal reasoning, despite the significant impact of DeepSeek-R1's achievements, similar explorations in the multimodal domain remain relatively limited. Vision-R1 \cite{huang2025visionr1incentivizingreasoningcapability} introduced a visual reinforcement fine-tuning method, designing specific reward functions for different visual tasks (such as IoU rewards and classification rewards) and adopting GRPO policy parameter updates, an approach that significantly enhances model performance in few-shot learning and visual reasoning tasks. LMM-R1 \cite{peng2025lmmr1empowering3blmms} enhances the reasoning capabilities of a 3B-scale multimodal model through a two-stage rule-reward reinforcement learning framework, which extends the OpenRLHF framework to support multimodal model training, thus enabling small models to demonstrate powerful reasoning and adaptability in complex tasks.

\subsubsection{Task Planning}
\label{sec:taskplanning}
Within the components of GUI Agents, the planning capability is of paramount importance, which assists the GUI Agents in task planning, task verification, and providing feedback.
During the planning phase, GUI Agents typically follow the CoT paradigm \cite{agashe_agent_2024,wu_os-copilot_2024,zhang_appagent_2023,li_appagent_2024,zhang_ufo_2024}. 
In the planning phase, the reasoning process often adopts the ReAct \cite{yao2023reactsynergizingreasoningacting} style. For example, CoA \cite{zhang_you_2024} improves CoT (Chain-of-Thought) reasoning by incorporating action history. Building on this, CoAT \cite{zhang_android_2024} further enhances CoT reasoning by integrating screen descriptions and previous action results, thereby improving reasoning accuracy and contextual awareness.

\textbf{Task Planner.}
GUI Agents typically utilize a task planner to plan complex tasks, obtaining a series of executable schemes to facilitate task completion. 
Generally, the Task Planner module of GUI Agents can be classified into two types—iterative planning and task decomposition planning.

\textit{Iterative Planning.} As shown in Figure~\ref{fig:planning} (a), the GUI Agents completes instruction tasks iteratively by dynamically adjusting the task plan based on the current state and feedback at each step, and inferring the next action strategy.
For instance, AppAgent \cite{li_appagent_2024} progressively completes instruction tasks through continuous iteration.
Additionally, some GUI Agents first generate an overall plan before proceeding with iterations.
For example, Mobile-Agent \cite{wang_mobile-agent_2024} employs a self-planning method, initially generating a system prompt based on user instructions and then gradually completing each operation step by step, with the prompt format following the ReAct style.
SheetCopilot \cite{li2023sheetcopilotbringingsoftwareproductivity} uses a state machine-based planner, which revises the plan through feedback from either LMs or software.
The iterative planning process is highly dynamic, allowing the system to continuously adjust task strategies based on feedback at each step, effectively adapting to environmental changes and ensuring the reliability and effectiveness of the plan.
However, when dealing with complex tasks, the reasoning steps of the GUI Agents may become excessively lengthy, which can lead to the LLMs generating hallucinations, causing subsequent planning to deviate from the intended objectives.

\textit{Task Decomposition Planning.}
As shown in Figure~\ref{fig:planning} (b), the GUI Agents decomposes complex tasks into a series of subtasks and then infers action strategies for these subtasks, allowing the Agents to complete the instruction task by successfully executing all of the subtasks \cite{gao_assistgui_2024}.
Compared to directly addressing complex tasks, decomposing tasks into simpler subtasks allows the language model to better maintain close connections between subtasks and the overall task, thereby reducing the risk of catastrophic forgetting or hallucination in large language models \cite{touvron2023llama2openfoundation,huang2024understandingplanningllmagents}. 
However, task decomposition-based planning lacks flexibility because the subtasks are fixed from the outset, which makes it difficult to adapt to dynamic changes in the environment or unforeseen obstacles.
Agent S \cite{agashe_agent_2024} employs a task planner to decompose complex tasks into detailed, topologically sorted subtasks in order to accomplish user instructions.
Existing planners, regardless of generating linear or nonlinear structured plans, require the agent to execute tasks in a sequential manner \cite{wu_os-copilot_2024}.
OS-Copilot utilizes a planner based on Directed Acyclic Graphs (DAG), allowing many independent tasks to be parallelized, thus minimizing execution time. 
Similarly, VillagerAgent \cite{dong2024villageragentgraphbasedmultiagentframework} constructs the decomposed subtasks into a DAG to achieve structured task management.
Additionally, MobA \cite{zhu2024mobatwolevelagentsystem} breaks down long-term tasks into a series of subtasks, allowing each subtask to be completed in a single step within the action module.
If a subtask encounters an error or cannot be completed in one step, it is further decomposed until the task is fully completed.

\textbf{Verifier.}
The successful execution of GUI Agents relies on two critical components: precise planning and robust verification mechanisms \cite{zhu2024mobatwolevelagentsystem,wang2023voyageropenendedembodiedagent,masterman2024landscapeemergingaiagent}. Among these, the verifier is particularly essential, as it directly assesses the quality of the outputs generated by GUI Agents, ensuring their reliability and effectiveness.
The verifier plays a key role in ensuring the quality of outputs generated by GUI agents, with the main function being to examine the reasoning steps and task execution states produced by the GUI Agents to ensure reliability and accuracy.
Generally, the verifier utilizes inputs such as generated trajectories and the environment before and after action execution to compute performance score for the agent within the given task context.
The verifier in GUI Agents generally leverages its own LLMs for both generation and evaluation purposes.
Voyager \cite{wang2023voyageropenendedembodiedagent} introduces a self-improving iterative prompting mechanism to perform self-verification of task states. 
Agent S \cite{agashe_agent_2024} uses the verifier to summary strategies from completed subtasks as experiences, which are then used as textual rewards to offer better strategic suggestions for subsequent tasks.
Additionally, Reflexion \cite{shinn2023reflexionlanguageagentsverbal} conducts a more comprehensive evaluation by generating detailed textual outputs based on the trajectories generated by the agent.
Moreover, the Self-Consistency \cite{wang2023selfconsistencyimproveschainthought} method is frequently used in evaluation tasks, selecting the most common answer from multiple generated outputs as the final solution.
LATS \cite{zhou2024languageagenttreesearch} proposes a mechanism that combines scores generated by the agent's own LLM with self-consistency evaluation to assign values to inference nodes.

\textbf{Feedback.}
In practical application scenarios, GUI Agents often need to handle long-term planning and execution of complex tasks.
During task execution, the agent is likely to encounter unforeseen errors, which could lead to task failure.
For instance, after executing a certain operation, the actual state of the UI may not match expectations, or the outcome of the operation may not satisfy the intended objectives.
Therefore, feedback mechanisms are used to monitor the entire task execution process and provide timely semantic feedback to the agent, allowing for rapid adjustment and correction of the task plan when issues arise.
In general, feedback can be categorized into two types: external feedback and internal feedback.

External Feedback primarily refers to environmental feedback, which captures the direct response to the agent's task execution, such as the observation results following the agent's action.
This feedback enables the agent to evaluate whether the current execution step has achieved the expected outcome, offering crucial guidance for subsequent actions.
Environmental feedback is usually obtained by the Agent through specific methods such as screenshots \cite{hong_cogagent_2023,cheng2024seeclickharnessingguigrounding,zhang_android_2024,baechler2024screenaivisionlanguagemodelui,shaw2023pixelsuiactionslearning}. For example, ReAct \cite{yao2023reactsynergizingreasoningacting} proposed a triplet $\langle {thought, action, observation} \rangle$, utilizing observations of the current environmental state after each action to provide feedback to the agent.
Specifically, \textit{thought} represents the process of reasoning and deliberation, \textit{action} denotes the action taken by the agent, and \textit{observation} is the information obtained from the environment.
Voyager \cite{wang2023voyageropenendedembodiedagent} integrates environmental feedback into the task execution process.
Similarly, MP5 \cite{qin2024mp5multimodalopenendedembodied} analyzes environmental feedback to identify the root causes of task execution failures and uses this information to guide the planner in refining subsequent task planning.
This mechanism prevents the system from continuing along incorrect paths and allows for flexible adjustment of the execution strategy according to the environmental state.
LATS \cite{zhou2024languageagenttreesearch} uses environmental feedback as observation expansion nodes and utilizes this feedback to directly provide evaluation rewards, thereby assisting in task assessment.

Internal Feedback typically refers to self-feedback.
Generally, GUI Agents utilize their own LLMs to generate feedback, which is used either as part of the prompt for the next action \cite{li2023sheetcopilotbringingsoftwareproductivity} or stored in a memory module \cite{wang2024mobileagentv2mobiledeviceoperation}, enabling the Agent to rapidly refine its decision-making plans and effectively solve problems by leveraging experience through linguistic feedback signals.
Feedback mechanisms can significantly enhance the accuracy and reliability of the agent in task execution \cite{ma_coco-agent_2024}
LATS \cite{zhou2024languageagenttreesearch} provides additional semantic signals to the agents through self-reflection.
Similarly, the plan reflection module in UFO \cite{zhang_ufo_2024} prompts the agent to continuously modify its plan at each decision step, allowing deviations from the original route as needed. 
Agent S \cite{agashe_agent_2024} adopts a similar approach by offering reflective suggestions based on the observation of the entire execution trajectory of subtasks, helping the agent consider alternative strategies and avoid repetitive actions.
Moreover, the input to the feedback model is not limited to action trajectories.
For instance, Reflexion \cite{shinn2023reflexionlanguageagentsverbal} generates linguistic self-feedback based on reward signals, trajectory information, and relevant memories, offering detailed and valuable guidance for future attempts of the agent.
Hu et al. \cite{hu2024selfevolvingmultiagentcollaborationnetworks} offer concise summary feedback of historical actions and current step information in each iteration, which reduces feedback overhead and mitigates performance degradation of the agent caused by information overload.
\subsection{Interaction}
\label{sec:interaction}
GUI Agents are intelligent systems designed to interact with graphical user interfaces to complete various tasks to help users. These agents rely on sophisticated interaction mechanisms to effectively navigate and manipulate interface elements. The interaction capabilities of these agents can be refined into three closely related components: Action Goal, Action Generation, and Action Space. Action Goal defines the specific objectives that guide the agent's task execution, directly aligning with user requirements to help complete daily tasks more efficiently and enhance productivity (\S\ref{sec:action_goal}). Action Generation involves the agent's ability to accurately interpret user instructions, deeply understand the GUI environment, and generate executable actions through processes like GUI Grounding and various Generation Strategies (\S\ref{sec:action_gen}). Action Space represents the complete set of possible operations available to the agent within interface constraints, including User Action Simulation and API Invocation approaches (\S\ref{sec:action_space}).

\subsubsection{Action Goal}
\label{sec:action_goal}
Action Goal defines the specific objectives that guide the GUI Agent's task execution. These objectives directly align with user requirements and ensure the agent efficiently and accurately completes the requested operations. Action goal encompasses diverse application areas including smartphone interaction \cite{zhang_appagent_2023}, work task automation, table data processing \cite{li2023sheetcopilotbringingsoftwareproductivity}, online shopping assistance \cite{2023arXiv230908172M}, gaming automation \cite{tan_cradle_2024}, and web navigation. Fundamentally, Action goal enables GUI Agents to help users complete daily tasks more efficiently, enhancing productivity and simplifying complex processes.
By achieving these goals, the GUI Agents becomes a capable assistant for enhancing productivity and simplifying task handling.

To achieve different goals, action generation is a critical step. Action generation involves the GUI Agent's ability to accurately interpret user instructions, deeply understand the GUI environment, and generate executable actions based on this understanding. This process forms the core of the Agent's task completion capabilities and encompasses various methods and strategies, which will be elaborated in the following sections.

\subsubsection{Action Generation}
Action Generation is the core process through which GUI Agents transform user intentions into executable operations. Unlike traditional intelligent agents that operate solely through API calls, GUI Agents distinctively need to interact directly with graphical interface elements, particularly through precise clicking operations on the screen \cite{zhang2025apiagentsvsgui}. Through our systematic review, we have categorized Action Generation into two key aspects: GUI Grounding and Action Strategy.
\label{sec:action_gen}
\paragraph{\textbf{GUI Grounding}} refers to the specialized process by which GUI Agents establish connections between user instructions and interface elements for the purpose of generating precise click operations. GUI Agents must directly interact with graphical environments through screen-based interactions, making GUI Grounding a distinctive and critical capability. This process enables agents to determine exact coordinates for clicks by mapping abstract instructions to specific interface elements.

For GUI Agents, effective grounding represents a fundamental requirement for successful interaction, as it allows them to navigate and manipulate interfaces through targeted clicking operations. This element localization process—essential for any screen-based interaction—serves as the foundation upon which other actions can be built. Current approaches to GUI grounding can be categorized into three methodologies:

\begin{itemize}
\item \textit{MLLMs-based Grounding} leverages both visual and textual information to connect commands with interface elements. \cite{2024arXiv240207945N} utilizes screen captures alongside user commands to decompose complex tasks into sequential subtasks. \cite{zhang_appagent_2023} enhances this approach by incorporating detailed XML descriptions of interaction elements. Further advancements by \cite{wang_mobile-agent_2024} and \cite{wang_mobile-agent-v2_2024} integrate specialized tools like OCR for text localization, icon detection, and CLIP for visual element identification. \cite{2023arXiv231107562Y} employs numerical labeling of elements detected through OCR and IconNet, while \cite{2024arXiv240113919H} uses GPT-4V-ACT to extract and annotate interaction elements with supplementary HTML analysis.
\item \textit{LLMs-based Grounding} converts GUI structures into formats that large language models can effectively process. \cite{wen_autodroid_2024} transforms GUIs into simplified HTML with dedicated tags, systematically exploring all interactive elements by scrolling through components and recording visible UI elements. \cite{lai_autowebglm_2024} combines HTML information with OCR processing to create more accurate element descriptions. Taking a different approach, \cite{wu_os-copilot_2024} focuses on command interpretation rather than direct GUI understanding, using directed acyclic graphs for task decomposition while incorporating external knowledge and internal memory.
\item \textit{Domain-specific Grounding} approaches target particular application contexts with specialized techniques. \cite{li2023sheetcopilotbringingsoftwareproductivity} focuses on spreadsheet software, analyzing column names and row counts to enhance task parsing. \cite{2024arXiv240204476K} employs an innovative ranking system for HTML elements, enriching representation by identifying visual neighbors of candidate elements and combining their markup and visual features to create contextually rich representations.
\end{itemize}

\paragraph{\textbf{Generation Strategy}}
While GUI Grounding focuses specifically on element-targeting clicks, GUI Agents must also generate a broader range of operations including scrolling, navigation, typing, and multi-step interactions. These non-click operations require different generation approaches that go beyond simple element localization. Based on our systematic review, we identify two primary strategies for generating these complex operation sequences:
\begin{itemize}
    \item \textit{Memory-Based Action Generation} leverages information from previous interactions or external knowledge. Several approaches maintain operational history to inform future actions. \cite{wang_mobile-agent_2024}, \cite{2023arXiv231107562Y}, and \cite{cheng_seeclick_2024} maintain operation flows that record task history, generating next actions based on instructions and previous operations. \cite{wang_mobile-agent-v2_2024} enhances this approach by storing focused content from historical screens. \cite{zhang_appagent_2023} and \cite{wen_autodroid_2024} build memories through autonomous exploration, retrieving similar scenarios to guide action generation for new tasks. External knowledge integration appears in works like \cite{li2023sheetcopilotbringingsoftwareproductivity}, which provides detailed documentation for spreadsheet operations, and \cite{2024arXiv241008164A}, which combines web-searched knowledge with historical experience to generate subtasks.
    
    \item \textit{Planning-Based Action Generation} relies on structured plans created before execution. \cite{2023arXiv230911436Z} generates not only immediate actions but also future action plans as references. \cite{zhang_ufo_2024} distinguishes between global and local planning, where a HostAgent generates coarse-grained action plans and selects appropriate applications, while an AppAgent \cite{zhang_appagent_2023} executes specific operations within those applications. Task decomposition approaches like \cite{2024arXiv240207945N} and \cite{2024arXiv241008164A} break complex tasks into sequential subtasks. These planning strategies enable GUI Agents to systematically approach complex tasks, adapting to environmental changes while maintaining focus on overall objectives.
\end{itemize}

\subsubsection{Action Space}
\label{sec:action_space}
\begin{figure}[h]
    \centering
    \includegraphics[width=0.50\textwidth]{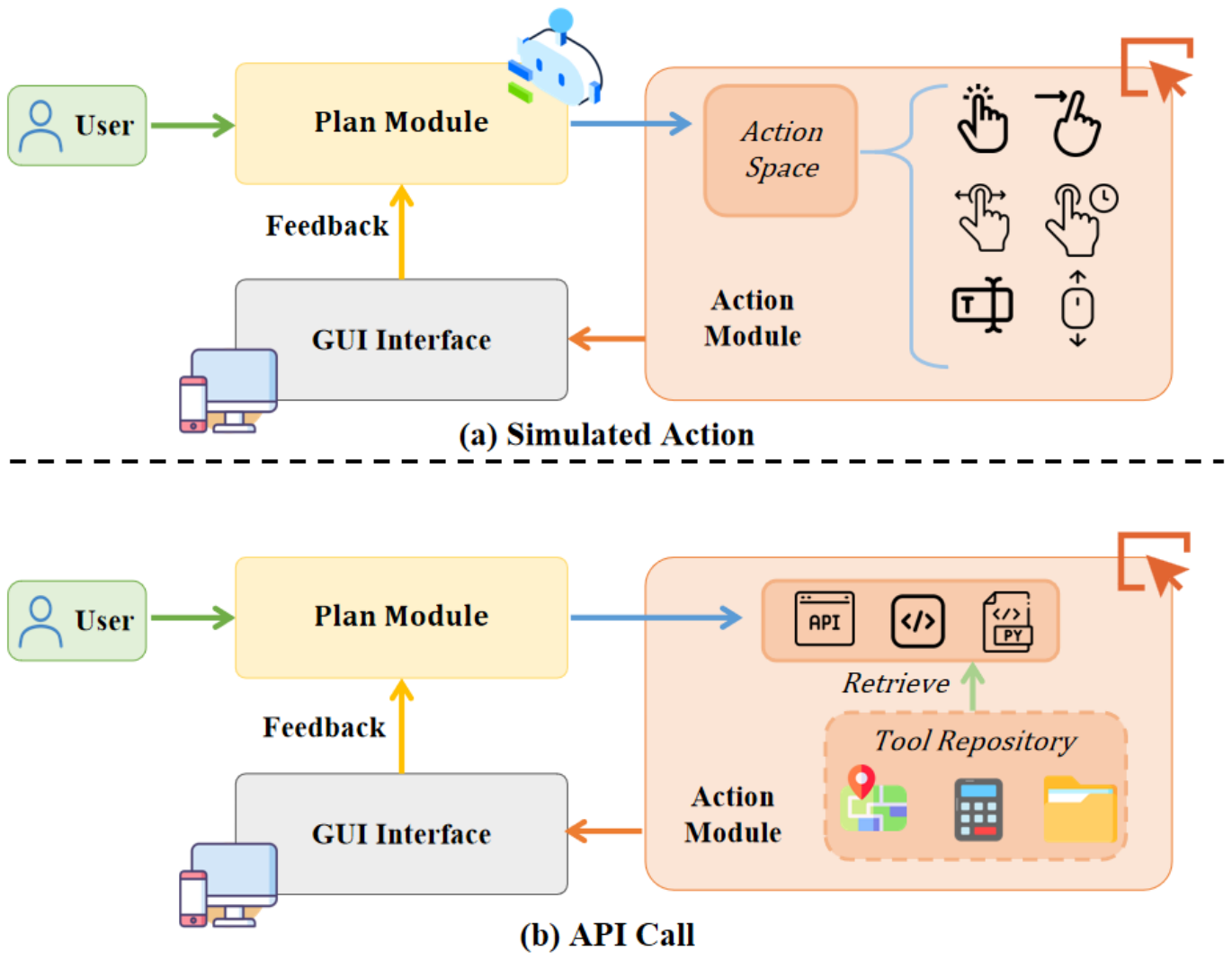}
    \caption{Different action space for GUI agents: (a) Simulated Action, and (b) API Call.}
    \label{fig:action}
\end{figure}

After completing action generation, GUI Agents require a well-defined action space to execute operations effectively. The action space represents the complete set of possible operations available to the agent within interface constraints. These constraints vary across different environments, limiting what actions can be performed in specific contexts, as shown in Figure \ref{fig:action}. For GUI Agents, action capabilities can be divided into two major categories:

(1) \textit{User Action Simulation} involves simulating typical user interactions with graphical interfaces. These actions mimic human behaviors such as clicking, typing, and swiping. \cite{zhang_appagent_2023} defines a basic set of operations including (Tap, Long\_Press, Swipe, Text, Back, and Exit) to replicate human-screen interactions. Some systems like \cite{zhang_appagent_2023}, \cite{2023arXiv231107562Y}, and \cite{zhang_ufo_2024} use numerical element labeling, directing actions to specific numbered elements (e.g., Tap(5)). Others, such as \cite{2024arXiv240302713Z} and \cite{cheng_seeclick_2024}, rely on coordinate-based targeting (e.g., Click(0.3,0.9)). These differences stem from varying GUI understanding methods: either annotating elements directly on screenshots or describing element positions textually.

(2) \textit{API Invocation Actions} leverages external or custom APIs rather than simulating direct user interactions. In \cite{wu_os-copilot_2024}, an executor generates executable function calls with appropriate parameters based on configuration hints. These might include command line instructions like "mkdir new folder" or other system calls. When suitable tools aren't available, a tool generator creates customized solutions for specific tasks.
Through these varied action spaces, GUI Agents can flexibly perform diverse tasks across different scenarios while maintaining operational safety and accuracy.

\section{Application}
\label{sec:application}
The development of GUI Agents has enabled automation across diverse computing environments, each presenting unique challenges and opportunities. This section examines four major application domains: mobile devices, desktop computers, web browsers, and games, analyzing their distinct characteristics and technical requirements. These four application domains demonstrate both the versatility of GUI Agents and the specific technical challenges each environment presents, as shown in Figure \ref{fig:enter-label}. The continued development of these applications drives innovation in core technologies while revealing new opportunities for automation and human-computer interaction.

\begin{figure}[h]
    \centering
    \includegraphics[width=1.0\linewidth]{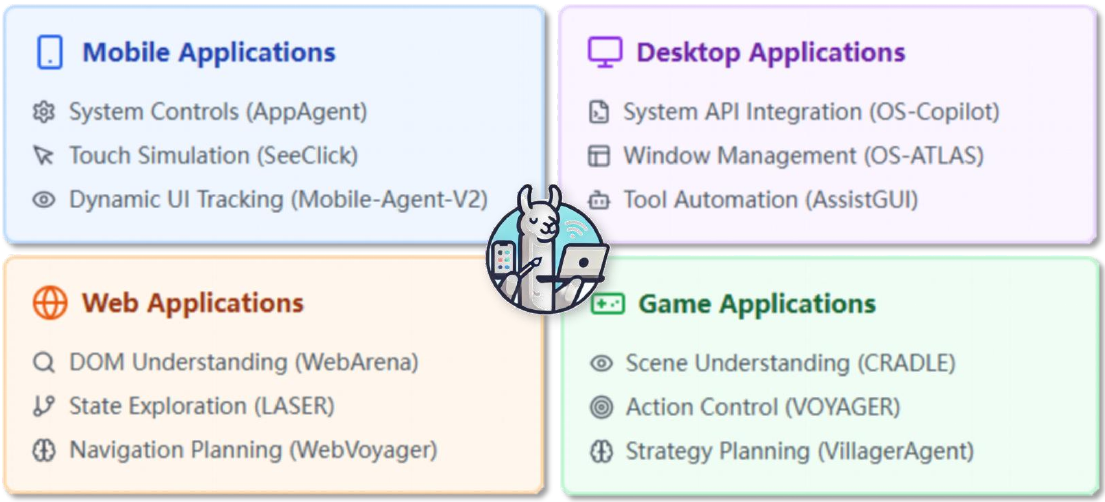}
    \caption{Applications of GUI Agents}
    \label{fig:enter-label}
\end{figure}

\subsection{Mobile}

Mobile environments represent a primary application domain for GUI Agents, characterized by touch-based interactions and dynamic interfaces. The proliferation of mobile applications has created significant opportunities for automation while introducing unique technical challenges.

Mobile GUI Agents have demonstrated remarkable capabilities across various tasks. In system operations, AppAgent \cite{zhang_appagent_2023} achieves sophisticated control over device settings, handling brightness adjustments, volume control, and other system configurations. Mobile-Agent \cite{wang_mobile-agent_2024} extends these capabilities by implementing comprehensive application management, including installation, updates, and removal procedures. For application interactions, CoCo-Agent \cite{ma_coco-agent_2024} enables automated social media engagement, while MobileFlow \cite{ nong_mobileflow_2024} specializes in productivity applications, offering particular advantages for Chinese-language interfaces.

The technical implementation of mobile GUI Agents faces several distinct challenges. First, the touch-based interaction model requires precise spatial understanding and gesture simulation. SeeClick \cite{cheng_seeclick_2024} addresses this through advanced element localization techniques, achieving high-precision touch event simulation. Second, mobile interfaces are highly dynamic, with frequent state changes and transitions. Mobile-Agent-V2 \cite{wang_mobile-agent-v2_2024} tackles this challenge through a sophisticated state tracking system, maintaining contextual awareness across application switches and screen updates.

Security considerations are particularly crucial in mobile environments. Recent implementations like AppAgent-V2 \cite{li_appagent_2024} incorporate comprehensive safety protocols, requiring explicit user confirmation for sensitive operations and maintaining detailed operation logs. This security-first approach ensures reliable automation while protecting user data and device integrity.

\subsection{Desktop}

Desktop environments offer GUI Agents extensive control capabilities through rich APIs and diverse interaction methods. The complexity of desktop applications demands sophisticated automation strategies and robust error handling mechanisms.

Desktop GUI Agents have achieved significant breakthroughs in system-level automation. OS-Copilot \cite{wu_os-copilot_2024} demonstrates comprehensive control over operating system functions, including file management, application control, and system configuration. UFO \cite{zhang_ufo_2024} specializes in Windows environments, providing precise interaction with native applications and system utilities. In productivity applications, AssistGUI \cite{gao2023assistgui} enables sophisticated automation of office suites, while SheetCopilot \cite{li2023sheetcopilotbringingsoftwareproductivity} offers specialized capabilities for spreadsheet manipulation.

These implementations leverage multiple technical approaches to achieve reliable automation. OS-Copilot \cite{wu_os-copilot_2024} utilizes a hybrid approach, combining direct API calls with GUI interaction when necessary. UFO \cite{zhang_ufo_2024} implements a sophisticated element detection system that maintains accuracy across different Windows versions and display configurations. OS-ATLAS \cite{wu_os-atlas_2024} addresses the challenge of multi-window environments through a hierarchical attention mechanism, effectively managing complex desktop layouts.

A notable advancement in desktop automation is the integration of natural language understanding with system operations. Recent work like OS-Copilot \cite{wu_os-copilot_2024} demonstrates the ability to translate high-level user instructions into precise sequences of system operations, bridging the gap between user intent and technical execution.

\subsection{Web Browser}

Web browsers present a standardized yet dynamic environment for GUI Agents, with DOM structures providing a consistent interface layer while supporting diverse interactive experiences.

Web-focused GUI Agents have demonstrated remarkable versatility. WebVoyager \cite{2024arXiv240113919H} executes complex web navigation tasks through sophisticated page understanding and interaction planning. LASER \cite{2023arXiv230908172M} implements a state-space exploration approach for robust e-commerce automation, handling dynamic content and variable page layouts. AutoWebGLM \cite{lai_autowebglm_2024} specializes in form interactions and data extraction, with particular emphasis on maintaining consistency across different website designs.

The technical implementation of web GUI Agents leverages several key innovations. WebArena's \cite{zhou_webarena_2024} DOM-based understanding system provides robust element identification and interaction planning. LASER's \cite{2023arXiv230908172M} state tracking mechanism enables reliable navigation through complex web applications, maintaining context across page loads and state transitions. These agents must handle AJAX updates, dynamic content loading, and varying network conditions, requiring sophisticated state management and error recovery mechanisms.

\subsection{Game}

Game environments present unique challenges for GUI Agents, requiring real-time decision-making and precise control execution in dynamic, visually complex environments.

Game-oriented GUI Agents have achieved impressive results across different genres. VOYAGER \cite{wang2023voyageropenendedembodiedagent} demonstrates sophisticated resource management and planning in Minecraft, combining visual understanding with strategic decision-making. Cradle \cite{tan_cradle_2024} implements precise control mechanisms for action games, while VillagerAgent \cite{dong2024villageragentgraphbasedmultiagentframework} manages complex task dependencies in multi-agent scenarios.

The technical implementation of game GUI Agents requires several innovative approaches. Real-time visual understanding is achieved through specialized perception models, often combining multiple analysis techniques. Cradle \cite{tan_cradle_2024} integrates GPT-4V with specialized game state understanding models, enabling robust scene interpretation. Action planning systems must handle both strategic decisions and tactical execution, often operating under strict time constraints.

A significant innovation in game automation is the development of skill learning systems. VOYAGER \cite{wang2023voyageropenendedembodiedagent} implements a progressive skill acquisition mechanism, allowing the agent to learn and combine basic actions into complex operation sequences. This approach enables adaptation to new game scenarios and the development of sophisticated playing strategies.

·\section{Datasets and Benchmarks}
\label{sec:datasets}
A comprehensive understanding of datasets and benchmarks is essential for advancing research in GUI Agents. In this section, we analyze existing benchmarks and datasets from three critical dimensions: the evaluation environments (\S\ref{sec:envir}) in which agents operate, the specific tasks (\S\ref{sec:task}) they are designed to accomplish, and the evaluation strategies (\S\ref{sec:eval_stra}) employed to assess their performance. Additionally, we critically examine the significant challenges (\S\ref{sec:Challenges}) facing current datasets and benchmarks.
\begin{table*}[htbp]
\centering
\begin{tabular}{p{3cm}p{0.6cm}p{3cm}p{2cm}p{1.8cm}p{0.6cm}p{0.6cm}p{0.6cm}p{1.0cm}}
\hline
\multirow{2}{*}{Datasets} & \multirow{2}{*}{\makecell{Env.}} & \multirow{2}{*}{Eval Method}  & \multirow{2}{*}{Eval Strategy} & \multirow{2}{*}{Platform} & \multicolumn{3}{c}{Task}          & \multirow{2}{*}{Time} \\ \cline{6-8}
                                     &                                                 &                                     &                                      &                           & Per. & Ope. & Dial. &                       \\ \hline
WoB\cite{shi2017world}                                  & ×                                               & HTML/JS state                       & Goal                                 & web                       & \checkmark          & \checkmark         &          & 2017.08               \\
Rico\cite{deka2017rico}                                 & ×                                               & /                                   & /                                    & mobile                    &            & \checkmark         &          & 2017.10               \\
MiniWoB++\cite{liu2018reinforcement}                            & ×                                               & HTML/JS state                       & Goal                                 & web                       &            & \checkmark         &          & 2018.02               \\
PIXELHELP\cite{li2020mapping}                            & ×                                               & action-match                        & Trajectory                           & mobile                    &            & \checkmark         &          & 2020.06               \\
WebSRC\cite{chen2021websrc}                               & ×                                               & UI/text match                       & Goal                                 & web                       & \checkmark          &           &          & 2021.11               \\
MoTIF\cite{burns2022dataset}                                & ×                                               & action/UI-match                     & Trajectory                           & mobile                    &            & \checkmark         &          & 2022.08               \\
META-GUI\cite{sun2022meta}                             & ×                                               & action-match                        & Trajectory                           & mobile                    &            & \checkmark         & \checkmark        & 2022.11               \\
WebShop\cite{yao2022webshop}                              & ×                                               & product attribution match           & Goal                                 & web                       &            & \checkmark         &          & 2023.02               \\
AITW\cite{rawles2024androidinthewild}                                 & ×                                               & action-match                        & Trajectory                           & mobile                    &            & \checkmark         &          & 2023.10               \\
Mind2Web\cite{deng2024mind2web}                             & ×                                               & action/UI-match                     & Trajectory                           & web                       &            & \checkmark         &          & 2023.12               \\
AssistGUI\cite{gao2023assistgui}                            & \checkmark                                               & image match/device state            & Goal                                 & desktop                   &            & \checkmark         &          & 2024.01               \\
MT-Mind2Web\cite{deng2024multi}                          & ×                                               & action/UI-match                     & Trajectory                           & web                       &            & \checkmark         & \checkmark        & 2024.02               \\
WebVLN\cite{chen2024webvln}                               & ×                                               & url/text match                      & Goal/Graph                           & web                       & \checkmark          & \checkmark         &          & 2024.03               \\
WebArena\cite{zhou2023webarena}                             & ×                                               & url/text match/LLM                  & Goal/Trajectory                      & web                       & \checkmark          & \checkmark         &          & 2024.04               \\
MMInA\cite{zhang2024mmina}                                & \checkmark                                               & text-match/LLM                      & Trajectory                           & web                       &            & \checkmark         &          & 2024.04               \\
VisualWebBench\cite{liu2024visualwebbench}                       & ×                                               & action/UI/text-match                & Goal                                 & web                       & \checkmark          & \checkmark         &          & 2024.04               \\
OSWorld\cite{xie2024osworld}                              & \checkmark                                               & device state                        & Goal                                 & desktop                   &            & \checkmark         &          & 2024.05               \\
Mobile-Env\cite{zhang2023mobile}                           & \checkmark                                               & device state                        & Goal                                 & mobile                    &            & \checkmark         &          & 2024.06               \\
VisualWebArena\cite{koh2024visualwebarena}                       & \checkmark                                               & url/text/image match/HTML state/LLM & Goal                                 & web                       & \checkmark          & \checkmark         &          & 2024.06               \\
VideoGUI\cite{lin2024videogui}                             & \checkmark                                               & action-match/LLM                    & Goal                                 & desktop                   &            & \checkmark         &          & 2024.06               \\
MobileAgentBench\cite{wang2024mobileagentbench}                     & \checkmark                                               & UI match                            & Goal                                 & mobile                    &            & \checkmark         &          & 2024.06               \\
GUICourse\cite{chen2024guicourse}                            & ×                                               & /                                   & /                                    & Web\&mobile               & \checkmark          & \checkmark         & \checkmark        & 2024.06               \\
GUI-WORLD\cite{chen2024gui}                            & ×                                               & text-match/LLM                      & Goal                                 & Multi Platforms           & \checkmark          & \checkmark         & \checkmark        & 2024.06               \\
GUI Odyssey\cite{lu2024gui}                          & ×                                               & action-match                        & Trajectory                           & mobile                    &            & \checkmark         &          & 2024.06               \\
WebCanvas\cite{pan2024webcanvas}                            & \checkmark                                               & UI/url match                        & Trajectory                           & web                       &            & \checkmark         &          & 2024.07               \\
Spider2-V\cite{cao2024spider2}                            & \checkmark                                               & device state                        & Goal                                 & desktop                   &            & \checkmark         &          & 2024.07               \\
OmniACT\cite{kapoor2024omniact}                              & ×                                               & action-match                        & Trajectory                           & desktop                   &            & \checkmark         &          & 2024.07               \\
AMEX\cite{chai2024amex}                                 & ×                                               & action-match                        & Trajectory                           & mobile                    & \checkmark          & \checkmark         &          & 2024.07               \\
LlamaTouch\cite{zhang2024llamatouch}                           & \checkmark                                               & UI/action-match/device state        & Trajectory                           & mobile                    &            & \checkmark         &          & 2024.08               \\
AndroidArena\cite{xing2024understanding}                         & \checkmark                                               & action-match/LLM                    & Trajectory                           & mobile                    &            & \checkmark         &          & 2024.08               \\
WindowsAgentArena\cite{bonatti2024windows}                    & \checkmark                                               & device state                        & Goal                                 & desktop                   &            & \checkmark         &          & 2024.09               \\
WebLINX\cite{lu2024weblinx}                              & ×                                               & action/UI/text-match                & Goal                                 & web                       &            & \checkmark         & \checkmark        & 2024.09               \\
AgentStudio\cite{zheng2024agentstudio}                          & \checkmark                                               & device state                        & Goal                                 & desktop                   &            & \checkmark         &          & 2024.10               \\
B-MOCA\cite{lee2024benchmarking}                               & \checkmark                                               & device state                        & Goal                                 & mobile                    &            & \checkmark         &          & 2024.10               \\
AndroidWorld\cite{rawles2024androidworld}                         & \checkmark                                               & device state                        & Goal                                 & mobile                    &            & \checkmark         &          & 2024.10               \\
CRAB\cite{xu2024crab}                                 & \checkmark                                               & UI/image-match/device state         & Graph                                & Multi Platforms           &            & \checkmark         &          & 2024.10               \\
ScreenPR\cite{fan2024read}                             & ×                                               & human/LLM/text match                & Goal                                 & mobile                    & \checkmark          &           &          & 2024.10               \\
SPA-BENCH \cite{chen2025spabenchcomprehensivebenchmarksmartphone} & $\checkmark$ & LLM & Goal/Trajectory & Mobile &  & $\checkmark$ &  & 2024.10 \\
Android Lab \cite{xu_androidlab_2024} & $\checkmark$ & device state & device state & Mobile &  & $\checkmark$ &  & 2024.10 \\
ANDROIDCONTROL\cite{li2024effects}                       & ×                                               & action-match                        & Trajectory                           & mobile                    &            & \checkmark         &          & 2024.11               \\
MultiUI\cite{liu2024harnessing}                              & ×                                               & /                                   & /                                    & web                       & \checkmark          & \checkmark         &          & 2024.11               \\ 

GUI Testing Arena \cite{zhao2024guitestingarenaunified} & $\checkmark$ & Action match/LLM & Goal/Trajectory & Mobile & $\checkmark$ & $\checkmark$ &  & 2024.12 \\

A3 \cite{chai2025a3androidagentarena} & $\checkmark$ & UI/Action match/LLM & Goal/Trajectory & Mobile &  & $\checkmark$ &  & 2025.01 \\

WebWalkerQA \cite{wu2025webwalkerbenchmarkingllmsweb} & $\checkmark$ & LLM & Goal & Web &  & $\checkmark$ & $\checkmark$ & 2025.01 \\

ScreenSpot-Pro \cite{li2024screenspot-pro} & $\times$ & UI match & Goal & Desktop & $\checkmark$ &  &  & 2025.01 \\

WorldGUI \cite{zhao2025worldguidynamictestingcomprehensive} & $\checkmark$ & UI match/device state & Goal & Web\&desktop &  & $\checkmark$ &  & 2025.02 \\

UI-Vision \cite{nayak2025uivisiondesktopcentricguibenchmark} & $\times$ & UI/action match & Goal & Desktop & $\checkmark$ & $\checkmark$ &  & 2025.03 \\
\hline
\end{tabular}
\caption{A Comprehensive Survey of GUI Interaction Datasets (2017-2024): Comparison of Real Environment Support, Evaluation Methods, Platforms, and Task Types (Perception, Operation, and Dialogue)}
\label{tab:xhl1}
\end{table*}

\subsection{Environment}
\label{sec:envir}
Effective evaluation is critical for assessing and advancing GUI Agents' capabilities in real-world scenarios. Through a comprehensive review of existing literature, we identify that the evaluation environments provided by current benchmarks and datasets related to GUI Agents can be categorized into three types: static replicas, simulated environments, and real-world environments.
\subsubsection{Static Replicas}
Static replicas are evaluation environments that capture GUI interfaces as fixed, non-interactive representations such as screenshots or HTML snapshots. These environments record interactions with the original GUI as static data, allowing for reproducible evaluation without requiring the actual application to be running. The static replica approach records interactions with the environment as static data, as shown in Table \ref{tab:xhl1}.

AitW \cite{rawles2024androidinthewild} constructed a large-scale dataset of human-annotated action sequences and associated application state data based on static screenshots, comprising 715,142 episodes, 5,689,993 screenshots, and 30,378 unique task instructions. PIXELHELP \cite{li2020mapping} documented touch event types during interactions, the UI objects manipulated, and the view hierarchy. Mind2Web \cite{deng2024mind2web} preserved web interactions in the form of static HTML snapshots, covering 31 domains and including over 2,000 tasks. These works facilitate the generation of large-scale datasets. However, since annotators cannot exhaustively cover all possible execution paths, it is impossible to explore or replay uncached operational paths during evaluation. Consequently, this method constitutes only a "pseudo-interaction", which may result in false negatives during evaluation. This limitation is discussed in detail in Section (\S\ref{subsec:Trajectory}).
\subsubsection{Simulated Environments}
Simulated environments replicate the real world to create fully isolated and controllable interaction scenarios, which eliminates variables introduced by dynamic online content, thereby improving the repeatability of evaluations.

MiniWoB in WoB \cite{shi2017world} and MiniWoB++ \cite{liu2018reinforcement} manually designed small-scale HTML pages to simulate interactions in real-world web tasks. FormWoB in WoB \cite{shi2017world} converted four real flight booking websites into web tasks, recording all HTTP requests and responses to create offline approximations of these websites. WebShop \cite{yao2022webshop} scraped product information from amazon.com to create a simulated e-commerce site. WebArena \cite{zhou2023webarena} enhanced realism by developing four fully operational, self-hosted web applications, each representing a distinct common domain of the internet. VisualWebArena \cite{koh2024visualwebarena} built upon WebArena \cite{zhou2023webarena} by incorporating visual modality information. Most tasks in the MMInA \cite{zhang2024mmina} dataset were derived from real websites. However, for tasks requiring image extraction from HTML files, offline and open-source websites were used as substitutes to bypass web protection mechanisms.

Despite these advances, applying GUI Agents to real-world scenarios still requires dynamic and authentic environments for evaluation. Online evaluation often involves numerous uncertainties. For instance, flight booking tasks may be closely tied to weather conditions that cannot be accounted for by static evaluation methods. Additionally, factors such as network fluctuations, security verifications, and popup advertisements must also be considered. Avoiding uncertainties entirely would hinder the ability of GUI Agents to handle complex real-world tasks effectively.
\subsubsection{Real-World Environments}
In real-world environments, tasks are executed on actual websites or applications, requiring GUI Agents to interact directly with the real world.

Benchmarks like MobileAgentBench \cite{wang2024mobileagentbench}, AndroidWorld \cite{rawles2024androidworld} and B-MOCA \cite{lee2024benchmarking} provide evaluation environments for real tasks on mobile devices. OSWorld \cite{xie2024osworld} and WindowsAgentArena \cite{bonatti2024windows} establish testing frameworks for desktop platforms. CRAB \cite{xu2024crab} goes a step further by designing cross-platform tasks, such as taking photos with a smartphone and transferring the images to a desktop computer for editing. Constructing web-related tasks is particularly challenging due to the rapid evolution of the online environment, which can render previously annotated tasks obsolete. For instance, many tasks in Mind2Web \cite{deng2024mind2web} are no longer functional on corresponding live websites. To address this, WebCanvas \cite{pan2024webcanvas} offers an efficient maintenance mechanism that quickly identifies the validity of annotated action sequences and key nodes through periodic monitoring and automated alerts, ensuring that tasks remain operable in real-world environments.

Although the inherent uncertainties of real-world environments pose significant challenges to task creation and maintenance, constructing tasks within such environments has become a prevailing research trend. This development is crucial for advancing GUI Agents capabilities.

\subsection{Task}
\label{sec:task}
Evaluating the performance of GUI Agents requires a clear understanding of the specific tasks they are designed to accomplish. Current benchmarks and datasets for GUI Agents can be categorized into three fundamental task types: perception, operation, and dialogue, each targeting different aspects of agent capabilities.
\subsubsection{Perception}
Perception serves a fundamental cognitive process essential for GUI Agents' interaction with graphical user interfaces, encompassing the recognition, extraction, and interpretation of interface elements. This comprehensive understanding of the interface architecture provides the foundation for subsequent decision-making processes and user interactions.
The research community has developed numerous benchmarks and datasets specifically annotated for perception tasks. WebArena \cite{zhou2023webarena} and VisualWebArena \cite{koh2024visualwebarena} feature web information retrieval challenges that test agents' visual comprehension capabilities. WebSRC \cite{chen2021websrc} introduces Web-Based Structural Reading Comprehension, which requires agents to analyze both spatial layouts and logical hierarchies of web pages to respond to queries accurately. ScreenPR \cite{fan2024read} evaluates agents' ability to generate precise explanations for specific screen regions based on user interaction points. MultiUI \cite{liu2024harnessing} focuses on embedded image interpretation within web environments. VisualWebBench \cite{liu2024visualwebbench} assesses webpage comprehension through captioning tasks that measure an agent's capacity to synthesize and summarize content holistically. GUICourse \cite{chen2024guicourse} contributes GUIEnv, a comprehensive dataset for optical character recognition and visual grounding. Finally, AMEX \cite{chai2024amex} enhances the evaluation landscape by providing detailed screen descriptions coupled with functional annotations for interface components.
\subsubsection{Operation}
Operation tasks represent a critical stage in GUI Agents interactions, where agents execute specific actions to achieve designated goals. These tasks encompass scenarios such as complex software operations and web navigation, rigorously assessing an agent's planning and execution capabilities.

The operational complexity spectrum begins with simple single-step operations, where agents address straightforward, well-defined tasks. For instance, MultiUI \cite{liu2024harnessing} provides single-step operation prediction tasks that test basic action selection. Similarly, the GUIAct dataset within GUICourse \cite{chen2024guicourse} offers 67,000 single-step operation instructions specifically designed to evaluate an agent's fundamental decision-making capabilities.
As complexity increases, many benchmarks focus on multi-step operations within specific software environments. AssistGUI \cite{gao2023assistgui} emphasizes productivity scenarios by collecting 100 tasks across nine commonly used applications, including Premiere Pro, After Effects, and PowerPoint. The evaluation landscape extends into professional domains through benchmarks like Spider2-V \cite{cao2024spider2}, which features data science and engineering software tasks in platforms such as BigQuery and Airbyte, and VideoGUI \cite{lin2024videogui}, which concentrates on video editing in professional tools like Adobe Photoshop and Stable Diffusion.

Another critical dimension of operational complexity emerges in web-based environments, where navigation tasks demand intricate interaction sequences across interconnected pages. WebShop \cite{yao2022webshop} simulates e-commerce scenarios requiring product search and selection, while MMInA \cite{zhang2024mmina} advances the challenge through multi-hop navigation tasks where agents must traverse multiple independent websites and integrate information from diverse sources to accomplish complex goals.
Comprehensive ecosystem evaluation spanning both application-specific and web-based operations is provided by benchmarks such as Mobile-Env \cite{zhang2023mobile}, OSWorld \cite{xie2024osworld}, AndroidWorld \cite{rawles2024androidworld}, and WindowsAgentArena \cite{bonatti2024windows}, which cover broader scenarios including application operations, system settings, programming tasks, and multimedia applications. The GUI Testing Arena \cite{zhao2024guitestingarenaunified} requires the Agent to discover GUI defects through exploration.
\subsubsection{Dialogue}
Most existing benchmarks and datasets assume that agents can complete tasks end-to-end upon receiving instructions, without requiring interaction with users during execution. However, in real-world scenarios, users may need to modify or augment their requirements mid-task, or agents may need to seek confirmation from users during critical decision-making stages. To address this, some works extend tasks to dialogue-based scenarios, aligning closer with real-world application demands.

For instance, the GUIChat modules in MT-Mind2Web \cite{deng2024multi}, WebLINX \cite{lu2024weblinx}, and GUICourse \cite{chen2024guicourse} design multi-turn dialogue tasks for web navigation and comprehension. META-GUI \cite{sun2022meta} provides task execution trajectories and associated dialogue data on the Android platform. GUI-WORLD \cite{chen2024gui} further expands to multi-platform environments, encompassing dialogue-based data for dynamic GUI understanding.
\subsection{Evaluation Strategy}
\label{sec:eval_stra}
\begin{figure*}[htbp]
    \centering
    \includegraphics[width=1\textwidth]{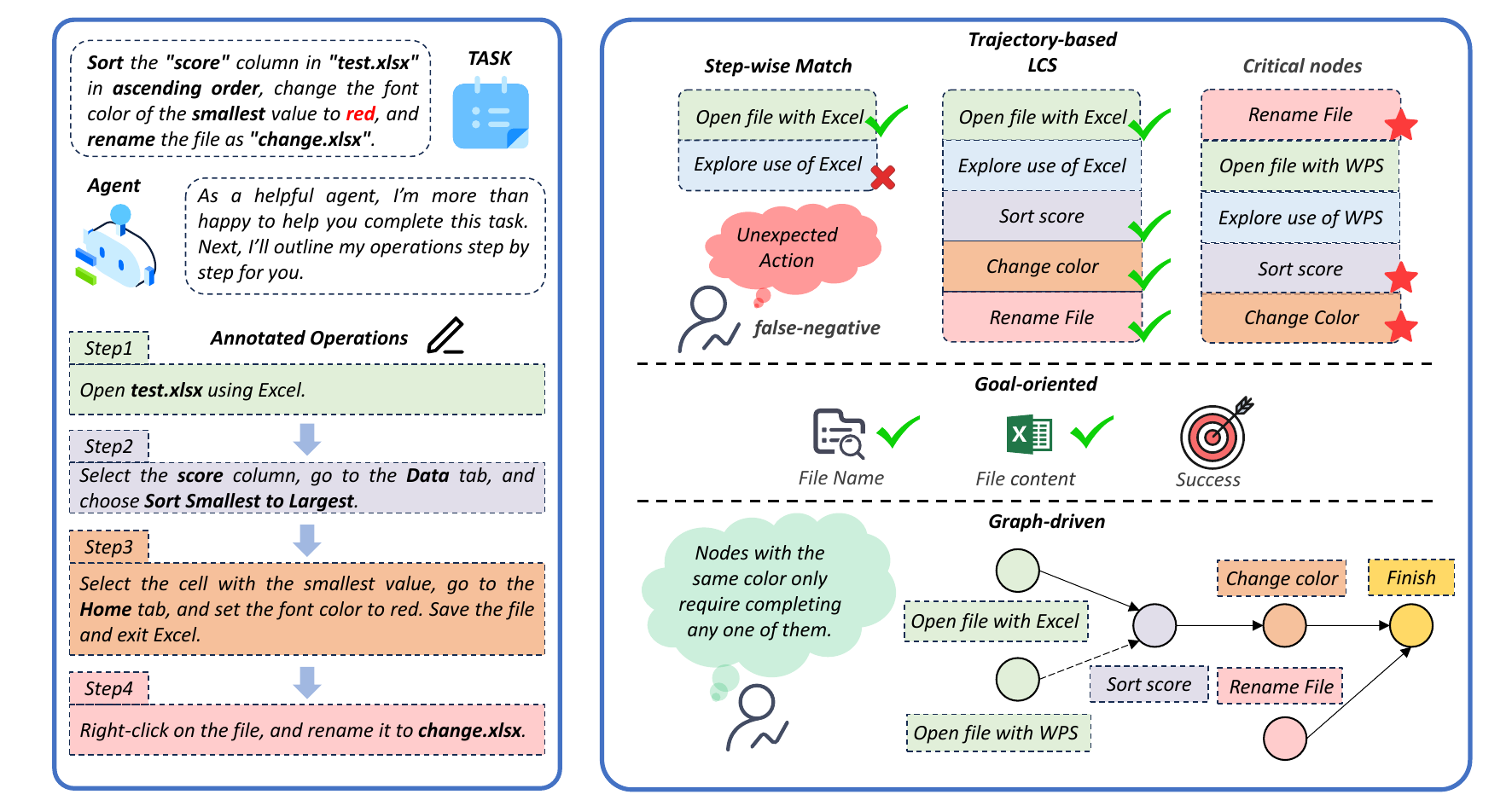}
    \caption{Comparison between different evaluation strategies. The left shows a sample Excel task with step-by-step operations, while the right contrasts three evaluation approaches - Trajectory-based (matching steps and critical nodes), Goal-oriented (assessing final outcomes), and Graph-driven (representing operations as connected nodes with alternative paths).}
    \label{fig:ikun}
\end{figure*}

Effective evaluation strategies are essential for accurately measuring the performance of GUI Agents across different tasks and environments. Through our literature review, we identify two predominant approaches to evaluation: trajectory-based methods that focus on the sequence of actions taken, and goal-oriented approaches that emphasize final outcomes regardless of the specific path. Figure \ref{fig:ikun} illustrates the fundamental differences between these evaluation strategies, highlighting how each method assesses agent performance from different perspectives. The trajectory-based methods track step-by-step actions against predefined paths, while goal-oriented approaches verify whether the final objective has been accomplished regardless of the specific steps taken.
\subsubsection{Trajectory-based methods}
\label{subsec:Trajectory}
Trajectory-based methods focus on analyzing the sequence of actions or states generated by an agent during task execution. Step-wise Action Match, a widely adopted evaluation approach  \cite{sun2022meta, rawles2024androidinthewild}, compares the action sequence generated by the agent with ground-truth to assess task completion. A task is considered successfully completed if the two sequences match exactly. However, this method may yield unfair evaluation outcomes since agents often require exploratory steps during task execution. If the agent performs actions not accounted for in the dataset, the task is deemed a failure.

To address this limitation, AndroidArena \cite{xing2024understanding} introduced the use of the Longest Common Subsequence (LCS) to compare sequences. Unlike traditional step-wise matching, LCS accommodates exploratory and redundant actions by tolerating deviations. Specifically, a match is considered successful if the ground-truth is a subsequence of the agent-generated sequence. Despite its advantages, such methods often lead to high false-negative rates in real-world evaluations. Annotators cannot exhaustively account for all possible execution paths, and predefined datasets typically provide only a single reference path. Given that multiple alternative solutions may effectively accomplish the task, human verification scores are often higher than those produced by trajectory-based methods, as noted in prior studies \cite{rawles2024androidinthewild, zhang2024llamatouch}.

To mitigate this issue, LlamaTouch \cite{zhang2024llamatouch}, WebCanvas \cite{pan2024webcanvas} and WebArena \cite{zhou2023webarena} focus solely on the critical nodes within the sequence. Critical nodes refer to actions or states that must be executed or encountered regardless of the method employed. A task is considered complete if the agent's trajectory contains all the annotated critical states. This approach adapts well to dynamic real-world environments and reduces the false-negative problem prevalent in earlier evaluation strategies.

In summary, trajectory-based methods emphasize the process of task execution by the agent. While they suffer from reliability issues in evaluation outcomes, they enable researchers to design fine-grained metrics, such as accuracy for specific action categories, by leveraging trajectory details.
\subsubsection{Goal-oriented approaches}
Goal-oriented approaches prioritize assessing whether an agent achieves the final objective, rather than the specific steps taken to do so. For tasks with straightforward evaluation criteria, completion can often be determined by a single signal match. For instance, GUI-WORLD \cite{chen2024gui} evaluates task success by comparing the similarity between the agent's output text and the ground-truth text. Mobile-Env \cite{zhang2023mobile} extends this approach by incorporating multiple signals and defining evaluation rules in a logic expression style. This allows flexibly handling scenarios requiring either the satisfaction of multiple conditions simultaneously or the fulfillment of any single condition.

For more complex and diverse tasks, such as system configuration or software operation, relying solely on surface-level signals like UI elements or screenshots often lacks robustness. For example, in a task involving editing a note and saving it, pressing the "Save" button might trigger only a brief success notification without causing any noticeable changes in the UI state. Furthermore, surface-level signals can sometimes be inaccessible. For instance, using UIAutomator may fail to dump the view hierarchy for dynamic screen content. To enhance the reliability of success evaluations, MobileAgentBench \cite{wang2024mobileagentbench} leverages Android Accessibility Service to capture application events (e.g., click events) as supplementary signals. This ensures a more accurate assessment of task success. Other approaches, such as MOBILE-ENV \cite{zhang2023mobile}, B-MOCA \cite{lee2024benchmarking} and AndroidWorld \cite{rawles2024androidworld}, utilize the Android Debug Bridge (ADB) to access underlying system states and generate reward signals. By gaining full access to system resources including the file system, application databases, and system settings, these methods provide a more comprehensive evaluation. Additionally, benchmarks such as OSWorld \cite{xie2024osworld} and WindowsAgentArena \cite{bonatti2024windows} extract critical components from the system's final state and design task-specific evaluation scripts. Compared to surface-level signal matching, this strategy achieves significantly higher accuracy in determining task success.

\subsubsection{Graph-driven techniques}
Recent research has introduced graph-driven evaluation techniques as a promising middle ground between trajectory-based and goal-oriented approaches. Crab \cite{xu2024crab} models tasks as directed graphs, encoding both sequential and parallel relationships between sub-nodes. In this framework, multiple paths to the target node elegantly represent alternative solutions, allowing for diverse yet valid execution strategies. The graph-based evaluator systematically decomposes complex tasks into verifiable sub-goals, with each sub-goal represented as a node and associated with a specific validation function that confirms its completion. Similarly, WebVLN \cite{chen2024webvln} applies graph theory to Vision-and-Language Navigation tasks by constructing a navigation graph where webpages form nodes and hyperlinks serve as edges. Task performance is quantitatively assessed by measuring the shortest path distance between the agent's final destination and the target webpage in the navigation graph. By combining the structural rigor of trajectory analysis with the flexibility to accommodate multiple solution paths, graph-driven methods offer a balanced evaluation approach. They provide the granular assessment metrics characteristic of trajectory-based methods while maintaining the solution flexibility inherent in goal-oriented approaches, effectively addressing limitations of both traditional evaluation paradigms.

\subsection{Challenges}
\label{sec:Challenges}
Current GUI Agents benchmarks and datasets face several critical challenges that impact their effectiveness and applicability. The evaluation process encounters significant cost barriers in terms of computational resources, time consumption, and financial investment, particularly for complex multi-step tasks requiring specialized models or human judgment. Additionally, existing benchmarks lack sufficient diversity in their content, interface representation, and cultural context, often being limited to English language, optimized interfaces, and regionally-specific examples. Finally, these evaluation environments raise important security and privacy concerns as they require extensive system access and may potentially expose sensitive information.
\subsubsection{Cost}
The evaluation of GUI Agent benchmarks faces significant cost challenges in terms of time, computational resources, and financial investment. For example, completing each task in OSWorld \cite{xie2024osworld} requires approximately 10–20 steps. Assuming that each step involves processing through local models and cloud API calls, taking 10–30 seconds per step, the serial evaluation of hundreds of tasks can take a day or longer. OSWorld \cite{xie2024osworld} mitigates this by allowing parallel simulation on a single machine to accelerate the evaluation process. WindowsAgentArena \cite{bonatti2024windows} introduces native cloud parallelization to further reduce memory and CPU demands. 

Additionally, some benchmarks involve expensive evaluations, particularly those requiring human judgment, such as ScreenPR \cite{fan2024read}. Benchmarks like VideoGUI \cite{lin2024videogui} and GUI-WORLD \cite{chen2024gui} rely on commercial LLMs for evaluation, adding monetary costs from API calls. Benchmark deployment also introduces a learning curve. User-friendly benchmarks, such as MobileAgentBench \cite{wang2024mobileagentbench}, offer evaluation parameter configuration with as few as ten lines of code. However, while this may be straightforward for researchers in the AI/ML community, it remains challenging for individuals without a technical background.

\subsubsection{Diversity}
The limited diversity of current datasets and benchmarks poses a significant challenge to developing robust and generalizable GUI Agents. Some datasets provide only textual content, failing to exploit multimodal information. Most datasets are currently limited to English, although multilingual models can be employed for translation. In Mind2Web \cite{deng2024mind2web}, the dataset primarily consists of websites frequently visited by U.S. users, which may not fully capture the browsing habits of global users. 

AndroidWorld \cite{rawles2024androidworld} includes only open-source applications with over one million downloads. These popular applications often have highly optimized user interfaces to enhance user experience, offering more features and shortcuts. This makes it more challenging to test agents on less-optimized interfaces. AITW \cite{rawles2024androidinthewild} captures UI drift across different versions of Android for Google applications. A promising solution is the environment randomization functionality in B-MoCA \cite{lee2024benchmarking}, which adjusts icon positions and sizes, wallpapers, language settings, and device types to simulate diverse real-world scenarios.

\subsubsection{Security}
The proliferation of GUI Agent benchmarks and datasets raises critical security and privacy concerns that must be addressed for responsible development and deployment. Collected data may contain sensitive and private information such as personal details, financial records, browsing histories, and authentication credentials. If datasets are misused, they could facilitate malicious purposes, such as bypassing security measures or enabling harmful activities.

The issue is compounded by the fact that GUI Agents require extensive access privileges to function effectively across applications and platforms. Benchmarks like OSWorld \cite{xie2024osworld} and WindowsAgentArena \cite{bonatti2024windows} operate with system-level permissions that could be exploited if not properly secured. Furthermore, as agents become more capable of executing complex sequences of actions, the potential for security vulnerabilities increases. For instance, agents designed to automate form-filling could be repurposed for credential harvesting or phishing attacks if their datasets are compromised.

Current security measures in benchmarks often focus on isolated environments but lack comprehensive threat modeling for real-world deployment scenarios. The absence of standardized security protocols for GUI Agent evaluation environments represents a significant gap in the field. Future benchmarks must incorporate robust anonymization techniques, permission management systems, and adversarial testing frameworks to ensure that GUI Agents can be evaluated thoroughly without compromising user security or privacy. Additionally, transparent documentation of data collection methodologies and explicit consent mechanisms are essential for maintaining ethical standards in benchmark development.
\section{Future Directions and Challenges}
\label{sec:future}
We have systematically organized GUI Agents into four core modules and examined existing datasets and benchmarks, we now discuss key challenges and future directions in this field. We explore limitations in data collection and benchmark development (\S\ref{subsec:data}), challenges in multimodal perception and visual grounding (\S\ref{subsec:grounding}), complexities in strategic planning (\S\ref{subsec:planning}), and the potential of reinforcement learning approaches (\S\ref{subsec:rl}) for enhancing GUI Agent capabilities in complex, real-world scenarios.
\subsection{Data Collection and Benchmark Development}
\label{subsec:data}
%
%
Data collection and comprehensive evaluation of GUI Agents remain critical challenges for future research. Current GUI Agent datasets primarily consist of multi-step interaction sequences, but existing datasets often fail to reflect real-world scenarios. For instance, they rarely capture how agents should resume tasks after interruptions, handle unexpected system prompts, or adapt to changing interface states. Future research should focus on developing more realistic multi-step datasets that incorporate these challenging scenarios, such as task resumption after interruptions, adaptation to dynamic interface changes, and handling of unexpected system responses.

The development of efficient methods to gather high-quality, diverse interaction data that accurately captures real-world usage patterns is essential. This includes creating benchmarks that test an agent's ability to maintain task context across interruptions and environmental changes. Equally important is establishing standardized evaluation frameworks that can holistically assess GUI Agent performance across dimensions such as accuracy, robustness, efficiency, and contextual understanding. Addressing these challenges will be crucial for advancing GUI Agent capabilities beyond simplified testing environments to truly functional real-world applications.
\subsection{Multimodal Perception and Visual Grounding}
\label{subsec:grounding}
Current GUI Agents still face significant limitations in screen understanding and visual grounding \cite{li2024screenspot-pro,cheng_seeclick_2024}. The fundamental challenge stems from the unique properties of GUI images—unlike natural images \cite{li_ferret-ui_2024},  they feature hierarchical layouts with numerous small elements that require precise comprehension. GUI screenshots present complex resolution challenges: while the overall image is large, individual interactive elements like icons and buttons are often extremely small, making accurate detection difficult \cite{cheng_seeclick_2024,li2024screenspot-pro}. This resolution discrepancy frequently leads multimodal language models to hallucinate interface elements or make critical localization errors during navigation tasks, severely limiting their reliability in real-world applications.
Future research can focus on developing specialized approaches to address these visual perception challenges. Promising directions include chain-of-thought reasoning to enhance contextual understanding of interface hierarchies and the integration of specialized visual parsing tools that can decompose complex screens into manageable components. Methods like SoM \cite{yang_set--mark_2023} and OmniParser \cite{lu_omniparser_2024} have demonstrated impressive results by tackling these specific limitations. While tool-based approaches offer immediate improvements, the ultimate goal remains advancing toward end-to-end GUI Agents that seamlessly integrate perception and action. Achieving this will require novel architectures and training paradigms specifically designed to handle the unique visual characteristics of GUI environments, potentially incorporating adaptive attention mechanisms that can efficiently process both macro-level layout and micro-level interface details.
\subsection{Strategic Planning and Decision Making}
\label{subsec:planning}
%
%
GUI Agents face critical planning and decision-making challenges when navigating complex interface environments. When unexpected interruptions occur—such as system pop-ups, network errors, or dynamic interface changes—current agents often fail catastrophically, either terminating tasks prematurely or becoming trapped in loops of repeated failed actions. This problem is exacerbated by the unpredictability of real-world interfaces, which frequently update their elements and information architecture, creating a fundamental mismatch between training environments and deployment scenarios.

Beyond handling interruptions, GUI Agents struggle with fundamental decision-making challenges in uncertain environments. When encountering ambiguous UI elements or multiple viable interaction paths, current agents lack sophisticated reasoning capabilities to determine optimal actions that align with specific user contexts and preferences. The ephemeral nature of modern interfaces further complicates this challenge, as transient elements such as tooltips, modal dialogs, and animated transitions create a constantly shifting interaction landscape that static planning models fail to navigate effectively. These temporal aspects of GUI environments demand more sophisticated understanding of state transitions and timing constraints than current perception frameworks can provide.
\subsection{Reinforcement Learning for GUI Agents}
\label{subsec:rl}
%
%
With the development of DeepSeek R1 \cite{deepseekai2025deepseekr1incentivizingreasoningcapability}, researchers have recognized the potential of reinforcement learning with rule-based rewards. An increasing number of studies building upon the R1 paradigm have achieved notable advances in the field. GUI Agents require multi-step reasoning and decision making in practical scenarios, which inherently aligns with the exploratory nature of reinforcement learning, making it a promising approach for enhancing these agents' capabilities.

Whether using reinforcement learning to train GUI Agents can achieve better performance remains an open question worth exploring. The high-dimensional state space represented by screen images presents computational challenges, while defining appropriate reward functions that balance task completion with user experience considerations remains difficult. Despite these challenges, the integration of reinforcement learning with GUI Agents represents a promising direction that could potentially lead to more adaptive agents capable of handling complex real-world interfaces.
\section{Conclusion}
\label{sec:conclusion}
In this survey, we have systematically reviewed the recent advancements in GUI Agents research, organizing their architecture into four fundamental modules: perception, exploration, planning, and interaction. We provided detailed explanations of each module, examining how they collectively enable agents to interpret, navigate, and manipulate graphical user interfaces across diverse platforms. Additionally, we explored various application scenarios of GUI Agents and comprehensively described existing benchmarks, highlighting their potential challenges related to computational cost, linguistic and interface diversity, and security concerns. Finally, we identified the remaining challenges in current GUI Agent systems and outlined promising future research directions that could advance this rapidly evolving field.


\ifCLASSOPTIONcaptionsoff
  \newpage
\fi


\bibliographystyle{IEEEtran}
\bibliography{zotero}

\begin{thebibliography}{100}
\providecommand{\url}[1]{#1}
\csname url@samestyle\endcsname
\providecommand{\newblock}{\relax}
\providecommand{\bibinfo}[2]{#2}
\providecommand{\BIBentrySTDinterwordspacing}{\spaceskip=0pt\relax}
\providecommand{\BIBentryALTinterwordstretchfactor}{4}
\providecommand{\BIBentryALTinterwordspacing}{\spaceskip=\fontdimen2\font plus
\BIBentryALTinterwordstretchfactor\fontdimen3\font minus \fontdimen4\font\relax}
\providecommand{\BIBforeignlanguage}[2]{{%
\expandafter\ifx\csname l@#1\endcsname\relax
\typeout{** WARNING: IEEEtran.bst: No hyphenation pattern has been}%
\typeout{** loaded for the language `#1'. Using the pattern for}%
\typeout{** the default language instead.}%
\else
\language=\csname l@#1\endcsname
\fi
#2}}
\providecommand{\BIBdecl}{\relax}
\BIBdecl

\bibitem{wang2025guiagentsfoundationmodels}
\BIBentryALTinterwordspacing
S.~Wang, W.~Liu, J.~Chen, Y.~Zhou, W.~Gan, X.~Zeng, Y.~Che, S.~Yu, X.~Hao, K.~Shao, B.~Wang, C.~Wu, Y.~Wang, R.~Tang, and J.~Hao, ``Gui agents with foundation models: A comprehensive survey,'' 2025. [Online]. Available: \url{https://arxiv.org/abs/2411.04890}
\BIBentrySTDinterwordspacing

\bibitem{zhang2025largelanguagemodelbrainedgui}
\BIBentryALTinterwordspacing
C.~Zhang, S.~He, J.~Qian, B.~Li, L.~Li, S.~Qin, Y.~Kang, M.~Ma, G.~Liu, Q.~Lin, S.~Rajmohan, D.~Zhang, and Q.~Zhang, ``Large language model-brained gui agents: A survey,'' 2025. [Online]. Available: \url{https://arxiv.org/abs/2411.18279}
\BIBentrySTDinterwordspacing

\bibitem{nguyen2024guiagentssurvey}
\BIBentryALTinterwordspacing
D.~Nguyen, J.~Chen, Y.~Wang, G.~Wu, N.~Park, Z.~Hu, H.~Lyu, J.~Wu, R.~Aponte, Y.~Xia, X.~Li, J.~Shi, H.~Chen, V.~D. Lai, Z.~Xie, S.~Kim, R.~Zhang, T.~Yu, M.~Tanjim, N.~K. Ahmed, P.~Mathur, S.~Yoon, L.~Yao, B.~Kveton, T.~H. Nguyen, T.~Bui, T.~Zhou, R.~A. Rossi, and F.~Dernoncourt, ``Gui agents: A survey,'' 2024. [Online]. Available: \url{https://arxiv.org/abs/2412.13501}
\BIBentrySTDinterwordspacing

\bibitem{yang2025magmafoundationmodelmultimodal}
\BIBentryALTinterwordspacing
J.~Yang, R.~Tan, Q.~Wu, R.~Zheng, B.~Peng, Y.~Liang, Y.~Gu, M.~Cai, S.~Ye, J.~Jang, Y.~Deng, L.~Liden, and J.~Gao, ``Magma: A foundation model for multimodal ai agents,'' 2025. [Online]. Available: \url{https://arxiv.org/abs/2502.13130}
\BIBentrySTDinterwordspacing

\bibitem{wang_mobile-agent_2024}
\BIBentryALTinterwordspacing
J.~Wang, H.~Xu, J.~Ye, M.~Yan, W.~Shen, J.~Zhang, F.~Huang, and J.~Sang, ``\BIBforeignlanguage{en-US}{Mobile-{Agent}: {Autonomous} {Multi}-{Modal} {Mobile} {Device} {Agent} with {Visual} {Perception}},'' Apr. 2024, arXiv:2401.16158. [Online]. Available: \url{http://arxiv.org/abs/2401.16158}
\BIBentrySTDinterwordspacing

\bibitem{zhang_appagent_2023}
\BIBentryALTinterwordspacing
C.~Zhang, Z.~Yang, J.~Liu, Y.~Han, X.~Chen, Z.~Huang, B.~Fu, and G.~Yu, ``\BIBforeignlanguage{en}{{AppAgent}: multimodal agents as smartphone users},'' Dec. 2023, arXiv:2312.13771 TLDR: A novel LLM-based multimodal agent framework designed to operate smartphone applications through a simplified action space, mimicking human-like interactions such as tapping and swiping, thereby broadening its applicability across diverse apps. [Online]. Available: \url{http://arxiv.org/abs/2312.13771}
\BIBentrySTDinterwordspacing

\bibitem{li_appagent_2024}
\BIBentryALTinterwordspacing
Y.~Li, C.~Zhang, W.~Yang, B.~Fu, P.~Cheng, X.~Chen, L.~Chen, and Y.~Wei, ``\BIBforeignlanguage{en}{{AppAgent} v2: {Advanced} {Agent} for {Flexible} {Mobile} {Interactions}},'' Aug. 2024. [Online]. Available: \url{https://arxiv.org/abs/2408.11824v3}
\BIBentrySTDinterwordspacing

\bibitem{cheng_seeclick_2024}
\BIBentryALTinterwordspacing
K.~Cheng, Q.~Sun, Y.~Chu, F.~Xu, Y.~Li, J.~Zhang, and Z.~Wu, ``\BIBforeignlanguage{en-US}{{SeeClick}: {Harnessing} {GUI} {Grounding} for {Advanced} {Visual} {GUI} {Agents}},'' Feb. 2024, arXiv:2401.10935 TLDR: This work proposes a novel visual GUI agent -- SeeClick, which only relies on screenshots for task automation and creates ScreenSpot, the first realistic GUI grounding benchmark that encompasses mobile, desktop, and web environments. [Online]. Available: \url{http://arxiv.org/abs/2401.10935}
\BIBentrySTDinterwordspacing

\bibitem{zhang_ufo_2024}
\BIBentryALTinterwordspacing
C.~Zhang, L.~Li, S.~He, X.~Zhang, B.~Qiao, S.~Qin, M.~Ma, Y.~Kang, Q.~Lin, S.~Rajmohan, D.~Zhang, and Q.~Zhang, ``\BIBforeignlanguage{en-US}{{UFO}: {A} {UI}-{Focused} {Agent} for {Windows} {OS} {Interaction}},'' May 2024, arXiv:2402.07939 TLDR: UFO, an innovative UI-Focused agent to fulfill user requests tailored to applications on Windows OS, harnessing the capabilities of GPT-Vision is introduced, standing as the first UI agent specifically tailored for task completion within the Windows OS environment. [Online]. Available: \url{http://arxiv.org/abs/2402.07939}
\BIBentrySTDinterwordspacing

\bibitem{lin2024showuivisionlanguageactionmodelgui}
\BIBentryALTinterwordspacing
K.~Q. Lin, L.~Li, D.~Gao, Z.~Yang, S.~Wu, Z.~Bai, W.~Lei, L.~Wang, and M.~Z. Shou, ``Showui: One vision-language-action model for gui visual agent,'' 2024. [Online]. Available: \url{https://arxiv.org/abs/2411.17465}
\BIBentrySTDinterwordspacing

\bibitem{jiang2025appagentxevolvingguiagents}
\BIBentryALTinterwordspacing
W.~Jiang, Y.~Zhuang, C.~Song, X.~Yang, and C.~Zhang, ``Appagentx: Evolving gui agents as proficient smartphone users,'' 2025. [Online]. Available: \url{https://arxiv.org/abs/2503.02268}
\BIBentrySTDinterwordspacing

\bibitem{he2024pcagentsleepai}
\BIBentryALTinterwordspacing
Y.~He, J.~Jin, S.~Xia, J.~Su, R.~Fan, H.~Zou, X.~Hu, and P.~Liu, ``Pc agent: While you sleep, ai works -- a cognitive journey into digital world,'' 2024. [Online]. Available: \url{https://arxiv.org/abs/2412.17589}
\BIBentrySTDinterwordspacing

\bibitem{zhang2025apiagentsvsgui}
\BIBentryALTinterwordspacing
C.~Zhang, S.~He, L.~Li, S.~Qin, Y.~Kang, Q.~Lin, and D.~Zhang, ``Api agents vs. gui agents: Divergence and convergence,'' 2025. [Online]. Available: \url{https://arxiv.org/abs/2503.11069}
\BIBentrySTDinterwordspacing

\bibitem{bai_uibert_2021}
\BIBentryALTinterwordspacing
C.~Bai, X.~Zang, Y.~Xu, S.~Sunkara, A.~Rastogi, J.~Chen, and B.~A.~y. Arcas, ``\BIBforeignlanguage{en-US}{{UIBert}: {Learning} {Generic} {Multimodal} {Representations} for {UI} {Understanding}},'' Aug. 2021, arXiv:2107.13731. [Online]. Available: \url{http://arxiv.org/abs/2107.13731}
\BIBentrySTDinterwordspacing

\bibitem{touvron2023llamaopenefficientfoundation}
\BIBentryALTinterwordspacing
H.~Touvron, T.~Lavril, G.~Izacard, X.~Martinet, M.-A. Lachaux, T.~Lacroix, B.~Rozière, N.~Goyal, E.~Hambro, F.~Azhar, A.~Rodriguez, A.~Joulin, E.~Grave, and G.~Lample, ``Llama: Open and efficient foundation language models,'' 2023. [Online]. Available: \url{https://arxiv.org/abs/2302.13971}
\BIBentrySTDinterwordspacing

\bibitem{bai2023qwentechnicalreport}
\BIBentryALTinterwordspacing
J.~Bai, S.~Bai, Y.~Chu, Z.~Cui, K.~Dang, X.~Deng, Y.~Fan, W.~Ge, Y.~Han, F.~Huang, B.~Hui, L.~Ji, M.~Li, J.~Lin, R.~Lin, D.~Liu, G.~Liu, C.~Lu, K.~Lu, J.~Ma, R.~Men, X.~Ren, X.~Ren, C.~Tan, S.~Tan, J.~Tu, P.~Wang, S.~Wang, W.~Wang, S.~Wu, B.~Xu, J.~Xu, A.~Yang, H.~Yang, J.~Yang, S.~Yang, Y.~Yao, B.~Yu, H.~Yuan, Z.~Yuan, J.~Zhang, X.~Zhang, Y.~Zhang, Z.~Zhang, C.~Zhou, J.~Zhou, X.~Zhou, and T.~Zhu, ``Qwen technical report,'' 2023. [Online]. Available: \url{https://arxiv.org/abs/2309.16609}
\BIBentrySTDinterwordspacing

\bibitem{openai2024gpt4technicalreport}
\BIBentryALTinterwordspacing
OpenAI, ``Gpt-4 technical report,'' 2024. [Online]. Available: \url{https://arxiv.org/abs/2303.08774}
\BIBentrySTDinterwordspacing

\bibitem{glm2024chatglmfamilylargelanguage}
\BIBentryALTinterwordspacing
T.~GLM, ``Chatglm: A family of large language models from glm-130b to glm-4 all tools,'' 2024. [Online]. Available: \url{https://arxiv.org/abs/2406.12793}
\BIBentrySTDinterwordspacing

\bibitem{deepseekai2025deepseekv3technicalreport}
\BIBentryALTinterwordspacing
DeepSeek-AI, ``Deepseek-v3 technical report,'' 2025. [Online]. Available: \url{https://arxiv.org/abs/2412.19437}
\BIBentrySTDinterwordspacing

\bibitem{wang2024qwen2vlenhancingvisionlanguagemodels}
\BIBentryALTinterwordspacing
P.~Wang, S.~Bai, S.~Tan, S.~Wang, Z.~Fan, J.~Bai, K.~Chen, X.~Liu, J.~Wang, W.~Ge, Y.~Fan, K.~Dang, M.~Du, X.~Ren, R.~Men, D.~Liu, C.~Zhou, J.~Zhou, and J.~Lin, ``Qwen2-vl: Enhancing vision-language model's perception of the world at any resolution,'' 2024. [Online]. Available: \url{https://arxiv.org/abs/2409.12191}
\BIBentrySTDinterwordspacing

\bibitem{liu2023visualinstructiontuning}
\BIBentryALTinterwordspacing
H.~Liu, C.~Li, Q.~Wu, and Y.~J. Lee, ``Visual instruction tuning,'' 2023. [Online]. Available: \url{https://arxiv.org/abs/2304.08485}
\BIBentrySTDinterwordspacing

\bibitem{bai2025qwen25vltechnicalreport}
\BIBentryALTinterwordspacing
S.~Bai, K.~Chen, X.~Liu, J.~Wang, W.~Ge, S.~Song, K.~Dang, P.~Wang, S.~Wang, J.~Tang, H.~Zhong, Y.~Zhu, M.~Yang, Z.~Li, J.~Wan, P.~Wang, W.~Ding, Z.~Fu, Y.~Xu, J.~Ye, X.~Zhang, T.~Xie, Z.~Cheng, H.~Zhang, Z.~Yang, H.~Xu, and J.~Lin, ``Qwen2.5-vl technical report,'' 2025. [Online]. Available: \url{https://arxiv.org/abs/2502.13923}
\BIBentrySTDinterwordspacing

\bibitem{lu2024deepseekvlrealworldvisionlanguageunderstanding}
\BIBentryALTinterwordspacing
H.~Lu, W.~Liu, B.~Zhang, B.~Wang, K.~Dong, B.~Liu, J.~Sun, T.~Ren, Z.~Li, H.~Yang, Y.~Sun, C.~Deng, H.~Xu, Z.~Xie, and C.~Ruan, ``Deepseek-vl: Towards real-world vision-language understanding,'' 2024. [Online]. Available: \url{https://arxiv.org/abs/2403.05525}
\BIBentrySTDinterwordspacing

\bibitem{ye2024mplugowlmodularizationempowerslarge}
\BIBentryALTinterwordspacing
Q.~Ye, H.~Xu, G.~Xu, J.~Ye, M.~Yan, Y.~Zhou, J.~Wang, A.~Hu, P.~Shi, Y.~Shi, C.~Li, Y.~Xu, H.~Chen, J.~Tian, Q.~Qian, J.~Zhang, F.~Huang, and J.~Zhou, ``mplug-owl: Modularization empowers large language models with multimodality,'' 2024. [Online]. Available: \url{https://arxiv.org/abs/2304.14178}
\BIBentrySTDinterwordspacing

\bibitem{chen2024internvlscalingvisionfoundation}
\BIBentryALTinterwordspacing
Z.~Chen, J.~Wu, W.~Wang, W.~Su, G.~Chen, S.~Xing, M.~Zhong, Q.~Zhang, X.~Zhu, L.~Lu, B.~Li, P.~Luo, T.~Lu, Y.~Qiao, and J.~Dai, ``Internvl: Scaling up vision foundation models and aligning for generic visual-linguistic tasks,'' 2024. [Online]. Available: \url{https://arxiv.org/abs/2312.14238}
\BIBentrySTDinterwordspacing

\bibitem{tang2025thinktwiceclickonce}
\BIBentryALTinterwordspacing
F.~Tang, Y.~Shen, H.~Zhang, S.~Chen, G.~Hou, W.~Zhang, W.~Zhang, K.~Song, W.~Lu, and Y.~Zhuang, ``Think twice, click once: Enhancing gui grounding via fast and slow systems,'' 2025. [Online]. Available: \url{https://arxiv.org/abs/2503.06470}
\BIBentrySTDinterwordspacing

\bibitem{shen2023hugginggptsolvingaitasks}
\BIBentryALTinterwordspacing
Y.~Shen, K.~Song, X.~Tan, D.~Li, W.~Lu, and Y.~Zhuang, ``Hugginggpt: Solving ai tasks with chatgpt and its friends in hugging face,'' 2023. [Online]. Available: \url{https://arxiv.org/abs/2303.17580}
\BIBentrySTDinterwordspacing

\bibitem{deepseekai2025deepseekr1incentivizingreasoningcapability}
\BIBentryALTinterwordspacing
DeepSeek-AI, D.~Guo, D.~Yang, H.~Zhang, J.~Song, R.~Zhang, R.~Xu, Q.~Zhu, S.~Ma, P.~Wang, X.~Bi, X.~Zhang, X.~Yu, Y.~Wu, Z.~F. Wu, Z.~Gou, Z.~Shao, Z.~Li, Z.~Gao, A.~Liu, B.~Xue, B.~Wang, B.~Wu, B.~Feng, C.~Lu, C.~Zhao, C.~Deng, C.~Zhang, C.~Ruan, D.~Dai, D.~Chen, D.~Ji, E.~Li, F.~Lin, F.~Dai, F.~Luo, G.~Hao, G.~Chen, G.~Li, H.~Zhang, H.~Bao, H.~Xu, H.~Wang, H.~Ding, H.~Xin, H.~Gao, H.~Qu, H.~Li, J.~Guo, J.~Li, J.~Wang, J.~Chen, J.~Yuan, J.~Qiu, J.~Li, J.~L. Cai, J.~Ni, J.~Liang, J.~Chen, K.~Dong, K.~Hu, K.~Gao, K.~Guan, K.~Huang, K.~Yu, L.~Wang, L.~Zhang, L.~Zhao, L.~Wang, L.~Zhang, L.~Xu, L.~Xia, M.~Zhang, M.~Zhang, M.~Tang, M.~Li, M.~Wang, M.~Li, N.~Tian, P.~Huang, P.~Zhang, Q.~Wang, Q.~Chen, Q.~Du, R.~Ge, R.~Zhang, R.~Pan, R.~Wang, R.~J. Chen, R.~L. Jin, R.~Chen, S.~Lu, S.~Zhou, S.~Chen, S.~Ye, S.~Wang, S.~Yu, S.~Zhou, S.~Pan, S.~S. Li, S.~Zhou, S.~Wu, S.~Ye, T.~Yun, T.~Pei, T.~Sun, T.~Wang, W.~Zeng, W.~Zhao, W.~Liu, W.~Liang, W.~Gao, W.~Yu, W.~Zhang, W.~L. Xiao, W.~An, X.~Liu, X.~Wang, X.~Chen, X.~Nie,
  X.~Cheng, X.~Liu, X.~Xie, X.~Liu, X.~Yang, X.~Li, X.~Su, X.~Lin, X.~Q. Li, X.~Jin, X.~Shen, X.~Chen, X.~Sun, X.~Wang, X.~Song, X.~Zhou, X.~Wang, X.~Shan, Y.~K. Li, Y.~Q. Wang, Y.~X. Wei, Y.~Zhang, Y.~Xu, Y.~Li, Y.~Zhao, Y.~Sun, Y.~Wang, Y.~Yu, Y.~Zhang, Y.~Shi, Y.~Xiong, Y.~He, Y.~Piao, Y.~Wang, Y.~Tan, Y.~Ma, Y.~Liu, Y.~Guo, Y.~Ou, Y.~Wang, Y.~Gong, Y.~Zou, Y.~He, Y.~Xiong, Y.~Luo, Y.~You, Y.~Liu, Y.~Zhou, Y.~X. Zhu, Y.~Xu, Y.~Huang, Y.~Li, Y.~Zheng, Y.~Zhu, Y.~Ma, Y.~Tang, Y.~Zha, Y.~Yan, Z.~Z. Ren, Z.~Ren, Z.~Sha, Z.~Fu, Z.~Xu, Z.~Xie, Z.~Zhang, Z.~Hao, Z.~Ma, Z.~Yan, Z.~Wu, Z.~Gu, Z.~Zhu, Z.~Liu, Z.~Li, Z.~Xie, Z.~Song, Z.~Pan, Z.~Huang, Z.~Xu, Z.~Zhang, and Z.~Zhang, ``Deepseek-r1: Incentivizing reasoning capability in llms via reinforcement learning,'' 2025. [Online]. Available: \url{https://arxiv.org/abs/2501.12948}
\BIBentrySTDinterwordspacing

\bibitem{xu_androidlab_2024}
\BIBentryALTinterwordspacing
Y.~Xu, X.~Liu, X.~Sun, S.~Cheng, H.~Yu, H.~Lai, S.~Zhang, D.~Zhang, J.~Tang, and Y.~Dong, ``\BIBforeignlanguage{en}{{AndroidLab}: training and systematic benchmarking of android autonomous agents},'' Oct. 2024, arXiv:2410.24024. [Online]. Available: \url{http://arxiv.org/abs/2410.24024}
\BIBentrySTDinterwordspacing

\bibitem{nong_mobileflow_2024}
\BIBentryALTinterwordspacing
S.~Nong, J.~Zhu, R.~Wu, J.~Jin, S.~Shan, X.~Huang, and W.~Xu, ``\BIBforeignlanguage{en}{{MobileFlow}: a multimodal {LLM} for mobile {GUI} agent},'' Aug. 2024, arXiv:2407.04346. [Online]. Available: \url{http://arxiv.org/abs/2407.04346}
\BIBentrySTDinterwordspacing

\bibitem{liu2025infiguiagentmultimodalgeneralistgui}
\BIBentryALTinterwordspacing
Y.~Liu, P.~Li, Z.~Wei, C.~Xie, X.~Hu, X.~Xu, S.~Zhang, X.~Han, H.~Yang, and F.~Wu, ``Infiguiagent: A multimodal generalist gui agent with native reasoning and reflection,'' 2025. [Online]. Available: \url{https://arxiv.org/abs/2501.04575}
\BIBentrySTDinterwordspacing

\bibitem{wen2024autodroidllmpoweredtaskautomation}
\BIBentryALTinterwordspacing
H.~Wen, Y.~Li, G.~Liu, S.~Zhao, T.~Yu, T.~J.-J. Li, S.~Jiang, Y.~Liu, Y.~Zhang, and Y.~Liu, ``Autodroid: Llm-powered task automation in android,'' Mar. 2024. [Online]. Available: \url{http://arxiv.org/abs/2308.15272}
\BIBentrySTDinterwordspacing

\bibitem{zhu2024mobatwolevelagentsystem}
\BIBentryALTinterwordspacing
Z.~Zhu, H.~Tang, Y.~Li, K.~Lan, Y.~Jiang, H.~Zhou, Y.~Wang, S.~Zhang, L.~Sun, L.~Chen, and K.~Yu, ``Moba: A two-level agent system for efficient mobile task automation,'' Oct. 2024. [Online]. Available: \url{http://arxiv.org/abs/2410.13757}
\BIBentrySTDinterwordspacing

\bibitem{agashe_agent_2024}
\BIBentryALTinterwordspacing
S.~Agashe, J.~Han, S.~Gan, J.~Yang, A.~Li, and X.~E. Wang, ``\BIBforeignlanguage{en}{Agent {S}: an open agentic framework that uses computers like a human},'' Oct. 2024, arXiv:2410.08164. [Online]. Available: \url{http://arxiv.org/abs/2410.08164}
\BIBentrySTDinterwordspacing

\bibitem{lu2024guiodysseycomprehensivedataset}
\BIBentryALTinterwordspacing
Q.~Lu, W.~Shao, Z.~Liu, F.~Meng, B.~Li, B.~Chen, S.~Huang, K.~Zhang, Y.~Qiao, and P.~Luo, ``Gui odyssey: A comprehensive dataset for cross-app gui navigation on mobile devices,'' 2024. [Online]. Available: \url{https://arxiv.org/abs/2406.08451}
\BIBentrySTDinterwordspacing

\bibitem{wei2023chainthoughtpromptingelicitsreasoning}
\BIBentryALTinterwordspacing
J.~Wei, X.~Wang, D.~Schuurmans, M.~Bosma, B.~Ichter, F.~Xia, E.~Chi, Q.~Le, and D.~Zhou, ``Chain-of-thought prompting elicits reasoning in large language models,'' Jan. 2023. [Online]. Available: \url{http://arxiv.org/abs/2201.11903}
\BIBentrySTDinterwordspacing

\bibitem{zeng2022socraticmodelscomposingzeroshot}
\BIBentryALTinterwordspacing
A.~Zeng, M.~Attarian, B.~Ichter, K.~Choromanski, A.~Wong, S.~Welker, F.~Tombari, A.~Purohit, M.~Ryoo, V.~Sindhwani, J.~Lee, V.~Vanhoucke, and P.~Florence, ``Socratic models: Composing zero-shot multimodal reasoning with language,'' May 2022. [Online]. Available: \url{http://arxiv.org/abs/2204.00598}
\BIBentrySTDinterwordspacing

\bibitem{yao2023reactsynergizingreasoningacting}
\BIBentryALTinterwordspacing
S.~Yao, J.~Zhao, D.~Yu, N.~Du, I.~Shafran, K.~Narasimhan, and Y.~Cao, ``React: Synergizing reasoning and acting in language models,'' Mar. 2023. [Online]. Available: \url{http://arxiv.org/abs/2210.03629}
\BIBentrySTDinterwordspacing

\bibitem{yang2023mmreactpromptingchatgptmultimodal}
\BIBentryALTinterwordspacing
Z.~Yang, L.~Li, J.~Wang, K.~Lin, E.~Azarnasab, F.~Ahmed, Z.~Liu, C.~Liu, M.~Zeng, and L.~Wang, ``Mm-react: Prompting chatgpt for multimodal reasoning and action,'' Mar. 2023. [Online]. Available: \url{http://arxiv.org/abs/2303.11381}
\BIBentrySTDinterwordspacing

\bibitem{wu_os-copilot_2024}
\BIBentryALTinterwordspacing
Z.~Wu, C.~Han, Z.~Ding, Z.~Weng, Z.~Liu, S.~Yao, T.~Yu, and L.~Kong, ``\BIBforeignlanguage{en}{{OS}-copilot: towards generalist computer agents with self-improvement},'' Feb. 2024, arXiv:2402.07456 TLDR: OS-Copilot is introduced, a framework to build generalist agents capable of interfacing with comprehensive elements in an operating system (OS), including the web, code terminals, files, multimedia, and various third-party applications, and FRIDAY, a self-improving embodied agent for automating general computer tasks. [Online]. Available: \url{http://arxiv.org/abs/2402.07456}
\BIBentrySTDinterwordspacing

\bibitem{wang2024mobileagentbench}
L.~Wang, Y.~Deng, Y.~Zha, G.~Mao, Q.~Wang, T.~Min, W.~Chen, and S.~Chen, ``Mobileagentbench: An efficient and user-friendly benchmark for mobile llm agents,'' \emph{arXiv preprint arXiv:2406.08184}, 2024.

\bibitem{deng2024mind2web}
X.~Deng, Y.~Gu, B.~Zheng, S.~Chen, S.~Stevens, B.~Wang, H.~Sun, and Y.~Su, ``Mind2web: Towards a generalist agent for the web,'' \emph{Advances in Neural Information Processing Systems}, vol.~36, 2024.

\bibitem{zhou2023webarena}
S.~Zhou, F.~F. Xu, H.~Zhu, X.~Zhou, R.~Lo, A.~Sridhar, X.~Cheng, T.~Ou, Y.~Bisk, D.~Fried \emph{et~al.}, ``Webarena: A realistic web environment for building autonomous agents,'' \emph{arXiv preprint arXiv:2307.13854}, 2023.

\bibitem{rawles_androidworld_2024}
\BIBentryALTinterwordspacing
C.~Rawles, S.~Clinckemaillie, Y.~Chang, J.~Waltz, G.~Lau, M.~Fair, A.~Li, W.~Bishop, W.~Li, F.~Campbell-Ajala, D.~Toyama, R.~Berry, D.~Tyamagundlu, T.~Lillicrap, and O.~Riva, ``{AndroidWorld}: {A} {Dynamic} {Benchmarking} {Environment} for {Autonomous} {Agents},'' Oct. 2024, arXiv:2405.14573. [Online]. Available: \url{http://arxiv.org/abs/2405.14573}
\BIBentrySTDinterwordspacing

\bibitem{rawles2024androidworld}
C.~Rawles, S.~Clinckemaillie, Y.~Chang, J.~Waltz, G.~Lau, M.~Fair, A.~Li, W.~Bishop, W.~Li, F.~Campbell-Ajala \emph{et~al.}, ``Androidworld: A dynamic benchmarking environment for autonomous agents,'' \emph{arXiv preprint arXiv:2405.14573}, 2024.

\bibitem{li2024screenspot-pro}
K.~Li, Z.~Meng, H.~Lin, Z.~Luo, Y.~Tian, J.~Ma, Z.~Huang, and T.-S. Chua, ``Screenspot-pro: Gui grounding for professional high-resolution computer use,'' 2025.

\bibitem{koh2024visualwebarena}
J.~Y. Koh, R.~Lo, L.~Jang, V.~Duvvur, M.~C. Lim, P.-Y. Huang, G.~Neubig, S.~Zhou, R.~Salakhutdinov, and D.~Fried, ``Visualwebarena: Evaluating multimodal agents on realistic visual web tasks,'' \emph{arXiv preprint arXiv:2401.13649}, 2024.

\bibitem{wen_autodroid_2024}
\BIBentryALTinterwordspacing
H.~Wen, Y.~Li, G.~Liu, S.~Zhao, T.~Yu, T.~J.-J. Li, S.~Jiang, Y.~Liu, Y.~Zhang, and Y.~Liu, ``\BIBforeignlanguage{en}{{AutoDroid}: {LLM}-powered task automation in android},'' Mar. 2024, arXiv:2308.15272 TLDR: This work introduces Auto-Droid, a mobile task automation system that can handle arbitrary tasks on any Android application without manual efforts and aims to combine the commonsense knowledge of LLMs and domain-specific knowledge of apps through automated dynamic analysis. [Online]. Available: \url{http://arxiv.org/abs/2308.15272}
\BIBentrySTDinterwordspacing

\bibitem{rawles2024androidinthewild}
C.~Rawles, A.~Li, D.~Rodriguez, O.~Riva, and T.~Lillicrap, ``Androidinthewild: A large-scale dataset for android device control,'' \emph{Advances in Neural Information Processing Systems}, vol.~36, 2024.

\bibitem{ding_mobileagent_2024}
\BIBentryALTinterwordspacing
T.~Ding, ``\BIBforeignlanguage{en-US}{{MobileAgent}: enhancing mobile control via human-machine interaction and {SOP} integration},'' Jan. 2024, arXiv:2401.04124. [Online]. Available: \url{http://arxiv.org/abs/2401.04124}
\BIBentrySTDinterwordspacing

\bibitem{gao2023assistgui}
D.~Gao, L.~Ji, Z.~Bai, M.~Ouyang, P.~Li, D.~Mao, Q.~Wu, W.~Zhang, P.~Wang, X.~Guo \emph{et~al.}, ``Assistgui: Task-oriented desktop graphical user interface automation,'' \emph{arXiv preprint arXiv:2312.13108}, 2023.

\bibitem{yang_set--mark_2023}
\BIBentryALTinterwordspacing
J.~Yang, H.~Zhang, F.~Li, X.~Zou, C.~Li, and J.~Gao, ``\BIBforeignlanguage{en-US}{Set-of-{Mark} {Prompting} {Unleashes} {Extraordinary} {Visual} {Grounding} in {GPT}-{4V}},'' Nov. 2023, arXiv:2310.11441 TLDR: The experiments show that GPT-4V with SoM in zero-shot setting outperforms the state-of-the-art fully-finetuned referring expression comprehension and segmentation model on RefCOCOg, and the effectiveness of SoM on a wide range of fine-grained vision and multimodal tasks is validated. [Online]. Available: \url{http://arxiv.org/abs/2310.11441}
\BIBentrySTDinterwordspacing

\bibitem{tan_cradle_2024}
\BIBentryALTinterwordspacing
W.~Tan, W.~Zhang, X.~Xu, H.~Xia, Z.~Ding, B.~Li, B.~Zhou, J.~Yue, J.~Jiang, Y.~Li, R.~An, M.~Qin, C.~Zong, L.~Zheng, Y.~Wu, X.~Chai, Y.~Bi, T.~Xie, P.~Gu, X.~Li, C.~Zhang, L.~Tian, C.~Wang, X.~Wang, B.~F. Karlsson, B.~An, S.~Yan, and Z.~Lu, ``\BIBforeignlanguage{en}{Cradle: empowering foundation agents towards general computer control},'' Jul. 2024, arXiv:2403.03186 TLDR: This work proposes the General Computer Control (GCC) setting: building foundation agents that can master any computer task by taking only screen images of the computer as input, and producing keyboard and mouse operations as output, similar to human-computer interaction. [Online]. Available: \url{http://arxiv.org/abs/2403.03186}
\BIBentrySTDinterwordspacing

\bibitem{hong_cogagent_2023}
\BIBentryALTinterwordspacing
W.~Hong, W.~Wang, Q.~Lv, J.~Xu, W.~Yu, J.~Ji, Y.~Wang, Z.~Wang, Y.~Zhang, J.~Li, B.~Xu, Y.~Dong, M.~Ding, and J.~Tang, ``\BIBforeignlanguage{en}{{CogAgent}: {A} {Visual} {Language} {Model} for {GUI} {Agents}},'' Dec. 2023, arXiv:2312.08914 [cs]. [Online]. Available: \url{http://arxiv.org/abs/2312.08914}
\BIBentrySTDinterwordspacing

\bibitem{qin2025uitarspioneeringautomatedgui}
\BIBentryALTinterwordspacing
Y.~Qin, Y.~Ye, J.~Fang, H.~Wang, S.~Liang, S.~Tian, J.~Zhang, J.~Li, Y.~Li, S.~Huang, W.~Zhong, K.~Li, J.~Yang, Y.~Miao, W.~Lin, L.~Liu, X.~Jiang, Q.~Ma, J.~Li, X.~Xiao, K.~Cai, C.~Li, Y.~Zheng, C.~Jin, C.~Li, X.~Zhou, M.~Wang, H.~Chen, Z.~Li, H.~Yang, H.~Liu, F.~Lin, T.~Peng, X.~Liu, and G.~Shi, ``Ui-tars: Pioneering automated gui interaction with native agents,'' 2025. [Online]. Available: \url{https://arxiv.org/abs/2501.12326}
\BIBentrySTDinterwordspacing

\bibitem{lu_omniparser_2024}
\BIBentryALTinterwordspacing
Y.~Lu, J.~Yang, Y.~Shen, and A.~Awadallah, ``\BIBforeignlanguage{en-US}{{OmniParser} for {Pure} {Vision} {Based} {GUI} {Agent}},'' Aug. 2024, arXiv:2408.00203. [Online]. Available: \url{http://arxiv.org/abs/2408.00203}
\BIBentrySTDinterwordspacing

\bibitem{wang2025mobileagenteselfevolvingmobileassistant}
\BIBentryALTinterwordspacing
Z.~Wang, H.~Xu, J.~Wang, X.~Zhang, M.~Yan, J.~Zhang, F.~Huang, and H.~Ji, ``Mobile-agent-e: Self-evolving mobile assistant for complex tasks,'' 2025. [Online]. Available: \url{https://arxiv.org/abs/2501.11733}
\BIBentrySTDinterwordspacing

\bibitem{wang_mobile-agent-v2_2024}
\BIBentryALTinterwordspacing
J.~Wang, H.~Xu, H.~Jia, X.~Zhang, M.~Yan, W.~Shen, J.~Zhang, F.~Huang, and J.~Sang, ``\BIBforeignlanguage{en-US}{Mobile-{Agent}-v2: {Mobile} {Device} {Operation} {Assistant} with {Effective} {Navigation} via {Multi}-{Agent} {Collaboration}},'' Jun. 2024, arXiv:2406.01014 TLDR: The proposed Mobile-Agent-v2, a multi-agent architecture for mobile device operation assistance, comprises three agents: planning agent, decision agent, and reflection agent that generates task progress, making the navigation of history operations more efficient. [Online]. Available: \url{http://arxiv.org/abs/2406.01014}
\BIBentrySTDinterwordspacing

\bibitem{zhou2023agentsopensourceframeworkautonomous}
\BIBentryALTinterwordspacing
W.~Zhou, Y.~E. Jiang, L.~Li, J.~Wu, T.~Wang, S.~Qiu, J.~Zhang, J.~Chen, R.~Wu, S.~Wang, S.~Zhu, J.~Chen, W.~Zhang, X.~Tang, N.~Zhang, H.~Chen, P.~Cui, and M.~Sachan, ``Agents: An open-source framework for autonomous language agents,'' 2023. [Online]. Available: \url{https://arxiv.org/abs/2309.07870}
\BIBentrySTDinterwordspacing

\bibitem{shinn2023reflexionlanguageagentsverbal}
\BIBentryALTinterwordspacing
N.~Shinn, F.~Cassano, E.~Berman, A.~Gopinath, K.~Narasimhan, and S.~Yao, ``Reflexion: Language agents with verbal reinforcement learning,'' Oct. 2023. [Online]. Available: \url{http://arxiv.org/abs/2303.11366}
\BIBentrySTDinterwordspacing

\bibitem{wang2023voyageropenendedembodiedagent}
\BIBentryALTinterwordspacing
G.~Wang, Y.~Xie, Y.~Jiang, A.~Mandlekar, C.~Xiao, Y.~Zhu, L.~Fan, and A.~Anandkumar, ``Voyager: An open-ended embodied agent with large language models,'' Oct. 2023. [Online]. Available: \url{http://arxiv.org/abs/2305.16291}
\BIBentrySTDinterwordspacing

\bibitem{zhu2024knowagentknowledgeaugmentedplanningllmbased}
\BIBentryALTinterwordspacing
Y.~Zhu, S.~Qiao, Y.~Ou, S.~Deng, N.~Zhang, S.~Lyu, Y.~Shen, L.~Liang, J.~Gu, and H.~Chen, ``Knowagent: Knowledge-augmented planning for llm-based agents,'' 2024. [Online]. Available: \url{https://arxiv.org/abs/2403.03101}
\BIBentrySTDinterwordspacing

\bibitem{qiao2024agentplanningworldknowledge}
\BIBentryALTinterwordspacing
S.~Qiao, R.~Fang, N.~Zhang, Y.~Zhu, X.~Chen, S.~Deng, Y.~Jiang, P.~Xie, F.~Huang, and H.~Chen, ``Agent planning with world knowledge model,'' 2024. [Online]. Available: \url{https://arxiv.org/abs/2405.14205}
\BIBentrySTDinterwordspacing

\bibitem{wang2023selfconsistencyimproveschainthought}
\BIBentryALTinterwordspacing
X.~Wang, J.~Wei, D.~Schuurmans, Q.~Le, E.~Chi, S.~Narang, A.~Chowdhery, and D.~Zhou, ``Self-consistency improves chain of thought reasoning in language models,'' Mar. 2023. [Online]. Available: \url{http://arxiv.org/abs/2203.11171}
\BIBentrySTDinterwordspacing

\bibitem{yao2023treethoughtsdeliberateproblem}
\BIBentryALTinterwordspacing
S.~Yao, D.~Yu, J.~Zhao, I.~Shafran, T.~L. Griffiths, Y.~Cao, and K.~Narasimhan, ``Tree of thoughts: Deliberate problem solving with large language models,'' 2023. [Online]. Available: \url{https://arxiv.org/abs/2305.10601}
\BIBentrySTDinterwordspacing

\bibitem{zhang2024restmctsllmselftrainingprocess}
\BIBentryALTinterwordspacing
D.~Zhang, S.~Zhoubian, Z.~Hu, Y.~Yue, Y.~Dong, and J.~Tang, ``Rest-mcts*: Llm self-training via process reward guided tree search,'' Sep. 2024. [Online]. Available: \url{http://arxiv.org/abs/2406.03816}
\BIBentrySTDinterwordspacing

\bibitem{besta2024graph}
M.~Besta, N.~Blach, A.~Kubicek, R.~Gerstenberger, M.~Podstawski, L.~Gianinazzi, J.~Gajda, T.~Lehmann, H.~Niewiadomski, P.~Nyczyk \emph{et~al.}, ``Graph of thoughts: Solving elaborate problems with large language models,'' in \emph{Proceedings of the AAAI Conference on Artificial Intelligence}, vol.~38, no.~16, 2024, pp. 17\,682--17\,690.

\bibitem{zhang_you_2024}
\BIBentryALTinterwordspacing
Z.~Zhang and A.~Zhang, ``You {Only} {Look} at {Screens}: {Multimodal} {Chain}-of-{Action} {Agents},'' Jun. 2024, arXiv:2309.11436. [Online]. Available: \url{http://arxiv.org/abs/2309.11436}
\BIBentrySTDinterwordspacing

\bibitem{zhang_android_2024}
\BIBentryALTinterwordspacing
J.~Zhang, J.~Wu, Y.~Teng, M.~Liao, N.~Xu, X.~Xiao, Z.~Wei, and D.~Tang, ``\BIBforeignlanguage{en-US}{Android in the {Zoo}: {Chain}-of-{Action}-{Thought} for {GUI} {Agents}},'' Jul. 2024, arXiv:2403.02713 TLDR: This work presents Chain-of-Action-Thought (dubbed CoAT), which takes the description of the previous actions, the current screen, and more importantly the action thinking of what actions should be performed and the outcomes led by the chosen action. [Online]. Available: \url{http://arxiv.org/abs/2403.02713}
\BIBentrySTDinterwordspacing

\bibitem{dong2024villageragentgraphbasedmultiagentframework}
\BIBentryALTinterwordspacing
Y.~Dong, X.~Zhu, Z.~Pan, L.~Zhu, and Y.~Yang, ``Villageragent: A graph-based multi-agent framework for coordinating complex task dependencies in minecraft,'' Jun. 2024. [Online]. Available: \url{http://arxiv.org/abs/2406.05720}
\BIBentrySTDinterwordspacing

\bibitem{2024arXiv240207945N}
R.~{Niu}, J.~{Li}, S.~{Wang}, Y.~{Fu}, X.~{Hu}, X.~{Leng}, H.~{Kong}, Y.~{Chang}, and Q.~{Wang}, ``{ScreenAgent: A Vision Language Model-driven Computer Control Agent},'' \emph{arXiv e-prints}, p. arXiv:2402.07945, Feb. 2024.

\bibitem{ma_coco-agent_2024}
\BIBentryALTinterwordspacing
X.~Ma, Z.~Zhang, and H.~Zhao, ``\BIBforeignlanguage{en}{{CoCo}-agent: a comprehensive cognitive {MLLM} agent for smartphone {GUI} automation},'' in \emph{\BIBforeignlanguage{en}{Findings of the {Association} for {Computational} {Linguistics}: {ACL} 2024}}, L.-W. Ku, A.~Martins, and V.~Srikumar, Eds.\hskip 1em plus 0.5em minus 0.4em\relax Bangkok, Thailand: Association for Computational Linguistics, Aug. 2024, pp. 9097--9110, tLDR: A Comprehensive Cognitive LLM Agent, CoCo-Agent, with two novel approaches, comprehensive environment perception (CEP) and conditional action prediction (CAP), to systematically improve the GUI automation performance. [Online]. Available: \url{https://aclanthology.org/2024.findings-acl.539}
\BIBentrySTDinterwordspacing

\bibitem{2024arXiv240113919H}
H.~{He}, W.~{Yao}, K.~{Ma}, W.~{Yu}, Y.~{Dai}, H.~{Zhang}, Z.~{Lan}, and D.~{Yu}, ``{WebVoyager: Building an End-to-End Web Agent with Large Multimodal Models},'' \emph{arXiv e-prints}, p. arXiv:2401.13919, Jan. 2024.

\bibitem{2023arXiv230908172M}
K.~{Ma}, H.~{Zhang}, H.~{Wang}, X.~{Pan}, W.~{Yu}, and D.~{Yu}, ``{LASER: LLM Agent with State-Space Exploration for Web Navigation},'' \emph{arXiv e-prints}, p. arXiv:2309.08172, Sep. 2023.

\bibitem{wu_os-atlas_2024}
\BIBentryALTinterwordspacing
Z.~Wu, Z.~Wu, F.~Xu, Y.~Wang, Q.~Sun, C.~Jia, K.~Cheng, Z.~Ding, L.~Chen, P.~P. Liang, and Y.~Qiao, ``\BIBforeignlanguage{en}{{OS}-{ATLAS}: a foundation action model for generalist {GUI} agents},'' Oct. 2024, arXiv:2410.23218 version: 1. [Online]. Available: \url{http://arxiv.org/abs/2410.23218}
\BIBentrySTDinterwordspacing

\bibitem{li2023sheetcopilotbringingsoftwareproductivity}
\BIBentryALTinterwordspacing
H.~Li, J.~Su, Y.~Chen, Q.~Li, and Z.~Zhang, ``Sheetcopilot: Bringing software productivity to the next level through large language models,'' Oct. 2023. [Online]. Available: \url{http://arxiv.org/abs/2305.19308}
\BIBentrySTDinterwordspacing

\bibitem{zheng_gpt-4vision_2024}
\BIBentryALTinterwordspacing
B.~Zheng, B.~Gou, J.~Kil, H.~Sun, and Y.~Su, ``\BIBforeignlanguage{en-US}{{GPT}-{4V}(ision) is a {Generalist} {Web} {Agent}, if {Grounded}},'' Mar. 2024, arXiv:2401.01614. [Online]. Available: \url{http://arxiv.org/abs/2401.01614}
\BIBentrySTDinterwordspacing

\bibitem{zhou_webarena_2024}
\BIBentryALTinterwordspacing
S.~Zhou, F.~F. Xu, H.~Zhu, X.~Zhou, R.~Lo, A.~Sridhar, X.~Cheng, T.~Ou, Y.~Bisk, D.~Fried, U.~Alon, and G.~Neubig, ``\BIBforeignlanguage{en-US}{{WebArena}: {A} {Realistic} {Web} {Environment} for {Building} {Autonomous} {Agents}},'' Apr. 2024, arXiv:2307.13854 TLDR: This paper builds an environment for language-guided agents that is highly realistic and reproducible, and creates an environment with fully functional websites from four common domains: e-commerce, social forum discussions, collaborative software development, and content management. [Online]. Available: \url{http://arxiv.org/abs/2307.13854}
\BIBentrySTDinterwordspacing

\bibitem{li_ferret-ui_2024}
\BIBentryALTinterwordspacing
Z.~Li, K.~You, H.~Zhang, D.~Feng, H.~Agrawal, X.~Li, M.~P.~S. Moorthy, J.~Nichols, Y.~Yang, and Z.~Gan, ``\BIBforeignlanguage{en}{Ferret-{UI} 2: mastering universal user interface understanding across platforms},'' Oct. 2024, arXiv:2410.18967. [Online]. Available: \url{http://arxiv.org/abs/2410.18967}
\BIBentrySTDinterwordspacing

\bibitem{rawles_android_2023}
\BIBentryALTinterwordspacing
C.~Rawles, A.~Li, D.~Rodriguez, O.~Riva, and T.~Lillicrap, ``\BIBforeignlanguage{en-US}{Android in the {Wild}: {A} {Large}-{Scale} {Dataset} for {Android} {Device} {Control}},'' Oct. 2023, arXiv:2307.10088 TLDR: A dataset for device-control research, Android in the Wild (AITW), which is orders of magnitude larger than current datasets, and contains multi-step tasks that require semantic understanding of language and visual context. [Online]. Available: \url{http://arxiv.org/abs/2307.10088}
\BIBentrySTDinterwordspacing

\bibitem{chowdhery2023palm}
A.~Chowdhery, S.~Narang, J.~Devlin, M.~Bosma, G.~Mishra, A.~Roberts, P.~Barham, H.~W. Chung, C.~Sutton, S.~Gehrmann \emph{et~al.}, ``Palm: Scaling language modeling with pathways,'' \emph{Journal of Machine Learning Research}, vol.~24, no. 240, pp. 1--113, 2023.

\bibitem{gao_assistgui_2024}
\BIBentryALTinterwordspacing
D.~Gao, L.~Ji, Z.~Bai, M.~Ouyang, P.~Li, D.~Mao, Q.~Wu, W.~Zhang, P.~Wang, X.~Guo, H.~Wang, L.~Zhou, and M.~Z. Shou, ``\BIBforeignlanguage{en}{{ASSISTGUI}: task-oriented desktop graphical user interface automation},'' Jan. 2024, arXiv:2312.13108 TLDR: An advanced Actor-Critic Embodied Agent framework is proposed, which incorporates a sophisticated GUI parser driven by an LLM-agent and an enhanced reasoning mechanism adept at handling lengthy procedural tasks that outshine existing methods in performance. [Online]. Available: \url{http://arxiv.org/abs/2312.13108}
\BIBentrySTDinterwordspacing

\bibitem{openai2021chatgpt}
{OpenAI}, ``{ChatGPT},'' {\url{https://openai.com/research/chatgpt/}}, 2021.

\bibitem{liu2024groundingdinomarryingdino}
\BIBentryALTinterwordspacing
S.~Liu, Z.~Zeng, T.~Ren, F.~Li, H.~Zhang, J.~Yang, Q.~Jiang, C.~Li, J.~Yang, H.~Su, J.~Zhu, and L.~Zhang, ``Grounding dino: Marrying dino with grounded pre-training for open-set object detection,'' 2024. [Online]. Available: \url{https://arxiv.org/abs/2303.05499}
\BIBentrySTDinterwordspacing

\bibitem{kirillov2023segment}
\BIBentryALTinterwordspacing
A.~Kirillov, E.~Mintun, N.~Ravi, H.~Mao, C.~Rolland, L.~Gustafson, T.~Xiao, S.~Whitehead, A.~C. Berg, W.-Y. Lo, P.~Dollár, and R.~Girshick, ``Segment anything,'' 2023. [Online]. Available: \url{https://arxiv.org/abs/2304.02643}
\BIBentrySTDinterwordspacing

\bibitem{du2020ppocrpracticalultralightweight}
\BIBentryALTinterwordspacing
Y.~Du, C.~Li, R.~Guo, X.~Yin, W.~Liu, J.~Zhou, Y.~Bai, Z.~Yu, Y.~Yang, Q.~Dang, and H.~Wang, ``Pp-ocr: A practical ultra lightweight ocr system,'' 2020. [Online]. Available: \url{https://arxiv.org/abs/2009.09941}
\BIBentrySTDinterwordspacing

\bibitem{openai_2025_operator}
OpenAI, ``Operator system card,'' January 2025, released on January 23, 2025.

\bibitem{li2024ferretui2masteringuniversal}
\BIBentryALTinterwordspacing
Z.~Li, K.~You, H.~Zhang, D.~Feng, H.~Agrawal, X.~Li, M.~P.~S. Moorthy, J.~Nichols, Y.~Yang, and Z.~Gan, ``Ferret-ui 2: Mastering universal user interface understanding across platforms,'' 2024. [Online]. Available: \url{https://arxiv.org/abs/2410.18967}
\BIBentrySTDinterwordspacing

\bibitem{vaswani2023attentionneed}
\BIBentryALTinterwordspacing
A.~Vaswani, N.~Shazeer, N.~Parmar, J.~Uszkoreit, L.~Jones, A.~N. Gomez, L.~Kaiser, and I.~Polosukhin, ``Attention is all you need,'' 2023. [Online]. Available: \url{https://arxiv.org/abs/1706.03762}
\BIBentrySTDinterwordspacing

\bibitem{sun_meta-gui_2022}
\BIBentryALTinterwordspacing
L.~Sun, X.~Chen, L.~Chen, T.~Dai, Z.~Zhu, and K.~Yu, ``\BIBforeignlanguage{en-US}{{META}-{GUI}: {Towards} {Multi}-modal {Conversational} {Agents} on {Mobile} {GUI}},'' in \emph{\BIBforeignlanguage{en-US}{Proceedings of the 2022 {Conference} on {Empirical} {Methods} in {Natural} {Language} {Processing}}}, Y.~Goldberg, Z.~Kozareva, and Y.~Zhang, Eds.\hskip 1em plus 0.5em minus 0.4em\relax Abu Dhabi, United Arab Emirates: Association for Computational Linguistics, Dec. 2022, pp. 6699--6712. [Online]. Available: \url{https://aclanthology.org/2022.emnlp-main.449}
\BIBentrySTDinterwordspacing

\bibitem{ren2016fasterrcnnrealtimeobject}
\BIBentryALTinterwordspacing
S.~Ren, K.~He, R.~Girshick, and J.~Sun, ``Faster r-cnn: Towards real-time object detection with region proposal networks,'' 2016. [Online]. Available: \url{https://arxiv.org/abs/1506.01497}
\BIBentrySTDinterwordspacing

\bibitem{devlin2019bertpretrainingdeepbidirectional}
\BIBentryALTinterwordspacing
J.~Devlin, M.-W. Chang, K.~Lee, and K.~Toutanova, ``Bert: Pre-training of deep bidirectional transformers for language understanding,'' 2019. [Online]. Available: \url{https://arxiv.org/abs/1810.04805}
\BIBentrySTDinterwordspacing

\bibitem{wang2024qwen2}
P.~Wang, S.~Bai, S.~Tan, S.~Wang, Z.~Fan, J.~Bai, K.~Chen, X.~Liu, J.~Wang, W.~Ge \emph{et~al.}, ``Qwen2-vl: Enhancing vision-language model's perception of the world at any resolution,'' \emph{arXiv preprint arXiv:2409.12191}, 2024.

\bibitem{liu2024visual}
H.~Liu, C.~Li, Q.~Wu, and Y.~J. Lee, ``Visual instruction tuning,'' \emph{Advances in neural information processing systems}, vol.~36, 2024.

\bibitem{dosovitskiy2021imageworth16x16words}
\BIBentryALTinterwordspacing
A.~Dosovitskiy, L.~Beyer, A.~Kolesnikov, D.~Weissenborn, X.~Zhai, T.~Unterthiner, M.~Dehghani, M.~Minderer, G.~Heigold, S.~Gelly, J.~Uszkoreit, and N.~Houlsby, ``An image is worth 16x16 words: Transformers for image recognition at scale,'' 2021. [Online]. Available: \url{https://arxiv.org/abs/2010.11929}
\BIBentrySTDinterwordspacing

\bibitem{huang2022layoutlmv3pretrainingdocumentai}
\BIBentryALTinterwordspacing
Y.~Huang, T.~Lv, L.~Cui, Y.~Lu, and F.~Wei, ``Layoutlmv3: Pre-training for document ai with unified text and image masking,'' 2022. [Online]. Available: \url{https://arxiv.org/abs/2204.08387}
\BIBentrySTDinterwordspacing

\bibitem{bi2024deepseek}
X.~Bi, D.~Chen, G.~Chen, S.~Chen, D.~Dai, C.~Deng, H.~Ding, K.~Dong, Q.~Du, Z.~Fu \emph{et~al.}, ``Deepseek llm: Scaling open-source language models with longtermism,'' \emph{arXiv preprint arXiv:2401.02954}, 2024.

\bibitem{liu_autoglm_2024}
\BIBentryALTinterwordspacing
X.~Liu, B.~Qin, D.~Liang, G.~Dong, H.~Lai, H.~Zhang, H.~Zhao, I.~L. Iong, J.~Sun, J.~Wang, J.~Gao, J.~Shan, K.~Liu, S.~Zhang, S.~Yao, S.~Cheng, W.~Yao, W.~Zhao, X.~Liu, X.~Liu, X.~Chen, X.~Yang, Y.~Yang, Y.~Xu, Y.~Yang, Y.~Wang, Y.~Xu, Z.~Qi, Y.~Dong, and J.~Tang, ``\BIBforeignlanguage{en-US}{{AutoGLM}: {Autonomous} {Foundation} {Agents} for {GUIs}},'' Oct. 2024, arXiv:2411.00820. [Online]. Available: \url{http://arxiv.org/abs/2411.00820}
\BIBentrySTDinterwordspacing

\bibitem{glm2024chatglm}
T.~GLM, A.~Zeng, B.~Xu, B.~Wang, C.~Zhang, D.~Yin, D.~Zhang, D.~Rojas, G.~Feng, H.~Zhao \emph{et~al.}, ``Chatglm: A family of large language models from glm-130b to glm-4 all tools,'' \emph{arXiv preprint arXiv:2406.12793}, 2024.

\bibitem{xu2024aguvisunifiedpurevision}
\BIBentryALTinterwordspacing
Y.~Xu, Z.~Wang, J.~Wang, D.~Lu, T.~Xie, A.~Saha, D.~Sahoo, T.~Yu, and C.~Xiong, ``Aguvis: Unified pure vision agents for autonomous gui interaction,'' 2024. [Online]. Available: \url{https://arxiv.org/abs/2412.04454}
\BIBentrySTDinterwordspacing

\bibitem{yang2024ariauivisualgroundinggui}
\BIBentryALTinterwordspacing
Y.~Yang, Y.~Wang, D.~Li, Z.~Luo, B.~Chen, C.~Huang, and J.~Li, ``Aria-ui: Visual grounding for gui instructions,'' 2024. [Online]. Available: \url{https://arxiv.org/abs/2412.16256}
\BIBentrySTDinterwordspacing

\bibitem{deitke2024molmopixmoopenweights}
\BIBentryALTinterwordspacing
M.~Deitke, C.~Clark, S.~Lee, R.~Tripathi, Y.~Yang, J.~S. Park, M.~Salehi, N.~Muennighoff, K.~Lo, L.~Soldaini, J.~Lu, T.~Anderson, E.~Bransom, K.~Ehsani, H.~Ngo, Y.~Chen, A.~Patel, M.~Yatskar, C.~Callison-Burch, A.~Head, R.~Hendrix, F.~Bastani, E.~VanderBilt, N.~Lambert, Y.~Chou, A.~Chheda, J.~Sparks, S.~Skjonsberg, M.~Schmitz, A.~Sarnat, B.~Bischoff, P.~Walsh, C.~Newell, P.~Wolters, T.~Gupta, K.-H. Zeng, J.~Borchardt, D.~Groeneveld, C.~Nam, S.~Lebrecht, C.~Wittlif, C.~Schoenick, O.~Michel, R.~Krishna, L.~Weihs, N.~A. Smith, H.~Hajishirzi, R.~Girshick, A.~Farhadi, and A.~Kembhavi, ``Molmo and pixmo: Open weights and open data for state-of-the-art vision-language models,'' 2024. [Online]. Available: \url{https://arxiv.org/abs/2409.17146}
\BIBentrySTDinterwordspacing

\bibitem{zheng2024synapsetrajectoryasexemplarpromptingmemory}
\BIBentryALTinterwordspacing
L.~Zheng, R.~Wang, X.~Wang, and B.~An, ``Synapse: Trajectory-as-exemplar prompting with memory for computer control,'' 2024. [Online]. Available: \url{https://arxiv.org/abs/2306.07863}
\BIBentrySTDinterwordspacing

\bibitem{kim-etal-2024-rada}
\BIBentryALTinterwordspacing
M.~Kim, V.~Bursztyn, E.~Koh, S.~Guo, and S.-w. Hwang, ``{R}a{DA}: Retrieval-augmented web agent planning with {LLM}s,'' Bangkok, Thailand, pp. 13\,511--13\,525, Aug. 2024. [Online]. Available: \url{https://aclanthology.org/2024.findings-acl.802}
\BIBentrySTDinterwordspacing

\bibitem{zhao2023expelllmagentsexperiential}
\BIBentryALTinterwordspacing
A.~Zhao, D.~Huang, Q.~Xu, M.~Lin, Y.-J. Liu, and G.~Huang, ``Expel: Llm agents are experiential learners,'' 2023. [Online]. Available: \url{https://arxiv.org/abs/2308.10144}
\BIBentrySTDinterwordspacing

\bibitem{fu2024autoguideautomatedgenerationselection}
\BIBentryALTinterwordspacing
Y.~Fu, D.-K. Kim, J.~Kim, S.~Sohn, L.~Logeswaran, K.~Bae, and H.~Lee, ``Autoguide: Automated generation and selection of state-aware guidelines for large language model agents,'' 2024. [Online]. Available: \url{https://arxiv.org/abs/2403.08978}
\BIBentrySTDinterwordspacing

\bibitem{zhu2023ghostminecraftgenerallycapable}
\BIBentryALTinterwordspacing
X.~Zhu, Y.~Chen, H.~Tian, C.~Tao, W.~Su, C.~Yang, G.~Huang, B.~Li, L.~Lu, X.~Wang, Y.~Qiao, Z.~Zhang, and J.~Dai, ``Ghost in the minecraft: Generally capable agents for open-world environments via large language models with text-based knowledge and memory,'' 2023. [Online]. Available: \url{https://arxiv.org/abs/2305.17144}
\BIBentrySTDinterwordspacing

\bibitem{zhou2023recurrentgptinteractivegenerationarbitrarily}
\BIBentryALTinterwordspacing
W.~Zhou, Y.~E. Jiang, P.~Cui, T.~Wang, Z.~Xiao, Y.~Hou, R.~Cotterell, and M.~Sachan, ``Recurrentgpt: Interactive generation of (arbitrarily) long text,'' 2023. [Online]. Available: \url{https://arxiv.org/abs/2305.13304}
\BIBentrySTDinterwordspacing

\bibitem{qian2024investigateconsolidateexploitgeneralstrategyintertask}
\BIBentryALTinterwordspacing
C.~Qian, S.~Liang, Y.~Qin, Y.~Ye, X.~Cong, Y.~Lin, Y.~Wu, Z.~Liu, and M.~Sun, ``Investigate-consolidate-exploit: A general strategy for inter-task agent self-evolution,'' 2024. [Online]. Available: \url{https://arxiv.org/abs/2401.13996}
\BIBentrySTDinterwordspacing

\bibitem{Instructor-XL}
\BIBentryALTinterwordspacing
H.~Su, W.~Shi, J.~Kasai, Y.~Wang, Y.~Hu, M.~Ostendorf, W.~tau Yih, N.~A. Smith, L.~Zettlemoyer, and T.~Yu, ``One embedder, any task: Instruction-finetuned text embeddings,'' 2023. [Online]. Available: \url{https://arxiv.org/abs/2212.09741}
\BIBentrySTDinterwordspacing

\bibitem{min2022metaicllearninglearncontext}
\BIBentryALTinterwordspacing
S.~Min, M.~Lewis, L.~Zettlemoyer, and H.~Hajishirzi, ``Metaicl: Learning to learn in context,'' May 2022. [Online]. Available: \url{http://arxiv.org/abs/2110.15943}
\BIBentrySTDinterwordspacing

\bibitem{xia-etal-2024-aligning}
\BIBentryALTinterwordspacing
Y.~Xia, T.~Yu, Z.~He, H.~Zhao, J.~McAuley, and S.~Li, ``Aligning as debiasing: Causality-aware alignment via reinforcement learning with interventional feedback,'' in \emph{Proceedings of the 2024 Conference of the North American Chapter of the Association for Computational Linguistics: Human Language Technologies (Volume 1: Long Papers)}, K.~Duh, H.~Gomez, and S.~Bethard, Eds.\hskip 1em plus 0.5em minus 0.4em\relax Mexico City, Mexico: Association for Computational Linguistics, Jun. 2024, pp. 4684--4695. [Online]. Available: \url{https://aclanthology.org/2024.naacl-long.262}
\BIBentrySTDinterwordspacing

\bibitem{yuan2024reversalthoughtenhancinglarge}
\BIBentryALTinterwordspacing
J.~Yuan, D.~Du, H.~Zhang, Z.~Di, and U.~Naseem, ``Reversal of thought: Enhancing large language models with preference-guided reverse reasoning warm-up,'' 2024. [Online]. Available: \url{https://arxiv.org/abs/2410.12323}
\BIBentrySTDinterwordspacing

\bibitem{feng2024alphazeroliketreesearchcanguide}
\BIBentryALTinterwordspacing
X.~Feng, Z.~Wan, M.~Wen, S.~M. McAleer, Y.~Wen, W.~Zhang, and J.~Wang, ``Alphazero-like tree-search can guide large language model decoding and training,'' Feb. 2024. [Online]. Available: \url{http://arxiv.org/abs/2309.17179}
\BIBentrySTDinterwordspacing

\bibitem{ding2024everythingthoughtsdefyinglaw}
\BIBentryALTinterwordspacing
R.~Ding, C.~Zhang, L.~Wang, Y.~Xu, M.~Ma, W.~Zhang, S.~Qin, S.~Rajmohan, Q.~Lin, and D.~Zhang, ``Everything of thoughts: Defying the law of penrose triangle for thought generation,'' Feb. 2024. [Online]. Available: \url{http://arxiv.org/abs/2311.04254}
\BIBentrySTDinterwordspacing

\bibitem{gao2024interpretablecontrastivemontecarlo}
\BIBentryALTinterwordspacing
Z.~Gao, B.~Niu, X.~He, H.~Xu, H.~Liu, A.~Liu, X.~Hu, and L.~Wen, ``Interpretable contrastive monte carlo tree search reasoning,'' Oct. 2024. [Online]. Available: \url{http://arxiv.org/abs/2410.01707}
\BIBentrySTDinterwordspacing

\bibitem{zhou2024languageagenttreesearch}
\BIBentryALTinterwordspacing
A.~Zhou, K.~Yan, M.~{Shlapentokh-Rothman}, H.~Wang, and Y.-X. Wang, ``Language agent tree search unifies reasoning acting and planning in language models,'' Jun. 2024. [Online]. Available: \url{http://arxiv.org/abs/2310.04406}
\BIBentrySTDinterwordspacing

\bibitem{hao2023reasoninglanguagemodelplanning}
\BIBentryALTinterwordspacing
S.~Hao, Y.~Gu, H.~Ma, J.~J. Hong, Z.~Wang, D.~Z. Wang, and Z.~Hu, ``Reasoning with language model is planning with world model,'' Oct. 2023. [Online]. Available: \url{http://arxiv.org/abs/2305.14992}
\BIBentrySTDinterwordspacing

\bibitem{sun2024fastbestndecodingspeculative}
\BIBentryALTinterwordspacing
H.~Sun, M.~Haider, R.~Zhang, H.~Yang, J.~Qiu, M.~Yin, M.~Wang, P.~Bartlett, and A.~Zanette, ``Fast best-of-n decoding via speculative rejection,'' Oct. 2024. [Online]. Available: \url{http://arxiv.org/abs/2410.20290}
\BIBentrySTDinterwordspacing

\bibitem{wang2024mathshepherdverifyreinforcellms}
\BIBentryALTinterwordspacing
P.~Wang, L.~Li, Z.~Shao, R.~X. Xu, D.~Dai, Y.~Li, D.~Chen, Y.~Wu, and Z.~Sui, ``Math-shepherd: Verify and reinforce llms step-by-step without human annotations,'' Feb. 2024. [Online]. Available: \url{http://arxiv.org/abs/2312.08935}
\BIBentrySTDinterwordspacing

\bibitem{lightman2023letsverifystepstep}
\BIBentryALTinterwordspacing
H.~Lightman, V.~Kosaraju, Y.~Burda, H.~Edwards, B.~Baker, T.~Lee, J.~Leike, J.~Schulman, I.~Sutskever, and K.~Cobbe, ``Let's verify step by step,'' May 2023. [Online]. Available: \url{http://arxiv.org/abs/2305.20050}
\BIBentrySTDinterwordspacing

\bibitem{shang2024synergythoughtselicitingefficientreasoning}
\BIBentryALTinterwordspacing
Y.~Shang, Y.~Li, F.~Xu, and Y.~Li, ``Synergy-of-thoughts: Eliciting efficient reasoning in hybrid language models,'' Aug. 2024. [Online]. Available: \url{http://arxiv.org/abs/2402.02563}
\BIBentrySTDinterwordspacing

\bibitem{setlur2024rewardingprogressscalingautomated}
\BIBentryALTinterwordspacing
A.~Setlur, C.~Nagpal, A.~Fisch, X.~Geng, J.~Eisenstein, R.~Agarwal, A.~Agarwal, J.~Berant, and A.~Kumar, ``Rewarding progress: Scaling automated process verifiers for llm reasoning,'' Oct. 2024. [Online]. Available: \url{http://arxiv.org/abs/2410.08146}
\BIBentrySTDinterwordspacing

\bibitem{zhang2024generativeverifiersrewardmodeling}
\BIBentryALTinterwordspacing
L.~Zhang, A.~Hosseini, H.~Bansal, M.~Kazemi, A.~Kumar, and R.~Agarwal, ``Generative verifiers: Reward modeling as next-token prediction,'' Oct. 2024. [Online]. Available: \url{http://arxiv.org/abs/2408.15240}
\BIBentrySTDinterwordspacing

\bibitem{zheng2023judgingllmjudgemtbenchchatbot}
\BIBentryALTinterwordspacing
L.~Zheng, W.-L. Chiang, Y.~Sheng, S.~Zhuang, Z.~Wu, Y.~Zhuang, Z.~Lin, Z.~Li, D.~Li, E.~P. Xing, H.~Zhang, J.~E. Gonzalez, and I.~Stoica, ``Judging llm-as-a-judge with mt-bench and chatbot arena,'' Dec. 2023. [Online]. Available: \url{http://arxiv.org/abs/2306.05685}
\BIBentrySTDinterwordspacing

\bibitem{qi2024mutualreasoningmakessmaller}
\BIBentryALTinterwordspacing
Z.~Qi, M.~Ma, J.~Xu, L.~L. Zhang, F.~Yang, and M.~Yang, ``Mutual reasoning makes smaller llms stronger problem-solvers,'' Aug. 2024. [Online]. Available: \url{http://arxiv.org/abs/2408.06195}
\BIBentrySTDinterwordspacing

\bibitem{zhou2023leastmostpromptingenablescomplex}
\BIBentryALTinterwordspacing
D.~Zhou, N.~Sch{\"a}rli, L.~Hou, J.~Wei, N.~Scales, X.~Wang, D.~Schuurmans, C.~Cui, O.~Bousquet, Q.~Le, and E.~Chi, ``Least-to-most prompting enables complex reasoning in large language models,'' Apr. 2023. [Online]. Available: \url{http://arxiv.org/abs/2205.10625}
\BIBentrySTDinterwordspacing

\bibitem{touvron2023llama2openfoundation}
\BIBentryALTinterwordspacing
H.~Touvron, L.~Martin, K.~Stone, P.~Albert, A.~Almahairi, Y.~Babaei, N.~Bashlykov, S.~Batra, P.~Bhargava, S.~Bhosale, D.~Bikel, L.~Blecher, C.~C. Ferrer, M.~Chen, G.~Cucurull, D.~Esiobu, J.~Fernandes, J.~Fu, W.~Fu, B.~Fuller, C.~Gao, V.~Goswami, N.~Goyal, A.~Hartshorn, S.~Hosseini, R.~Hou, H.~Inan, M.~Kardas, V.~Kerkez, M.~Khabsa, I.~Kloumann, A.~Korenev, P.~S. Koura, M.-A. Lachaux, T.~Lavril, J.~Lee, D.~Liskovich, Y.~Lu, Y.~Mao, X.~Martinet, T.~Mihaylov, P.~Mishra, I.~Molybog, Y.~Nie, A.~Poulton, J.~Reizenstein, R.~Rungta, K.~Saladi, A.~Schelten, R.~Silva, E.~M. Smith, R.~Subramanian, X.~E. Tan, B.~Tang, R.~Taylor, A.~Williams, J.~X. Kuan, P.~Xu, Z.~Yan, I.~Zarov, Y.~Zhang, A.~Fan, M.~Kambadur, S.~Narang, A.~Rodriguez, R.~Stojnic, S.~Edunov, and T.~Scialom, ``Llama 2: Open foundation and fine-tuned chat models,'' Jul. 2023. [Online]. Available: \url{http://arxiv.org/abs/2307.09288}
\BIBentrySTDinterwordspacing

\bibitem{huang2024understandingplanningllmagents}
\BIBentryALTinterwordspacing
X.~Huang, W.~Liu, X.~Chen, X.~Wang, H.~Wang, D.~Lian, Y.~Wang, R.~Tang, and E.~Chen, ``Understanding the planning of llm agents: A survey,'' 2024. [Online]. Available: \url{https://arxiv.org/abs/2402.02716}
\BIBentrySTDinterwordspacing

\bibitem{yuan2024selfrewardinglanguagemodels}
\BIBentryALTinterwordspacing
W.~Yuan, R.~Y. Pang, K.~Cho, X.~Li, S.~Sukhbaatar, J.~Xu, and J.~Weston, ``Self-rewarding language models,'' Feb. 2024. [Online]. Available: \url{http://arxiv.org/abs/2401.10020}
\BIBentrySTDinterwordspacing

\bibitem{huangqdmrbasedplanningsolvingpromptingcomplex}
J.~Huang, Q.~She, W.~Jiang, H.~Wu, Y.~Hao, T.~Xu, and F.~Wu, ``Qdmr-based planning-and-solving prompting for complex reasoning tasks.''

\bibitem{zelikman2022starbootstrappingreasoningreasoning}
\BIBentryALTinterwordspacing
E.~Zelikman, Y.~Wu, J.~Mu, and N.~D. Goodman, ``Star: Bootstrapping reasoning with reasoning,'' May 2022. [Online]. Available: \url{http://arxiv.org/abs/2203.14465}
\BIBentrySTDinterwordspacing

\bibitem{hosseini2024vstartrainingverifiersselftaught}
\BIBentryALTinterwordspacing
A.~Hosseini, X.~Yuan, N.~Malkin, A.~Courville, A.~Sordoni, and R.~Agarwal, ``V-star: Training verifiers for self-taught reasoners,'' Aug. 2024. [Online]. Available: \url{http://arxiv.org/abs/2402.06457}
\BIBentrySTDinterwordspacing

\bibitem{madaan2023selfrefineiterativerefinementselffeedback}
\BIBentryALTinterwordspacing
A.~Madaan, N.~Tandon, P.~Gupta, S.~Hallinan, L.~Gao, S.~Wiegreffe, U.~Alon, N.~Dziri, S.~Prabhumoye, Y.~Yang, S.~Gupta, B.~P. Majumder, K.~Hermann, S.~Welleck, A.~Yazdanbakhsh, and P.~Clark, ``Self-refine: Iterative refinement with self-feedback,'' May 2023. [Online]. Available: \url{http://arxiv.org/abs/2303.17651}
\BIBentrySTDinterwordspacing

\bibitem{yuan2023scalingrelationshiplearningmathematical}
\BIBentryALTinterwordspacing
Z.~Yuan, H.~Yuan, C.~Li, G.~Dong, K.~Lu, C.~Tan, C.~Zhou, and J.~Zhou, ``Scaling relationship on learning mathematical reasoning with large language models,'' Sep. 2023. [Online]. Available: \url{http://arxiv.org/abs/2308.01825}
\BIBentrySTDinterwordspacing

\bibitem{zelikman2024quietstarlanguagemodelscan}
\BIBentryALTinterwordspacing
E.~Zelikman, G.~Harik, Y.~Shao, V.~Jayasiri, N.~Haber, and N.~D. Goodman, ``Quiet-star: Language models can teach themselves to think before speaking,'' Mar. 2024. [Online]. Available: \url{http://arxiv.org/abs/2403.09629}
\BIBentrySTDinterwordspacing

\bibitem{zhang2024chainpreferenceoptimizationimproving}
\BIBentryALTinterwordspacing
X.~Zhang, C.~Du, T.~Pang, Q.~Liu, W.~Gao, and M.~Lin, ``Chain of preference optimization: Improving chain-of-thought reasoning in llms,'' Oct. 2024. [Online]. Available: \url{http://arxiv.org/abs/2406.09136}
\BIBentrySTDinterwordspacing

\bibitem{andukuri2024stargateteachinglanguagemodels}
\BIBentryALTinterwordspacing
C.~Andukuri, J.-P. Fr{\"a}nken, T.~Gerstenberg, and N.~D. Goodman, ``Star-gate: Teaching language models to ask clarifying questions,'' Aug. 2024. [Online]. Available: \url{http://arxiv.org/abs/2403.19154}
\BIBentrySTDinterwordspacing

\bibitem{zhang2024multimodalchainthoughtreasoninglanguage}
\BIBentryALTinterwordspacing
Z.~Zhang, A.~Zhang, M.~Li, H.~Zhao, G.~Karypis, and A.~Smola, ``Multimodal chain-of-thought reasoning in language models,'' May 2024. [Online]. Available: \url{http://arxiv.org/abs/2302.00923}
\BIBentrySTDinterwordspacing

\bibitem{wu2023rolechainthoughtcomplexvisionlanguage}
\BIBentryALTinterwordspacing
Y.~Wu, P.~Zhang, W.~Xiong, B.~Oguz, J.~C. Gee, and Y.~Nie, ``The role of chain-of-thought in complex vision-language reasoning task,'' Nov. 2023. [Online]. Available: \url{http://arxiv.org/abs/2311.09193}
\BIBentrySTDinterwordspacing

\bibitem{lulearnexplainmultimodalreasoning}
P.~Lu, S.~Mishra, T.~Xia, L.~Qiu, K.-W. Chang, S.-C. Zhu, O.~Tafjord, P.~Clark, and A.~Kalyan, ``Learn to explain: Multimodal reasoning via thought chains for science question answering.''

\bibitem{mondal2024kamcotknowledgeaugmentedmultimodal}
\BIBentryALTinterwordspacing
D.~Mondal, S.~Modi, S.~Panda, R.~Singh, and G.~S. Rao, ``Kam-cot: Knowledge augmented multimodal chain-of-thoughts reasoning,'' Jan. 2024. [Online]. Available: \url{http://arxiv.org/abs/2401.12863}
\BIBentrySTDinterwordspacing

\bibitem{mitracompositionalchainthoughtpromptinglarge}
C.~Mitra, B.~Huang, T.~Darrell, and R.~Herzig, ``Compositional chain-of-thought prompting for large multimodal models.''

\bibitem{shao2024visualcotadvancingmultimodal}
\BIBentryALTinterwordspacing
H.~Shao, S.~Qian, H.~Xiao, G.~Song, Z.~Zong, L.~Wang, Y.~Liu, and H.~Li, ``Visual cot: Advancing multi-modal language models with a comprehensive dataset and benchmark for chain-of-thought reasoning,'' Nov. 2024. [Online]. Available: \url{http://arxiv.org/abs/2403.16999}
\BIBentrySTDinterwordspacing

\bibitem{li2024vocotunleashingvisuallygrounded}
\BIBentryALTinterwordspacing
Z.~Li, R.~Luo, J.~Zhang, M.~Qiu, and Z.~Wei, ``Vocot: Unleashing visually grounded multi-step reasoning in large multi-modal models,'' May 2024. [Online]. Available: \url{http://arxiv.org/abs/2405.16919}
\BIBentrySTDinterwordspacing

\bibitem{zhang2024cocotcontrastivechainthoughtprompting}
\BIBentryALTinterwordspacing
D.~Zhang, J.~Yang, H.~Lyu, Z.~Jin, Y.~Yao, M.~Chen, and J.~Luo, ``Cocot: Contrastive chain-of-thought prompting for large multimodal models with multiple image inputs,'' Jan. 2024. [Online]. Available: \url{http://arxiv.org/abs/2401.02582}
\BIBentrySTDinterwordspacing

\bibitem{zheng2023ddcotdutydistinctchainthoughtprompting}
\BIBentryALTinterwordspacing
G.~Zheng, B.~Yang, J.~Tang, H.-Y. Zhou, and S.~Yang, ``Ddcot: Duty-distinct chain-of-thought prompting for multimodal reasoning in language models,'' Oct. 2023. [Online]. Available: \url{http://arxiv.org/abs/2310.16436}
\BIBentrySTDinterwordspacing

\bibitem{luan2024textcotzoomenhancedmultimodal}
\BIBentryALTinterwordspacing
B.~Luan, H.~Feng, H.~Chen, Y.~Wang, W.~Zhou, and H.~Li, ``Textcot: Zoom in for enhanced multimodal text-rich image understanding,'' Apr. 2024. [Online]. Available: \url{http://arxiv.org/abs/2404.09797}
\BIBentrySTDinterwordspacing

\bibitem{tanboostingpowersmallmultimodal}
C.~Tan, J.~Wei, Z.~Gao, L.~Sun, S.~Li, R.~Guo, B.~Yu, and S.~Z. Li, ``Boosting the power of small multimodal reasoning models to match larger models with self-consistency training.''

\bibitem{qi2024cogcomtrainlargevisionlanguage}
\BIBentryALTinterwordspacing
J.~Qi, M.~Ding, W.~Wang, Y.~Bai, Q.~Lv, W.~Hong, B.~Xu, L.~Hou, J.~Li, Y.~Dong, and J.~Tang, ``Cogcom: Train large vision-language models diving into details through chain of manipulations,'' May 2024. [Online]. Available: \url{http://arxiv.org/abs/2402.04236}
\BIBentrySTDinterwordspacing

\bibitem{menon2024whiteboardthoughtthinkingstepstepmodalities}
\BIBentryALTinterwordspacing
S.~Menon, R.~Zemel, and C.~Vondrick, ``Whiteboard-of-thought: Thinking step-by-step across modalities,'' Jun. 2024. [Online]. Available: \url{http://arxiv.org/abs/2406.14562}
\BIBentrySTDinterwordspacing

\bibitem{wei2024mccotmodularcollaborativecot}
\BIBentryALTinterwordspacing
L.~Wei, W.~Wang, X.~Shen, Y.~Xie, Z.~Fan, X.~Zhang, Z.~Wei, and W.~Chen, ``Mc-cot: A modular collaborative cot framework for zero-shot medical-vqa with llm and mllm integration,'' Oct. 2024. [Online]. Available: \url{http://arxiv.org/abs/2410.04521}
\BIBentrySTDinterwordspacing

\bibitem{xin2024rallerobustcodeclanguage}
\BIBentryALTinterwordspacing
D.~Xin, X.~Tan, K.~Shen, Z.~Ju, D.~Yang, Y.~Wang, S.~Takamichi, H.~Saruwatari, S.~Liu, J.~Li, and S.~Zhao, ``Rall-e: Robust codec language modeling with chain-of-thought prompting for text-to-speech synthesis,'' May 2024. [Online]. Available: \url{http://arxiv.org/abs/2404.03204}
\BIBentrySTDinterwordspacing

\bibitem{gong2024seamlessexpressivelmspeechlanguagemodel}
\BIBentryALTinterwordspacing
H.~Gong and B.~Veluri, ``Seamlessexpressivelm: Speech language model for expressive speech-to-speech translation with chain-of-thought,'' May 2024. [Online]. Available: \url{http://arxiv.org/abs/2405.20410}
\BIBentrySTDinterwordspacing

\bibitem{hu2024chainthoughtpromptingspeechtranslation}
\BIBentryALTinterwordspacing
K.~Hu, Z.~Chen, C.-H.~H. Yang, P.~{\.Z}elasko, O.~Hrinchuk, V.~Lavrukhin, J.~Balam, and B.~Ginsburg, ``Chain-of-thought prompting for speech translation,'' Sep. 2024. [Online]. Available: \url{http://arxiv.org/abs/2409.11538}
\BIBentrySTDinterwordspacing

\bibitem{shao2024deepseekmathpushinglimitsmathematical}
\BIBentryALTinterwordspacing
Z.~Shao, P.~Wang, Q.~Zhu, R.~Xu, J.~Song, X.~Bi, H.~Zhang, M.~Zhang, Y.~K. Li, Y.~Wu, and D.~Guo, ``Deepseekmath: Pushing the limits of mathematical reasoning in open language models,'' 2024. [Online]. Available: \url{https://arxiv.org/abs/2402.03300}
\BIBentrySTDinterwordspacing

\bibitem{schulman2017proximalpolicyoptimizationalgorithms}
\BIBentryALTinterwordspacing
J.~Schulman, F.~Wolski, P.~Dhariwal, A.~Radford, and O.~Klimov, ``Proximal policy optimization algorithms,'' 2017. [Online]. Available: \url{https://arxiv.org/abs/1707.06347}
\BIBentrySTDinterwordspacing

\bibitem{xie2025logicrlunleashingllmreasoning}
\BIBentryALTinterwordspacing
T.~Xie, Z.~Gao, Q.~Ren, H.~Luo, Y.~Hong, B.~Dai, J.~Zhou, K.~Qiu, Z.~Wu, and C.~Luo, ``Logic-rl: Unleashing llm reasoning with rule-based reinforcement learning,'' 2025. [Online]. Available: \url{https://arxiv.org/abs/2502.14768}
\BIBentrySTDinterwordspacing

\bibitem{song2025r1searcherincentivizingsearchcapability}
\BIBentryALTinterwordspacing
H.~Song, J.~Jiang, Y.~Min, J.~Chen, Z.~Chen, W.~X. Zhao, L.~Fang, and J.-R. Wen, ``R1-searcher: Incentivizing the search capability in llms via reinforcement learning,'' 2025. [Online]. Available: \url{https://arxiv.org/abs/2503.05592}
\BIBentrySTDinterwordspacing

\bibitem{huang2025visionr1incentivizingreasoningcapability}
\BIBentryALTinterwordspacing
W.~Huang, B.~Jia, Z.~Zhai, S.~Cao, Z.~Ye, F.~Zhao, Z.~Xu, Y.~Hu, and S.~Lin, ``Vision-r1: Incentivizing reasoning capability in multimodal large language models,'' 2025. [Online]. Available: \url{https://arxiv.org/abs/2503.06749}
\BIBentrySTDinterwordspacing

\bibitem{peng2025lmmr1empowering3blmms}
\BIBentryALTinterwordspacing
Y.~Peng, G.~Zhang, M.~Zhang, Z.~You, J.~Liu, Q.~Zhu, K.~Yang, X.~Xu, X.~Geng, and X.~Yang, ``Lmm-r1: Empowering 3b lmms with strong reasoning abilities through two-stage rule-based rl,'' 2025. [Online]. Available: \url{https://arxiv.org/abs/2503.07536}
\BIBentrySTDinterwordspacing

\bibitem{masterman2024landscapeemergingaiagent}
\BIBentryALTinterwordspacing
T.~Masterman, S.~Besen, M.~Sawtell, and A.~Chao, ``The landscape of emerging ai agent architectures for reasoning, planning, and tool calling: A survey,'' Apr. 2024. [Online]. Available: \url{http://arxiv.org/abs/2404.11584}
\BIBentrySTDinterwordspacing

\bibitem{cheng2024seeclickharnessingguigrounding}
\BIBentryALTinterwordspacing
K.~Cheng, Q.~Sun, Y.~Chu, F.~Xu, Y.~Li, J.~Zhang, and Z.~Wu, ``Seeclick: Harnessing gui grounding for advanced visual gui agents,'' 2024. [Online]. Available: \url{https://arxiv.org/abs/2401.10935}
\BIBentrySTDinterwordspacing

\bibitem{baechler2024screenaivisionlanguagemodelui}
\BIBentryALTinterwordspacing
G.~Baechler, S.~Sunkara, M.~Wang, F.~Zubach, H.~Mansoor, V.~Etter, V.~Cărbune, J.~Lin, J.~Chen, and A.~Sharma, ``Screenai: A vision-language model for ui and infographics understanding,'' 2024. [Online]. Available: \url{https://arxiv.org/abs/2402.04615}
\BIBentrySTDinterwordspacing

\bibitem{shaw2023pixelsuiactionslearning}
\BIBentryALTinterwordspacing
P.~Shaw, M.~Joshi, J.~Cohan, J.~Berant, P.~Pasupat, H.~Hu, U.~Khandelwal, K.~Lee, and K.~Toutanova, ``From pixels to ui actions: Learning to follow instructions via graphical user interfaces,'' 2023. [Online]. Available: \url{https://arxiv.org/abs/2306.00245}
\BIBentrySTDinterwordspacing

\bibitem{qin2024mp5multimodalopenendedembodied}
\BIBentryALTinterwordspacing
Y.~Qin, E.~Zhou, Q.~Liu, Z.~Yin, L.~Sheng, R.~Zhang, Y.~Qiao, and J.~Shao, ``Mp5: A multi-modal open-ended embodied system in minecraft via active perception,'' 2024. [Online]. Available: \url{https://arxiv.org/abs/2312.07472}
\BIBentrySTDinterwordspacing

\bibitem{wang2024mobileagentv2mobiledeviceoperation}
\BIBentryALTinterwordspacing
J.~Wang, H.~Xu, H.~Jia, X.~Zhang, M.~Yan, W.~Shen, J.~Zhang, F.~Huang, and J.~Sang, ``Mobile-agent-v2: Mobile device operation assistant with effective navigation via multi-agent collaboration,'' Jun. 2024. [Online]. Available: \url{http://arxiv.org/abs/2406.01014}
\BIBentrySTDinterwordspacing

\bibitem{hu2024selfevolvingmultiagentcollaborationnetworks}
\BIBentryALTinterwordspacing
Y.~Hu, Y.~Cai, Y.~Du, X.~Zhu, X.~Liu, Z.~Yu, Y.~Hou, S.~Tang, and S.~Chen, ``Self-evolving multi-agent collaboration networks for software development,'' 2024. [Online]. Available: \url{https://arxiv.org/abs/2410.16946}
\BIBentrySTDinterwordspacing

\bibitem{2023arXiv231107562Y}
A.~{Yan}, Z.~{Yang}, W.~{Zhu}, K.~{Lin}, L.~{Li}, J.~{Wang}, J.~{Yang}, Y.~{Zhong}, J.~{McAuley}, J.~{Gao}, Z.~{Liu}, and L.~{Wang}, ``{GPT-4V in Wonderland: Large Multimodal Models for Zero-Shot Smartphone GUI Navigation},'' \emph{arXiv e-prints}, p. arXiv:2311.07562, Nov. 2023.

\bibitem{lai_autowebglm_2024}
\BIBentryALTinterwordspacing
H.~Lai, X.~Liu, I.~L. Iong, S.~Yao, Y.~Chen, P.~Shen, H.~Yu, H.~Zhang, X.~Zhang, Y.~Dong, and J.~Tang, ``{AutoWebGLM}: {A} {Large} {Language} {Model}-based {Web} {Navigating} {Agent},'' Oct. 2024, arXiv:2404.03648 TLDR: Inspired by human browsing patterns, an HTML simplification algorithm is designed to represent webpages, preserving vital information succinctly and bootstrap the model by reinforcement learning and rejection sampling to further facilitate webpage comprehension, browser operations, and efficient task decomposition by itself. [Online]. Available: \url{http://arxiv.org/abs/2404.03648}
\BIBentrySTDinterwordspacing

\bibitem{2024arXiv240204476K}
J.~{Kil}, C.~H. {Song}, B.~{Zheng}, X.~{Deng}, Y.~{Su}, and W.-L. {Chao}, ``{Dual-View Visual Contextualization for Web Navigation},'' \emph{arXiv e-prints}, p. arXiv:2402.04476, Feb. 2024.

\bibitem{2024arXiv241008164A}
S.~{Agashe}, J.~{Han}, S.~{Gan}, J.~{Yang}, A.~{Li}, and X.~E. {Wang}, ``{Agent S: An Open Agentic Framework that Uses Computers Like a Human},'' \emph{arXiv e-prints}, p. arXiv:2410.08164, Oct. 2024.

\bibitem{2023arXiv230911436Z}
Z.~{Zhang} and A.~{Zhang}, ``{You Only Look at Screens: Multimodal Chain-of-Action Agents},'' \emph{arXiv e-prints}, p. arXiv:2309.11436, Sep. 2023.

\bibitem{2024arXiv240302713Z}
J.~{Zhang}, J.~{Wu}, Y.~{Teng}, M.~{Liao}, N.~{Xu}, X.~{Xiao}, Z.~{Wei}, and D.~{Tang}, ``{Android in the Zoo: Chain-of-Action-Thought for GUI Agents},'' \emph{arXiv e-prints}, p. arXiv:2403.02713, Mar. 2024.

\bibitem{shi2017world}
T.~Shi, A.~Karpathy, L.~Fan, J.~Hernandez, and P.~Liang, ``World of bits: An open-domain platform for web-based agents,'' in \emph{International Conference on Machine Learning}.\hskip 1em plus 0.5em minus 0.4em\relax PMLR, 2017, pp. 3135--3144.

\bibitem{deka2017rico}
B.~Deka, Z.~Huang, C.~Franzen, J.~Hibschman, D.~Afergan, Y.~Li, J.~Nichols, and R.~Kumar, ``Rico: A mobile app dataset for building data-driven design applications,'' in \emph{Proceedings of the 30th annual ACM symposium on user interface software and technology}, 2017, pp. 845--854.

\bibitem{liu2018reinforcement}
E.~Z. Liu, K.~Guu, P.~Pasupat, T.~Shi, and P.~Liang, ``Reinforcement learning on web interfaces using workflow-guided exploration,'' \emph{arXiv preprint arXiv:1802.08802}, 2018.

\bibitem{li2020mapping}
Y.~Li, J.~He, X.~Zhou, Y.~Zhang, and J.~Baldridge, ``Mapping natural language instructions to mobile ui action sequences,'' \emph{arXiv preprint arXiv:2005.03776}, 2020.

\bibitem{chen2021websrc}
X.~Chen, Z.~Zhao, L.~Chen, D.~Zhang, J.~Ji, A.~Luo, Y.~Xiong, and K.~Yu, ``Websrc: A dataset for web-based structural reading comprehension,'' \emph{arXiv preprint arXiv:2101.09465}, 2021.

\bibitem{burns2022dataset}
A.~Burns, D.~Arsan, S.~Agrawal, R.~Kumar, K.~Saenko, and B.~A. Plummer, ``A dataset for interactive vision-language navigation with unknown command feasibility,'' in \emph{European Conference on Computer Vision}.\hskip 1em plus 0.5em minus 0.4em\relax Springer, 2022, pp. 312--328.

\bibitem{sun2022meta}
L.~Sun, X.~Chen, L.~Chen, T.~Dai, Z.~Zhu, and K.~Yu, ``Meta-gui: Towards multi-modal conversational agents on mobile gui,'' \emph{arXiv preprint arXiv:2205.11029}, 2022.

\bibitem{yao2022webshop}
S.~Yao, H.~Chen, J.~Yang, and K.~Narasimhan, ``Webshop: Towards scalable real-world web interaction with grounded language agents,'' \emph{Advances in Neural Information Processing Systems}, vol.~35, pp. 20\,744--20\,757, 2022.

\bibitem{deng2024multi}
Y.~Deng, X.~Zhang, W.~Zhang, Y.~Yuan, S.-K. Ng, and T.-S. Chua, ``On the multi-turn instruction following for conversational web agents,'' \emph{arXiv preprint arXiv:2402.15057}, 2024.

\bibitem{chen2024webvln}
Q.~Chen, D.~Pitawela, C.~Zhao, G.~Zhou, H.-T. Chen, and Q.~Wu, ``Webvln: Vision-and-language navigation on websites,'' in \emph{Proceedings of the AAAI Conference on Artificial Intelligence}, vol.~38, no.~2, 2024, pp. 1165--1173.

\bibitem{zhang2024mmina}
Z.~Zhang, S.~Tian, L.~Chen, and Z.~Liu, ``Mmina: Benchmarking multihop multimodal internet agents,'' \emph{arXiv preprint arXiv:2404.09992}, 2024.

\bibitem{liu2024visualwebbench}
J.~Liu, Y.~Song, B.~Y. Lin, W.~Lam, G.~Neubig, Y.~Li, and X.~Yue, ``Visualwebbench: How far have multimodal llms evolved in web page understanding and grounding?'' \emph{arXiv preprint arXiv:2404.05955}, 2024.

\bibitem{xie2024osworld}
T.~Xie, D.~Zhang, J.~Chen, X.~Li, S.~Zhao, R.~Cao, T.~J. Hua, Z.~Cheng, D.~Shin, F.~Lei \emph{et~al.}, ``Osworld: Benchmarking multimodal agents for open-ended tasks in real computer environments,'' \emph{arXiv preprint arXiv:2404.07972}, 2024.

\bibitem{zhang2023mobile}
D.~Zhang, H.~Xu, Z.~Zhao, L.~Chen, R.~Cao, and K.~Yu, ``Mobile-env: an evaluation platform and benchmark for llm-gui interaction,'' \emph{arXiv preprint arXiv:2305.08144}, 2023.

\bibitem{lin2024videogui}
K.~Q. Lin, L.~Li, D.~Gao, Q.~WU, M.~Yan, Z.~Yang, L.~Wang, and M.~Z. Shou, ``Videogui: A benchmark for gui automation from instructional videos,'' \emph{arXiv preprint arXiv:2406.10227}, 2024.

\bibitem{chen2024guicourse}
W.~Chen, J.~Cui, J.~Hu, Y.~Qin, J.~Fang, Y.~Zhao, C.~Wang, J.~Liu, G.~Chen, Y.~Huo \emph{et~al.}, ``Guicourse: From general vision language models to versatile gui agents,'' \emph{arXiv preprint arXiv:2406.11317}, 2024.

\bibitem{chen2024gui}
D.~Chen, Y.~Huang, S.~Wu, J.~Tang, L.~Chen, Y.~Bai, Z.~He, C.~Wang, H.~Zhou, Y.~Li \emph{et~al.}, ``Gui-world: A dataset for gui-oriented multimodal llm-based agents,'' \emph{arXiv preprint arXiv:2406.10819}, 2024.

\bibitem{lu2024gui}
Q.~Lu, W.~Shao, Z.~Liu, F.~Meng, B.~Li, B.~Chen, S.~Huang, K.~Zhang, Y.~Qiao, and P.~Luo, ``Gui odyssey: A comprehensive dataset for cross-app gui navigation on mobile devices,'' \emph{arXiv preprint arXiv:2406.08451}, 2024.

\bibitem{pan2024webcanvas}
Y.~Pan, D.~Kong, S.~Zhou, C.~Cui, Y.~Leng, B.~Jiang, H.~Liu, Y.~Shang, S.~Zhou, T.~Wu \emph{et~al.}, ``Webcanvas: Benchmarking web agents in online environments,'' \emph{arXiv preprint arXiv:2406.12373}, 2024.

\bibitem{cao2024spider2}
R.~Cao, F.~Lei, H.~Wu, J.~Chen, Y.~Fu, H.~Gao, X.~Xiong, H.~Zhang, Y.~Mao, W.~Hu \emph{et~al.}, ``Spider2-v: How far are multimodal agents from automating data science and engineering workflows?'' \emph{arXiv preprint arXiv:2407.10956}, 2024.

\bibitem{kapoor2024omniact}
R.~Kapoor, Y.~P. Butala, M.~Russak, J.~Y. Koh, K.~Kamble, W.~Alshikh, and R.~Salakhutdinov, ``Omniact: A dataset and benchmark for enabling multimodal generalist autonomous agents for desktop and web,'' \emph{arXiv preprint arXiv:2402.17553}, 2024.

\bibitem{chai2024amex}
Y.~Chai, S.~Huang, Y.~Niu, H.~Xiao, L.~Liu, D.~Zhang, P.~Gao, S.~Ren, and H.~Li, ``Amex: Android multi-annotation expo dataset for mobile gui agents,'' \emph{arXiv preprint arXiv:2407.17490}, 2024.

\bibitem{zhang2024llamatouch}
L.~Zhang, S.~Wang, X.~Jia, Z.~Zheng, Y.~Yan, L.~Gao, Y.~Li, and M.~Xu, ``Llamatouch: A faithful and scalable testbed for mobile ui automation task evaluation,'' \emph{arXiv preprint arXiv:2404.16054}, 2024.

\bibitem{xing2024understanding}
M.~Xing, R.~Zhang, H.~Xue, Q.~Chen, F.~Yang, and Z.~Xiao, ``Understanding the weakness of large language model agents within a complex android environment,'' in \emph{Proceedings of the 30th ACM SIGKDD Conference on Knowledge Discovery and Data Mining}, 2024, pp. 6061--6072.

\bibitem{bonatti2024windows}
R.~Bonatti, D.~Zhao, F.~Bonacci, D.~Dupont, S.~Abdali, Y.~Li, Y.~Lu, J.~Wagle, K.~Koishida, A.~Bucker \emph{et~al.}, ``Windows agent arena: Evaluating multi-modal os agents at scale,'' \emph{arXiv preprint arXiv:2409.08264}, 2024.

\bibitem{lu2024weblinx}
X.~H. L{\`u}, Z.~Kasner, and S.~Reddy, ``Weblinx: Real-world website navigation with multi-turn dialogue,'' \emph{arXiv preprint arXiv:2402.05930}, 2024.

\bibitem{zheng2024agentstudio}
L.~Zheng, Z.~Huang, Z.~Xue, X.~Wang, B.~An, and S.~Yan, ``Agentstudio: A toolkit for building general virtual agents,'' \emph{arXiv preprint arXiv:2403.17918}, 2024.

\bibitem{lee2024benchmarking}
J.~Lee, T.~Min, M.~An, D.~Hahm, H.~Lee, C.~Kim, and K.~Lee, ``Benchmarking mobile device control agents across diverse configurations,'' \emph{arXiv preprint arXiv:2404.16660}, 2024.

\bibitem{xu2024crab}
T.~Xu, L.~Chen, D.-J. Wu, Y.~Chen, Z.~Zhang, X.~Yao, Z.~Xie, Y.~Chen, S.~Liu, B.~Qian \emph{et~al.}, ``Crab: Cross-environment agent benchmark for multimodal language model agents,'' \emph{arXiv preprint arXiv:2407.01511}, 2024.

\bibitem{fan2024read}
Y.~Fan, L.~Ding, C.-C. Kuo, S.~Jiang, Y.~Zhao, X.~Guan, J.~Yang, Y.~Zhang, and X.~E. Wang, ``Read anywhere pointed: Layout-aware gui screen reading with tree-of-lens grounding,'' \emph{arXiv preprint arXiv:2406.19263}, 2024.

\bibitem{chen2025spabenchcomprehensivebenchmarksmartphone}
\BIBentryALTinterwordspacing
J.~Chen, D.~Yuen, B.~Xie, Y.~Yang, G.~Chen, Z.~Wu, L.~Yixing, X.~Zhou, W.~Liu, S.~Wang, K.~Zhou, R.~Shao, L.~Nie, Y.~Wang, J.~Hao, J.~Wang, and K.~Shao, ``Spa-bench: A comprehensive benchmark for smartphone agent evaluation,'' 2025. [Online]. Available: \url{https://arxiv.org/abs/2410.15164}
\BIBentrySTDinterwordspacing

\bibitem{li2024effects}
W.~Li, W.~Bishop, A.~Li, C.~Rawles, F.~Campbell-Ajala, D.~Tyamagundlu, and O.~Riva, ``On the effects of data scale on computer control agents,'' \emph{arXiv preprint arXiv:2406.03679}, 2024.

\bibitem{liu2024harnessing}
J.~Liu, T.~Ou, Y.~Song, Y.~Qu, W.~Lam, C.~Xiong, W.~Chen, G.~Neubig, and X.~Yue, ``Harnessing webpage uis for text-rich visual understanding,'' \emph{arXiv preprint arXiv:2410.13824}, 2024.

\bibitem{zhao2024guitestingarenaunified}
\BIBentryALTinterwordspacing
K.~Zhao, J.~Song, L.~Sha, H.~Shen, Z.~Chen, T.~Zhao, X.~Liang, and J.~Yin, ``Gui testing arena: A unified benchmark for advancing autonomous gui testing agent,'' 2024. [Online]. Available: \url{https://arxiv.org/abs/2412.18426}
\BIBentrySTDinterwordspacing

\bibitem{chai2025a3androidagentarena}
\BIBentryALTinterwordspacing
Y.~Chai, H.~Li, J.~Zhang, L.~Liu, G.~Liu, G.~Wang, S.~Ren, S.~Huang, and H.~Li, ``A3: Android agent arena for mobile gui agents,'' 2025. [Online]. Available: \url{https://arxiv.org/abs/2501.01149}
\BIBentrySTDinterwordspacing

\bibitem{wu2025webwalkerbenchmarkingllmsweb}
\BIBentryALTinterwordspacing
J.~Wu, W.~Yin, Y.~Jiang, Z.~Wang, Z.~Xi, R.~Fang, L.~Zhang, Y.~He, D.~Zhou, P.~Xie, and F.~Huang, ``Webwalker: Benchmarking llms in web traversal,'' 2025. [Online]. Available: \url{https://arxiv.org/abs/2501.07572}
\BIBentrySTDinterwordspacing

\bibitem{zhao2025worldguidynamictestingcomprehensive}
\BIBentryALTinterwordspacing
H.~H. Zhao, D.~Gao, and M.~Z. Shou, ``Worldgui: Dynamic testing for comprehensive desktop gui automation,'' 2025. [Online]. Available: \url{https://arxiv.org/abs/2502.08047}
\BIBentrySTDinterwordspacing

\bibitem{nayak2025uivisiondesktopcentricguibenchmark}
\BIBentryALTinterwordspacing
S.~Nayak, X.~Jian, K.~Q. Lin, J.~A. Rodriguez, M.~Kalsi, R.~Awal, N.~Chapados, M.~T. Özsu, A.~Agrawal, D.~Vazquez, C.~Pal, P.~Taslakian, S.~Gella, and S.~Rajeswar, ``Ui-vision: A desktop-centric gui benchmark for visual perception and interaction,'' 2025. [Online]. Available: \url{https://arxiv.org/abs/2503.15661}
\BIBentrySTDinterwordspacing

\end{thebibliography}

 \end{document}